\documentclass[12pt,preprint]{aastex}

\newcommand{\Ampere}{\mbox{Amp\`{e}re}}
\newcommand{\etal}{\mbox{et al.}}
\newcommand{\udg}{\mbox{$^{\dag}$}}

\newcommand{\be}{\begin{equation}}
\newcommand{\ee}{\end{equation}}
\newcommand{\bea}{\begin{eqnarray}}
\newcommand{\eea}{\end{eqnarray}}

\newcommand{\vecb}{\mbox{${\bf b}$}}
\newcommand{\vecn}{\mbox{${\bf n}$}}
\newcommand{\vecu}{\mbox{${\bf u}$}}
\newcommand{\vecv}{\mbox{${\bf v}$}}
\newcommand{\vecx}{\mbox{${\bf x}$}}
\newcommand{\vecy}{\mbox{${\bf y}$}}
\newcommand{\vecz}{\mbox{${\bf z}$}}
\newcommand{\vecB}{\mbox{${\bf B}$}}
\newcommand{\vecD}{\mbox{${\bf D}$}}
\newcommand{\vecE}{\mbox{${\bf E}$}}
\newcommand{\vecF}{\mbox{${\bf F}$}}

\newcommand{\vecJ}{\mbox{${\bf J}$}}
\newcommand{\vecK}{\mbox{${\bf K}$}}
\newcommand{\vectau}{\mbox{\boldmath$\tau$}}

\newcommand{\vecphi}{\mbox{\boldmath$\phi$}}

\newcommand{\xhat}{\mbox{$\hat{\vecx}$}}
\newcommand{\yhat}{\mbox{$\hat{\vecy}$}}
\newcommand{\zhat}{\mbox{$\hat{\vecz}$}}
\newcommand{\nhat}{\mbox{$\hat{\vecn}$}}

\newcommand{\matrA}{\mbox{${\sf A}$}}
\newcommand{\matrh}{\mbox{${\sf h}$}}

\newcommand{\ltsim}{\mbox{$\la$}}
\newcommand{\gtsim}{\mbox{$\ga$}}
\newcommand{\vdot}{\mbox{\boldmath$\cdot$}}
\newcommand{\cross}{\mbox{\boldmath$\times$}}
\newcommand{\grad}{\mbox{\boldmath$\nabla$}}
\newcommand{\curl}{\mbox{\boldmath$\grad\cross$}}
\newcommand{\goesto}{\mbox{$\longrightarrow$}}

\newcommand{\cmMMM}{\mbox{cm$^{-3}$}}
\newcommand{\kms}{\mbox{km\,s$^{-1}$}}

\newcommand{\sM}{\mbox{s$^{-1}$}}
\newcommand{\gcmMM}{\mbox{g\,cm$^{-2}$}}
\newcommand{\gcmMMM}{\mbox{g\,cm$^{-3}$}}
\newcommand{\ergsM}{\mbox{erg\,s$^{-1}$}}

\newcommand{\Msol}{\mbox{$M_{\odot}$}}
\newcommand{\mH}{\mbox{$m_{\rm H}$}}
\newcommand{\kB}{\mbox{$k_{\rm B}$}}

\newcommand{\Htwo}{\mbox{H$_2$}}

\newcommand{\lamh}{\mbox{$\lambda_{\rm h}$}}

\newcommand{\eng}{\mbox{$n_{\rm g}$}}
\newcommand{\mug}{\mbox{$\mu_{\rm g}$}}
\newcommand{\enH}{\mbox{$n_{\rm H}$}}
\newcommand{\enn}{\mbox{$n_{\rm n}$}}
\newcommand{\emn}{\mbox{$m_{\rm n}$}}

\newcommand{\rhog}{\mbox{$\rho_{\rm g}$}}

\newcommand{\xe}{\mbox{$x_{\rm e}$}}

\newcommand{\mui}{\mbox{$\mu_{\rm i}$}}
\newcommand{\eni}{\mbox{$n_{\rm i}$}}
\newcommand{\rhoi}{\mbox{$\rho_{\rm i}$}}
\newcommand{\emi}{\mbox{$m_{\rm i}$}}

\newcommand{\Zi}{\mbox{$Z_{\rm i}$}}
\newcommand{\vecvi}{\mbox{$\vecv_{\rm i}$}}
\newcommand{\betai}{\mbox{$\beta_{\rm i}$}}

\newcommand{\betae}{\mbox{$\beta_{\rm e}$}}
\newcommand{\sigin}{\mbox{$\left<\sigma v\right>_{\rm in}$}}
\newcommand{\tauin}{\mbox{$\tau_{\rm in}$}}
\newcommand{\tauen}{\mbox{$\tau_{\rm en}$}}

\newcommand{\chid}{\mbox{$\chi_{\rm d}$}}
\newcommand{\chisd}{\mbox{$\chi_{\rm sd}$}}
\newcommand{\rhos}{\mbox{$\rho_{\rm s}$}}
\newcommand{\taudn}{\mbox{$\tau_{\rm dn}$}}

\newcommand{\Lsf}{\mbox{$L_{\rm sf}$}}
\newcommand{\Lb}{\mbox{$L_{\rm body}$}}
\newcommand{\tadv}{\mbox{$\tau_{\rm adv}$}}
\newcommand{\tdif}{\mbox{$\tau_{\rm dif}$}}
\newcommand{\tacc}{\mbox{$\tau_{\rm acc}$}}

\newcommand{\tdifeb}{\mbox{$\tau_{\rm EB}$}}

\newcommand{\etao}{\mbox{$\eta_{\rm o}$}}
\newcommand{\etaa}{\mbox{$\eta_{\rm a}$}}
\newcommand{\etah}{\mbox{$\eta_{\rm h}$}}
\newcommand{\mup}{\mbox{$\mu_{\rm p}$}}

\newcommand{\mub}{\mbox{$\mu_{\rm b}$}}
\newcommand{\epsb}{\mbox{$\epsilon_{\rm b}$}}
\newcommand{\sigb}{\mbox{$\sigma_{\rm b}$}}

\newcommand{\modx}{\mbox{$\left|\vecx\right|$}}
\newcommand{\zp}{\mbox{$z^{\prime}$}}

\newcommand{\vecvzero}{\mbox{$\vecv_0$}}
\newcommand{\vzero}{\mbox{$v_0$}} 
\newcommand{\vx}{\mbox{$v_x$}}
\newcommand{\vy}{\mbox{$v_y$}}

\newcommand{\vecEpprm}{\mbox{$\vecE_{\rm p}^{\prime}$}}

\newcommand{\vecBzero}{\mbox{$\vecB_0$}}
\newcommand{\vecBone}{\mbox{$\vecB_1$}}
\newcommand{\Bzero}{\mbox{$B_0$}}

\newcommand{\Bz}{\mbox{$B_z$}}
\newcommand{\Bonex}{\mbox{$B_{1x}$}}
\newcommand{\Boney}{\mbox{$B_{1y}$}}

\newcommand{\vecBp}{\mbox{$\vecB_{\rm p}$}}
\newcommand{\vecBb}{\mbox{$\vecB_{\rm b}$}}

\newcommand{\vecEzero}{\mbox{$\vecE_0$}}
\newcommand{\Ezero}{\mbox{$E_0$}}

\newcommand{\vecEp}{\mbox{$\vecE_{\rm p}$}}
\newcommand{\vecEb}{\mbox{$\vecE_{\rm b}$}}
\newcommand{\vecEm}{\mbox{$\vecE_{\rm m}$}}
\newcommand{\vecEo}{\mbox{$\vecE_{\rm o}$}}
\newcommand{\vecEh}{\mbox{$\vecE_{\rm h}$}}
\newcommand{\vecEa}{\mbox{$\vecE_{\rm a}$}}

\newcommand{\Bstar}{\mbox{$B_*$}}
\newcommand{\zbar}{\mbox{$\bar{z}$}}
\newcommand{\ux}{\mbox{$u_x$}}
\newcommand{\uy}{\mbox{$u_y$}}
\newcommand{\uxp}{\mbox{$u_x^{\prime}$}}
\newcommand{\uyp}{\mbox{$u_y^{\prime}$}}
\newcommand{\uxpp}{\mbox{$u_x^{\prime\prime}$}}
\newcommand{\uypp}{\mbox{$u_y^{\prime\prime}$}}

\newcommand{\bxp}{\mbox{$b_x^{\prime}$}}
\newcommand{\byp}{\mbox{$b_y^{\prime}$}}
\newcommand{\bxpp}{\mbox{$b_x^{\prime\prime}$}}
\newcommand{\bypp}{\mbox{$b_y^{\prime\prime}$}}

\newcommand{\eps}{\mbox{$\epsilon$}}
\newcommand{\km}{\mbox{$k_-$}}
\newcommand{\kp}{\mbox{$k_+$}}
\newcommand{\kR}{\mbox{$k_{\rm R}$}}
\newcommand{\kI}{\mbox{$k_{\rm I}$}}

\newcommand{\Cm}{\mbox{$C_-$}}
\newcommand{\Cp}{\mbox{$C_+$}}
\newcommand{\Am}{\mbox{$A_-$}}
\newcommand{\Ap}{\mbox{$A_+$}}

\slugcomment{To appear in \apj}

\shorttitle{Induction Heating in Protoplanetary Disks}
\shortauthors{Menzel \& Roberge}

\begin{document}

\title{
Reexamination of Induction Heating of Primitive Bodies in Protoplanetary Disks
}

\author{
Raymond L.\ Menzel\altaffilmark{1}
and
Wayne G.\ Roberge\altaffilmark{1}
}

\affil{New York Center for Astrobiology
\\ and \\
Department of Physics, Applied Physics and Astronomy,
Rensselaer Polytechnic Institute,  110 8th Street, Troy, NY 12180, USA}

\email{menzer@rpi.edu, roberw@rpi.edu}

\begin{abstract}
We reexamine the unipolar induction mechanism for heating asteroids originally 
proposed in a classic series of papers by Sonett and collaborators.  As 
originally conceived, induction heating is caused by the ``motional electric 
field'' which appears in the frame of an asteroid immersed in a fully-ionized, 
magnetized solar wind and drives currents through its interior.  However we 
point out that classical induction heating contains a subtle conceptual error, 
in consequence of which the electric field inside the asteroid was calculated 
incorrectly.  The problem is that the motional electric field used by Sonett et 
al.\ is the electric field in the freely streaming plasma far from the asteroid; 
in fact the motional field vanishes at the asteroid surface for realistic 
assumptions about the plasma density.  In this paper we revisit and improve the 
induction heating scenario by:  (1) correcting the conceptual error by self 
consistently calculating the electric field in and around the boundary layer at 
the asteroid-plasma interface; (2) considering weakly-ionized plasmas consistent 
with current ideas about protoplanetary disks; and (3) considering more 
realistic scenarios which do not require a fully ionized, powerful T~Tauri wind in
the disk midplane.  We present exemplary solutions 
for two highly idealized flows which show that the interior electric field can either 
vanish or be comparable to the fields predicted by classical induction 
depending on the flow geometry.  We term the heating driven by these flows 
``electrodynamic heating", calculate its upper limits, and compare them to 
heating produced by short-lived radionuclides. 
\end{abstract}


\section{Introduction}
\label{sec-intro}

The mineralogy of asteroids inferred from studies of meteorites and asteroid spectroscopy
implies that the asteroids were heated during the first few millions years of their lifetimes
(e.g., Ghosh \etal\ 2006 and references therein).
The degree of heating depended strongly on heliocentric distance.
Igneous-type asteroids at the inner edge of the asteroid belt show signs of heating to $>$\,1200\,K
and are thought to be pieces of the exposed metallic cores of differentiated asteroids (Keil 2000).
At the outer edge of the asteroid belt no heating occurred and the asteroids there are thought
to be composed of unaltered, primitive solar system materials (McKinnon 1989).
Near the center of the asteroid belt, metamorphic type asteroids show mineralogical evidence of aqueous
alteration at temperatures of 300--450\,K and
thermal metamorphism at temperatures up to 1200\,K (Keil 2000 and references therein).
Some of the asteroids in this region produce fragments which reach Earth
in the form of carbonaceous chondrites.
The latter may contain innumerable organic compounds, including small amounts of amino acids, the
distribution of which depends on the severity of thermal alteration inside
the parent body (Glavin et al.\ 2011).
In addition, excesses of L-type amino acids have been found in aqueously altered meteorites
(Cronin \& Pizzarello 1997; Glavin \& Dworkin 2009).
Because asteroids may have provided significant amounts
of organic material to the early Earth (Chyba \& Sagan 1992),
understanding how asteroids were heated may be central to understanding the origin of life.        

Two theories have been proposed to explain asteroid heating: the decay of short-lived
radionuclides (SLRs, Urey 1955) and unipolar induction heating (Sonett et al.\ 1970).
Heating by SLRs such as $^{\rm{26}}$Al is a widely accepted scenario.
The discovery of an excess abundance of the daughter nuclide $^{\rm{26}}$Mg in calcium-aluminum rich inclusions (CAIs)
from the Allende meteorite implies an $^{26}$Al abundance large enough to cause significant
heating or even melting (Wasserburg et al.\ 1977; Lee et al.\ 1977).
However the gradient of heating across the asteroid belt requires a finely tuned
dependence of asteroid accretion times on heliocentric distance.
For example Grimm $\&$ McSween (1993) were able to reproduce the observed gradient by assuming that all bodies at each specific heliocentric distance $R$ inside the asteroid belt accreted instantaneously, regardless of size,  at a time $\tau_{\rm{ac}}$ relative to CAI formation determined by the equation
\be
\tau_{\rm{ac}} \propto R\,^{n},
\ee
where $n$ is an adjustable parameter ranging from 1.5 to 3.
Other models have attempted to relax this assumption; however the later models predict that most
of the mass of the asteroid belt would be contained in small bodies that would not achieve melting or thermal metamorphism
(McSween et al.\ 2002).
This would imply an unobserved excess of small, unheated asteroids
in all parts of the asteroid belt.
Furthermore, although $^{26}$Al heating may have been important in our solar system,
the SLR scenario is unlikely to occur elsewhere: Ouellette et al.\ (2010) predict that the probability
of another protoplanetary disk receiving the same concentration of SLRs is only $\sim 10^{-3}$--$10^{-2}$.

For these reasons we revisit the unipolar induction heating mechanism described in Sonett et al.\ (1970).
In their scenario an asteroid or similarly unmagnetized body is immersed in a uniform, fully-ionized
solar wind with velocity \vecvzero, magnetic field $\vecB=\vecBzero$, and electric field $\vecE=0$ in the wind's rest frame.
Sonett et al.\ observed that in the frame of the asteroid there appears a motional electric field,
\be
\vecE_{\rm{m}} = - \frac{ \vecv_{0}}{c} \times \vecB_{0} \equiv \vecE_{0}.
\label{eq-SonettmotionEfield}
\ee
If the solid body is not a perfect insulator, a nonzero electric field inside it will drive currents 
which will generate heat via Ohmic dissipation.
Sonett et al.\ calculated the electric field inside the body 
by treating the latter as a dielectric sphere immersed in a uniform electric field \vecEzero.
They assumed that the interior currents are able to form a closed circuit through the plasma
and estimated the distortion of the ambient magnetic field \vecBzero\ by secondary magnetic
fields generated by currents in- and around the body.
However no attempt was made to account for distortions in the velocity field, \vecv, which
was assumed to be \vecvzero\ everywhere. 

Unfortunately the classical induction scenario is based on a subtle misconception.
Expression~($\ref{eq-SonettmotionEfield}$) is a {\em local}\/ relation, in the sense that
\vecEm\ is the electric field, as
measured in the body frame, {\em at the location where the velocity is equal to 
$\vecv_{0}$}.\/  Consequently, \vecEzero\ is the body frame electric field in
the freely streaming plasma.
If the velocity $\vecv$ depends on position, \vecx, then the motional field is
also position dependent. Thus,
\be
\vecEm(\vecx) = - \frac{\vecv(\vecx)}{c} \cross \vecBzero
\ee
is the motional field at position $\vecx$.
Strictly speaking, unipolar induction describes the heating of bodies that do
not perturb the velocity field of the surrounding plasma.
This is a good approximation in the solar system today,\footnote{The
unipolar induction mechanism was originally conceived to explain the magnetic
fields of the Moon and other bodies in the solar system today
(Sonett \& Colburn 1968; Schwartz et al.\ 1969) and later applied to asteroids.}
where the collision
mean free path exceeds the size of any solid body by many orders of magnitude.
However the collision mean free path was of order meters in the solar nebula
so the distortion of the velocity field cannot be ignored in calculations of asteroid heating.
In general, friction between the plasma and body surface will cause the formation of a shear layer,
in which \vecv\ decreases systematically to zero at the body surface.
It follows that the motional electric field also vanishes at the body surface.

In the following sections we show that, while the motional electric field 
vanishes on and inside a large body, the total electric field may not.
In Section~\ref{sec-goveq} we give the equations which govern the velocity and 
electromagnetic fields in and around a body immersed in a flowing plasma.
We adopt a multifluid, magnetohydrodynamic description appropriate for  weakly 
ionized protoplanetary disks.  In this paper we make no assumptions about the 
origin of the flow, which could be due to the body's orbital motion 
(Weidenschilling 1977a; Morris et al.\ 2012) or passing shock waves (e.g., Desch \&
Connolly\ 2002; Miura \& Nakamoto\ 2006; Hood et al.\ 2009).  
Section~\ref{sec-tcoeff} describes our disk model and the method used to
calculate the abundances of charged particles and transport coefficients in the
disk midplane.
In Section~{\ref{sec-slab} we present 
two simple examples which show that, although the motional electric field 
vanishes at the body surface, a non-zero total electric field can exist due to 
magnetic field gradients in the flow set up by Ohmic dissipation, the Hall 
effect, and ambipolar diffusion. The effects of small dust grains and possible 
relevance of our results to asteroid heating are discussed in 
Section~\ref{sec-disc} and our results are summarized in Section~\ref{sec-summ}.

\section{Multifluid, Magnetohydrodynamic Flow Past an Arbitrary Body}
\label{sec-goveq}

\subsection{Governing Equations for the Shear Flow}
\label{sec-govsf}

Consider the flow of plasma past a large, unmagnetized body in a weakly ionized protoplanetary disk.
Close to the body surface the plasma is distorted by shearing motions.
The dynamics of the shear layer are described by the equations for mass conservation,
\be
\frac{\partial\rho}{\partial t} + \grad\vdot\left(\rho\vecv\right) = 0,
\label{eq-mass}
\ee
momentum conservation,
\be
\rho\,\left[\,\frac{\partial\vecv}{\partial t} + \left(\vecv\vdot\grad\right)\vecv\,\right] ~=~
-\grad\left(\frac{B^2}{8\pi}\right) + \frac{1}{4\pi}\left(\vecB\vdot\grad\right)\vecB
-\grad P + \alpha\nabla^2\vecv,
\label{eq-momentum}
\ee
and energy conservation,
\be
\frac{\partial U}{\partial t} + \grad\vdot\left[\left(U+P\right)\vecv\right] = \Gamma - \Lambda,
\label{eq-energy}
\ee
plus the induction equation for a weakly-ionized multifluid plasma,
\be
\frac{\partial\vecB}{\partial t}
=
\curl \left(\vecv\cross\vecB\right) 
-\curl \left[ \etao \curl\vecB \right]
-\curl \left[ \etah \left( \curl\vecB \right) \cross \hat{\vecB} \right]
-\curl \left[ \etaa \left( \curl\vecB \right)_{\perp} \right]
\label{eq-induction}
\ee
(Wardle 2007; Balbus 2011), 
where $\rho$ is mass density, \vecv\ is fluid velocity, $P$ is thermal pressure,
and $U$ is the density of kinetic plus internal energy.
The quantity $\Gamma-\Lambda$ is the net rate of heating ($\Gamma$)
minus cooling ($\Lambda$) per unit volume due to the absorption and emission of photons,
x-ray absorption, cosmic-ray ionization, etc.
Strictly speaking, \vecv\ is the velocity of the {\em neutral}\/ particles
which form the bulk of the weakly-ionized plasma.
In Equation~(\ref{eq-induction}), $\hat{\vecB}$ is the unit vector parallel to \vecB\ and
$\left(\curl\vecB\right)_{\perp}$ is the component of $\curl\vecB$ perpendicular to \vecB.
The plasma is described by four transport coefficients: the shear
viscosity, $\alpha$, and the diffusivities \etao, \etah, and \etaa\ 
associated with Ohmic dissipation, the Hall effect, and ambipolar
diffusion, respectively.

The solution must satisfy boundary conditions at
infinity and also at the plasma-body interface.
We work in a frame whose origin lies somewhere inside the body and moves with it. 
Then the boundary conditions at infinity are
\be
\lim_{\modx\goesto\infty} ~ 
\left[ \,
\begin{array}{c}
\rho(\vecx,t) \\ P(\vecx,t) \\ U(\vecx,t) \\ \vecv(\vecx,t) \\ \vecB(\vecx,t)
\end{array}
\,\right]
~=~
\left[\,
\begin{array}{c}
\rho_0 \\ P_0 \\ U_0 \\ \vecvzero \\ \vecBzero 
\end{array}
\,\right],
\label{eq-infbc}
\ee
where the subscript ``$0$'' denotes values in the undisturbed plasma.
The free-stream velocity \vecvzero\ and ambient magnetic field \vecBzero\ are parameters.
The ambient density, pressure, and energy density must be obtained from a model of the protoplanetary disk.

At the plasma-body interface the velocity satisfies the no-slip condition,
\be
\vecv=0   
~~~~~~~~~\hbox{[plasma-body interface]},
\ee
and the magnetic induction satisfies 
\be
\nhat \cross \left(\vecB_{\rm p}/\mup - \vecB_{\rm b}/\mub \right)  = \frac{4\pi}{c}\vecK
~~~~~~~~~\hbox{[plasma-body interface]}
\label{eq-htanbc}
\ee
and
\be
\nhat \vdot  \left(\vecB_{\rm p} - \vecB_{\rm b} \right)  = 0
~~~~~~~~~\hbox{[plasma-body interface]}
\label{eq-bnormbc}
\ee
(e.g., Jackson 1975), where $\mu$ is the magnetic permeability and \vecK\ is the surface current.
The unit normal, \nhat, points away from the body and 
quantities with subscripts b (``body'') and p (``plasma'') are evaluated
just in- and outside of the body, respectively, in Equations~(\ref{eq-htanbc})--(\ref{eq-bnormbc}).

Dimensional analysis of Equations~(\ref{eq-mass})--(\ref{eq-induction})
yields the fundamental length- and time scales for the shear flow.
The thickness of the shear layer is predicted to be of order $L_{\rm{sf}}$, where
\be
\Lsf \equiv \frac{\left(\eta\alpha\right)^{1/2}}{B_0}
\sim 10\,
\left(\frac{\eta}{10^{15}\,{\rm cm}^2\,{\rm s}^{-1}}\right)^{1/2}\,
\left(\frac{\alpha}{10^{-5}\,{\rm Poise}}\right)^{1/2}\,
\left(\frac{B_0}{\rm 0.1\,G}\right)^{-1} ~~{\rm km},
\label{eq-lscale}
\ee
and\footnote{It's important to note that in some cases the Hall diffusivity $\etah$ 
can be negative (Wardle\ 2007), in which case the definition given in 
Equation~(\ref{eq-etdef}) would not be useful.  However we adopt this definition
because in our models $\etah$ is positive in the midplane of the disk for all radii of 
interest.
} 
\be
\eta \equiv \etao + \etaa + \etah.
\label{eq-etdef}
\ee
It is important to note that the scaling for $\eta$ in Equation~\ref{eq-lscale} is
appropriate for a dust-free plasma.  If dust is present in the disk, 
the diffusivities and thus $L_{\rm{sf}}$ will change (See Sections~\ref{sec-tcoeff} 
and \ref{sec-efanddust}).
There are three characteristic time scales:
the time scale for viscous forces to accelerate the plasma,
\be
\tacc \equiv \frac{\rho\Lsf^2}{\alpha}
\sim 10^6
\left(\frac{\enH}{10^{13}\,\cmMMM}\right)\,
\left(\frac{\eta}{10^{15}\,{\rm cm}^2\,{\rm s}^{-1}}\right)\,
\left(\frac{B_0}{\rm 0.1\,G}\right)^{-2} ~~~{\rm s},
\label{eq-tacc}
\ee
the advection time scale,
\be
\tadv \equiv \frac{\Lsf}{v} \sim
10\,\left(\frac{\eta}{10^{15}\,{\rm cm}^2\,{\rm s}^{-1}}\right)^{1/2}\,
\left(\frac{\alpha}{10^{-5}\,{\rm Poise}}\right)^{1/2}\,
\left(\frac{B_0}{\rm 0.1\,G}\right)^{-1}\,
\left(\frac{v_0}{\rm km\,s}\right)^{-1} ~~~{\rm s},
\label{eq-tadv}
\ee
and the magnetic diffusion time
\be
\tdif \equiv \frac{\Lsf^2}{\eta} \sim
10^{-3}\,\left(\frac{\alpha}{10^{-5}\,{\rm Poise}}\right)\,
\left(\frac{B_0}{\rm 0.1\,G}\right)^{-2} ~~~{\rm s}.
\label{eq-tdif}
\ee

The scaling in Equations~(\ref{eq-lscale}),~(\ref{eq-tadv}), and~(\ref{eq-tdif}) is appropriate
if $\alpha$ is the {\it molecular}\/ viscosity of the gas (e.g., Schaefer 2010); however the
effective viscosity could be much larger if the flow is turbulent.
Although the transition from laminar to turbulent flow is not well understood for
MHD flow over bodies, this transition has been widely studied for MHD channel flow
(Hartmann 1937; Thess et al.\ 2007 and references therein) and is generally determined by the ratio
\be
F \equiv \frac{{\rm Re}}{{\rm Ha}}
\ee
of the Reynold number,
\be
{\rm Re} \equiv \frac{\rho v L}{\alpha},
\ee
to the Hartmann number,
\be
{\rm Ha} \equiv B L \sqrt{\frac{\sigma_{\rm{o}}}{\alpha}},
\ee
where $v$ is the velocity at the center of the channel, 
$L$ is the channel width, and $\sigma_{\rm{o}}$
is the Ohmic conductivity of the plasma (See Equation~[\ref{eq-sigmao}]).
For MHD channel flow with a uniform magnetic field perpendicular to the walls,
the flow is found to remain laminar if the parameter $F$ remains less than a critical
value $F_{\rm{c}} \sim 100$ determined by both experiments and numerical simulations (Moresco $\&$ Alboussi\`{e}re 2004; Thess et al. 2007).
Although we obviously are not considering MHD channel flow, it seems plausible that the
idealized flows over planar body surfaces (See Section~\ref{sec-slab}}) considered in this paper may behave similarly.
For realistic bodies, $F_{\rm{c}}$ will probably depend greatly on the geometry of the flow;
however here we will assume that $F_{\rm{c}}$
is equal to the above value.
If $\alpha$ is the molecular viscosity of the plasma, we find that
\be
{\rm Re} = \frac{\rho v_{0} L_{\rm{sf}}}{\alpha} \sim 10^{5} 
\left(\frac{n_{\rm{H}}}{10^{13}\,\rm{cm}^{-3}} \right)
\left(\frac{v_{0}}{\rm{km}\, \rm{s}^{-1}} \right)
\left( \frac{\eta}{10^{15}\, \rm{cm} ^{2} \,\rm{s} ^{-1}} \right)^{1/2}
\left(\frac{B_{0}}{0.1\,\rm{G}} \right)^{-1} \left(\frac{\alpha}{10^{-5}\, \rm{Poise}} \right)^{-1/2} ,
\ee
\be
{\rm Ha} = B_{0} L_{\rm{sf}} \sqrt{\frac{\sigma_{\rm{p}}}{\alpha}} \sim 10^{10}
\left(\frac{\eta} {10^{15}\, \rm{cm}^{2}\, \rm{s}^{-1}} \right)^{1/2} 
\left( \frac{\sigma_{\rm{o}}}{10^{6}\, \rm{s}^{-1}} \right)^{1/2}  ~~~,
\ee
and hence that
\be
F = \frac{{\rm Re}}{{\rm Ha}} \sim 10^{-5} 
\left(\frac{n_{\rm{H}}}{10^{13}\,\rm{cm}^{-3}} \right) 
\left(\frac{v_{0}}{\rm{km}\, \rm{s}^{-1}} \right) 
\left( \frac{\sigma_{\rm{o}}}{10^{6}\, \rm{s} ^{-1}} \right)^{-1/2} 
\left(\frac{B_{0}}{0.1\,\rm{G}} \right)^{-1} \left(\frac{\alpha}{10^{-5} \, \rm{Poise}} \right)^{-1/2}  ,
\ee
which shows that $F \ll F_{\rm{c}}$, suggesting that the flows we study are indeed laminar.

If these flows are in fact turbulent, a larger effective viscosity could be implied.  
A probable upper limit on $\alpha$ is set by assuming that the hydromagnetic 
turbulence which likely dominates the large-scale dynamics of a protostellar 
disk extends all the way down to scales of order the body size, $L_{\rm{body}}$.  
If this were true, $\alpha$ would be comparable to
\be
\alpha_{\rm{t}} = \kappa_{\rm{t}} \rho c_{\rm{s}} H
\ee
(Shakura $\&$ Sunyaev 1973), where $\kappa_{\rm{t}}$ is a 
dimensionless number, $c_{\rm{s}}$ is the sound speed, $H=c_{\rm{s}}/\Omega$ is 
the local disk scale height, and $\Omega$ is the local angular velocity.  
Putting in numbers appropriate for the asteroid belt gives
\be
\alpha_{\rm{t}} \sim 10^{4} \left(\frac{\kappa_{\rm{t}}}{0.01} \right) 
\left(\frac{n_{\rm{H}}}{10^{13}\,\rm{cm}^{-3}} \right) 
\left(\frac{T}{100\,\rm{K}} \right)^{1/2} 
\left(\frac{H}{10^{12}\,\rm{cm}} \right) ~~\rm{Poise},
\ee
so that $\alpha_{\rm{t}} \sim 10^{9} \alpha$ and $L_{\rm{sf}} \gg L_{\rm{body}}$ 
for bodies of asteroidal size.  
However even in this highly speculative scenario
the time scales in Equations~(\ref{eq-tacc})--(\ref{eq-tdif}) would all be $\la 100$\,yr., 
and therefore steady flow can still be assumed.

%

The electric field inside the body is coupled to the electric field
in the shear layer via boundary conditions at the body-plasma interface
(see Equations~[\ref{eq-etanbc}]--[\ref{eq-dnormbc}]); thus it is
necessary to calculate the electric field \vecEp\ in the plasma.
Equation (\ref{eq-induction}), Faraday's Law, Helmholtz' Theorem\footnote{A
vector field is uniquely determined if its divergence and curl are both known.},
and macroscopic charge neutrality together imply that in a steady flow
\be
\vecEp = - \frac{\vecv}{c}\cross\vecB ~+~
\frac{\etao}{c}\,\curl\vecB ~+~
\frac{\etah}{c}\,\left[\left(\curl\vecB\right)\cross\hat{\vecB}\right] ~+~
\frac{\etaa}{c}\,\left(\curl\vecB\right)_{\perp}.
\label{eq-eplasma}
\ee
Equation (\ref{eq-eplasma}) says that in general the electric field
at each point in the shear flow is the sum of four contributions,
\be
\vecEp = \vecEm + \vecEo + \vecEh + \vecEa,
\label{eq-eparts}
\ee
of which the first is the motional E-field.
We refer to the other contributions as the Ohm field,
\be
\vecEo \equiv \frac{\etao}{c}\,\curl\vecB,
\label{eq-eohm}
\ee
the Hall field,
\be
\vecEh \equiv \frac{\etah}{c}\,\left[\left(\curl\vecB\right)\cross\hat{\vecB}\right],
\label{eq-ehall}
\ee
and the ambipolar field,
\be
\vecEa \equiv \frac{\etaa}{c}\,\left(\curl\vecB\right)_{\perp},
\label{eq-eamb}
\ee
because they are the electric fields required to maintain the magnetic field
gradients set up by Ohmic dissipation, etc.

The boundary condition on the electric field at infinity is
\be
\lim_{\modx\goesto\infty} \vecEp = \vecEzero.
\ee
The boundary conditions at the body-plasma interface are
\be
\nhat \cross \left(\vecEp-\vecEb\right) = 0,
\label{eq-etanbc}
\ee
and
\be
\nhat \vdot  \left(\vecD_{\rm p} - \vecD_{\rm b}\right)  = 4\pi\Sigma,
\label{eq-dnormbc}
\ee
where $\Sigma$ is the surface charge density, $\vecD=\eps\vecE$ is the dielectric
displacement and we assume, for the present, that the dielectric constant, \eps, is a scalar.
 
\subsection{The Body Interior}
\label{sec-govint}

To calculate the electric and magnetic fields inside the body one must
solve Maxwell's equations there.
The light-crossing time for an asteroid of size $L_{\rm{body}}$,
\be
\tau_{\rm \ell c} = 0.3\,\left(\frac{\Lb}{100\,{\rm km}}\right)  ~~{\rm ms}, 
\ee
is utterly negligible compared to the flow time scales (Section~\ref{sec-govsf}),
so the fields inside the body are well approximated by Faraday's Law,
\be
\curl\vecEb = -\frac{1}{c}\frac{\partial\vecBb}{\partial t},
\label{eq-faraday}
\ee
and {\Ampere}'s Law with zero displacement current,
\be
\curl\vecBb = \frac{4\pi\sigma_{\rm{b}}\mu_{\rm{b}}}{c}\vecEb ~+~ \frac{1}{\mu_{\rm{b}}}\,\grad\mu_{\rm{b}}\cross\vecBb.
\label{eq-ampere}
\ee
We have assumed that the macroscopic charge density vanishes inside the body and that Ohm's Law
holds with a scalar conductivity $\sigma_{\rm{b}}$.
However Equations~(\ref{eq-faraday})--(\ref{eq-ampere}) make no assumptions about the
position dependence of the material properties $\mu_{\rm{b}}$, $\sigma_{\rm{b}}$, and dielectric constant $\eps_{\rm{b}}$.

Equations (\ref{eq-faraday})--(\ref{eq-ampere}) describe the dissipation
of electromagnetic field energy by Ohmic heating of the body material.
This can be seen by momentarily neglecting the position dependence of $\mu_{\rm{b}}$.
Then Equations~(\ref{eq-faraday}) and (\ref{eq-ampere}) reduce to diffusion equations for
the electric field,
\be
\frac{\partial\vecEb}{\partial t} = \frac{c^2}{4\pi\sigma_{\rm{b}}\mu_{\rm{b}}}\,\nabla^2\vecEb,
\ee
and magnetic field,
\be
\frac{\partial\vecBb}{\partial t} = \frac{c^2}{4\pi\sigma_{\rm{b}}\mu_{\rm{b}}}\,\nabla^2\vecBb.
\ee
In an isolated body, Ohmic dissipation would eventually reduce the electromagnetic
field energy in the body to zero.
However the electric and magnetic fields in the shear layer and surrounding
plasma constitute an energy reservoir which tends to replenish losses
inside the body, via the diffusion of \vecE\ and \vecB\ from the plasma into
the body.
The time to reach a steady state in which diffusive gains balance Ohmic losses is
just the diffusion time,
\be
\tdifeb = \frac{4\pi\sigma_{\rm{b}}\mu_{\rm{b}} L_{\rm body}^2}{c^2}.
\ee

Estimates of \tdifeb\ depend on the very uncertain electrical conductivities of primitive
solar system materials.
For example, estimates of $\sigma_{\rm{b}}$ for plausible asteroidal constituents
span $\sim 16$ orders of magnitude (Ip \& Herbert 1983), and
the situation is further complicated if the body contains a spatially connected metallic
phase (H.\ Watson, private communication).
Direct measurements of $\sigma_{\rm{b}}$ have been carried out for a few
chondritic meteorites (Schwerer et al.\ 1971; Brecher 1973).
They are consistent with an Arrhenius-law temperature dependence,
\be
\sigma_{\rm{b}} = \sigma_0 \, \exp\left(-E_{\rm a}/k_{\rm{B}}T\right),
\label{eq-sigarr}
\ee
with $\sigma_0 \sim 10^{12}\,{\rm s}^{-1}$ and $E_{\rm a} \approx 0.1$\,eV.
In the absence of more extensive measurements we will adopt these values, giving
\be
\tdifeb \sim \mbox{10--10$^3$} \,\mu_{\rm{b}}\,\left(\frac{\Lb}{100\,{\rm km}}\right)^2 ~~~{\rm s}
\label{eq-tdifeb}
\ee
over the temperature range $100$--$200$\,K.
Given the shortness of the times in Equation~(\ref{eq-tdifeb}),
it seems reasonable to neglect the time dependence of \vecEb\ and \vecBb,
uncertainties in $\sigma_{\rm{b}}$ notwithstanding.
Then Faraday's Law reduces to 
\be
\curl\vecEb = 0,
\ee
i.e., finding \vecEb\ inside the body reduces to a boundary-value problem in electrostatics.
Once \vecEb\ is known, {\Ampere}'s Law reduces to a set of coupled ordinary differential
equations for \vecBb.
In Sections~\ref{sec-bpar}--\ref{sec-bperp} we show how this all works out for 
two extremely simple models.

\subsection{Self Heating}
\label{sec-govsh}

Dissipative processes inside the shear flow transform ordered kinetic
energy into heat.
Since the transport coefficients depend on temperature, it is
important to know whether the heating is large enough to
raise the temperature above ambient.
The friction associated with viscosity heats the plasma at a rate
\be
\Gamma_{\rm visc} = \vectau\,{\mathbf :}\,\grad\vecv,
\ee
where \vectau\ is the viscous stress tensor and $\grad\vecv$ is the velocity
gradient tensor.
In Cartesian coordinates
\be
\tau_{ij} = \alpha
\,\left(\,\frac{\partial v_i}{\partial x_j}+\frac{\partial v_j}{\partial x_i}\,\right)
~~~i,j = x,y,z
\ee
and
\be
\Gamma_{\rm visc} = \sum_{i,j}\ \tau_{ij}\,\frac{\partial v_i}{\partial x_j}.
\label{eq-gvisc}
\ee
Taking $v \sim v_0$ and $\partial v_i/\partial x_j \sim v_0/\Lsf$ gives the order-of-magnitude
estimate
\be
\Gamma_{\rm visc} \sim 10^{-7}
\,\left(\frac{\vzero}{\kms}\right)^2
\,\left(\frac{\Bzero}{\rm 0.1\,G}\right)^2
\,\left(\frac{\eta}{10^{15}\,\mbox{cm$^2$\,s$^{-1}$}}\right)^{-1}
~~~\mbox{erg\,cm$^{-3}$\,s$^{-1}$}.
\label{eq-gviscome}
\ee
Notice that the RHS of Equation~(\ref{eq-gviscome}) is independent of the viscosity.

The shear flow is also heated by the friction associated with streaming
motions between charged and neutral particles; however electron-neutral scattering 
can generally be neglected (Balbus\ 2011).
Let \eni\ be the number density of ions with mass \emi\ and charge $\Zi e$
and \enn\ be the number density of neutral particles with mass \emn.
If \vecE\ or \vecB\ is nonzero, the ions will be accelerated by the Lorentz force.
On a very short time scale they reach a terminal velocity with respect to the neutrals,
such that the Lorentz force is balanced by the drag force associated with
elastic ion-neutral scattering.
The terminal drift velocity is
\be
\vecvi - \vecv = c\, \left[\,
\betai\,\frac{\left(\vecBp\vdot\vecEpprm\right)\,\vecBp}{B_{\rm p}^3}
+\frac{\beta_{\rm i}^2}{1+\beta_{\rm i}^2}\,\frac{\vecEpprm\cross\vecBp}{B_{\rm p}^2}
+\frac{\beta_{\rm i}}{1+\beta_{\rm i}^2}\,\frac{\vecBp\cross\left(\vecEpprm\cross\vecBp\right)}{B_{\rm p}^3}
\right]
\label{eq-Wardleiondriftv}
\ee
(Wardle 1998), where \vecvi\ is the ion velocity and
\be
\vecEpprm = \vecEp + \frac{\vecv}{c} \cross \vecBp
\ee
is the electric field in the frame of the neutral particles (=the bulk plasma).
The drift velocity depends on the dimensionless ion Hall parameter,
\be
\betai \equiv \Omega_{\rm i}\,\tauin,
\label{eq-hallparam}
\ee
where
\be
\Omega_{\rm i} = \frac{\Zi eB}{\emi c}
\ee
is the ion cyclotron frequency,
\be
\tauin \equiv \frac{1+\emi/\emn}{\enn\sigin}
\label{eq-iondragt}
\ee
is the time scale for the ions to reach their terminal drift speed,
and \sigin\ is the momentum transfer rate coefficient for elastic ion-neutral scattering.
Ion-neutral streaming heats the neutral gas at a rate
\be
\Gamma_{\rm in} \approx \eni\,\enn\,\sigin\,\mu_{\rm in}\,\left|\vecvi-\vecv\right|^2
\ee
(Draine et al.\ 1983; Chernoff 1987),
where $\mu_{\rm in}$ is the ion-neutral reduced mass.

\section{Transport Coefficients}
\label{sec-tcoeff}

In Section~\ref{sec-slab} we present analytic solutions which describe
shear flow past an infinite slab.
These solutions are cast in dimensionless units which depend
only on certain combinations of the transport coefficients.
However to express the solutions in terms of ordinary units
we require numerical values of $\alpha$, \etao, \etah, and \etaa.
We estimate these as follows.
 
\subsection{Protoplanetary Disk Model}
\label{sec-ppd}

\begin{figure}
\figurenum{1}
\plotone{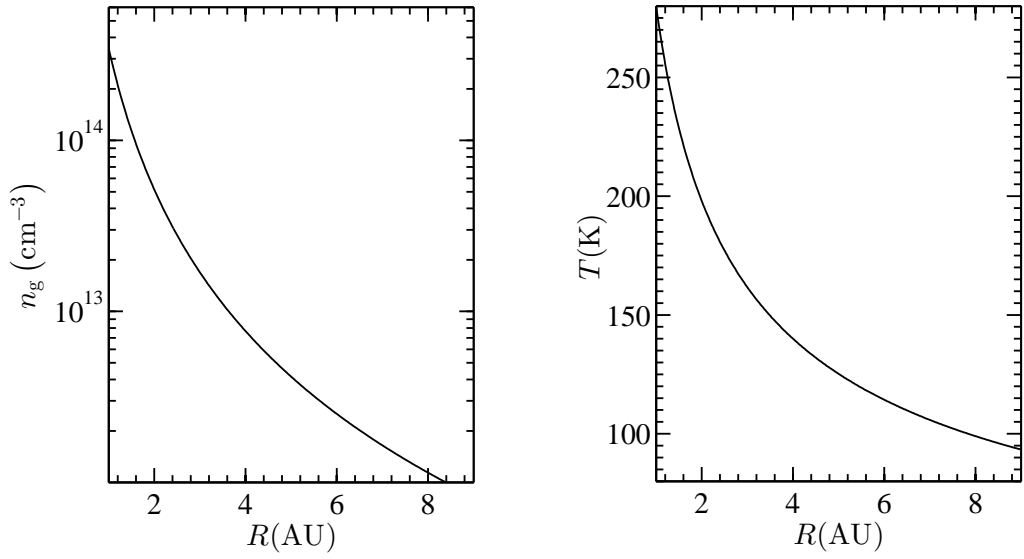}
\caption{
The midplane number density of neutral particles (left) and temperature
(right) plotted vs.\ distance $R$ from the central star.
}
\label{fig-disknT}
\end{figure}

We use the density and temperature predicted by the
minimum mass solar nebula (MMSN) model (Weidenschilling 1977b; Hayashi 1981) as
described by Sano \etal\ (2000).
The gas temperature is
\be
T(R) = 280\,\left(\frac{R}{\rm AU}\right)^{-1/2}
~~~~~{\rm K},
\label{eq-tdisk}
\ee
where $R$ is distance from the central star.
The mass density is
\be
\rhog(R,Z) = \rhog(R,0)\,\exp\left[-\left(\frac{Z}{H}\right)^2\right],
\label{eq-rhogrz}
\ee
where $Z$ is vertical distance from the midplane.
The midplane density of the MMSN is
\be
\rhog(R,0) = 1.4 \times 10^{-9} \, \left(\frac{R}{\rm AU}\right)^{-11/4}\,
\left(\frac{M_*}{\Msol}\right)^{1/2}\,
\left(\frac{\mug}{2.34}\right)^{1/2}
~~~~~\gcmMMM,
\label{eq-rhoc}
\ee
where $M_*$ is the mass of the central star, $\mu_{\rm{g}}m_{\rm{H}}$ is the
mean mass per neutral particle, and $m_{\rm{H}}$ is the hydrogen mass.
We set $M_*=1$\,\Msol\ and $\mug=2.33$, the latter corresponding
to a gas composed of molecular hydrogen and
helium with number densities $0.5\enH$ and $0.1\enH$, respectively,
where \enH\ is the number density of hydrogen nuclei.
Then the gas number density at the midplane is
\be
\eng(R,0) = 0.6\,\enH(R,0) = 3.5 \times 10^{14}\,\left(\frac{R}{\rm AU}\right)^{-11/4}
~~~~~\cmMMM.
\ee
The midplane temperature and number density are plotted versus $R$
in Figure~\ref{fig-disknT}.  We also require the ambient magnetic field, $B_{0}$.  It is not yet possible to directly measure the magnetic fields in
protoplanetary disks; however several lines of evidence suggest that $B_{0} \sim
0.1$--$1$\,G, including simulations of star formation (Desch \& Mouschovias 
2001) and constraints implied by protostellar mass accretions rates (Wardle 
2007; Bai \& Goodman 2009).  We will adopt the value $B_{0}=0.3$\,G for our 
calculations discussed in Section~\ref{sec-disc}
which is approximately the geometric mean of the extremes.

\subsection{Ionization Equilibrium}
\label{sec-ions}

The magnetic diffusivities depend on the number densities of ions,
electrons, and charged dust grains (Section~\ref{sec-eta}).
To estimate these we use the semianalytical model of ionization
equilibrium developed by Okuzumi (2009).
Okuzumi's model calculates the number densities of electrons, dust grains
in different charge states, plus the total ion number density,
\be
\eni \equiv \sum_k\ n_i^{(k)},
\ee
where the sum is over different ionic species $k$ $\left({=\rm Mg}^+, {\rm HCO}^+, {\rm H}_3^+, \ldots\right)$.
Okuzumi's results are in excellent agreement with full numerical calculations in which
different ionic species are treated explicitly.\footnote{See Okuzumi (2009) Figure~3 but note
that the figure has an error: the code which plotted Figure~3b multiplied every quantity by
a factor of $\approx 0.36$ (S. Okuzumi, private communication).}
The inputs to Okuzumi's model are listed in Table~\ref{table-ions}, where
\be
\beta_{\rm{r}} \equiv \frac{1}{n_{\rm i}}\,\sum_k\ n_i^{(k)}\,\beta_{\rm{r}}^{(k)}
\ee
is the average recombination rate coefficient of the ions,
\be
u_{\rm i} \equiv \frac{1}{n_{\rm i}}\,\sum_k\ n_i^{(k)}\,u_{\rm i}^{(k)}
\ee
is the average of the mean ion thermal speed, and
\be
\zeta \equiv \frac{1}{n_{\rm g}}\,\sum_j\ n_{\rm g}^{(j)}\,\zeta^{(j)},
\label{eq-zeta}
\ee
where $\zeta\eng$ is the number of ionizations per unit volume per unit time and
the sum is over all gas-phase species $j$.
More detailed treatments of ionization equilibrium in disks (e.g., Sano \etal\ 2000)
show that the abundance of atomic ions exceeds the abundance of molecular ions
by 2--3 orders of magnitude.
We will take Mg$^+$ to be a proxy for the ions and set
\be
u_{\rm i} = \left(\frac{8\kB T}{\pi\mui\mH}\right)^{1/2},
\label{eq-uion}
\ee
where $\mui=24$, and
\be
\beta_{\rm{r}} = 2.8 \times 10^{-12}\,\left(T/300\,{\rm K}\right)^{-0.7},
\label{eq-mgrr}
\ee
which is the rate coefficient for radiative recombination of Mg$^+$
(McElroy \etal\ 2013).

The ionization rate, $\zeta$, is crucial but uncertain.
In a protoplanetary disk it has
contributions from cosmic rays, x rays, and live radionuclides
so that
\be
\zeta = \zeta_{\rm cr} ~+~ \zeta_{\rm xr} ~+~ \zeta_{\rm ra}.
\ee
The cosmic ray ionization rate is
\be
\zeta_{\rm cr} = \zeta_{{\rm cr},0}\,F_{\rm cr}(R,Z),
\ee
where $\zeta_{\rm{cr},0}$ is the rate far from the midplane. The factor
\be
\begin{array}{rcl}
F_{\rm cr}(R,Z) & \equiv & \frac{1}{2}\left\{
\exp\left[-\Sigma_{\rm g}^+(R,Z)/\Sigma_{\rm cr}\right]\,
\left[1+\left(\Sigma_{\rm g}^+/\Sigma_{\rm cr}\right)^{3/4}\right]^{-4/3} \right.\\
 & & \\
 & + &
\left.
\exp\left[-\Sigma_{\rm g}^-(R,Z)/\Sigma_{\rm cr}\right]\,
\left[1+\left(\Sigma_{\rm g}^-/\Sigma_{\rm cr}\right)^{3/4}\right]^{-4/3}
\right\}\\
\end{array}
\ee
describes attenuation by the gaseous disk (Okuzumi 2009), with
\be
\Sigma_{\rm g}^+(R,Z) \equiv \int_Z^{+\infty}\ \rho_{\rm g}\left(R,Z^{\prime}\right)\,dZ^{\prime},
\ee
\be
\Sigma_{\rm g}^-(R,Z) \equiv \int_{-\infty}^Z\ \rho_{\rm g}\left(R,Z^{\prime}\right)\,dZ^{\prime},
\ee
and $\Sigma_{\rm cr}=96$\,\gcmMM.
Here we are interested only in points with $Z=0$ so that
\be
\Sigma_{\rm g}^+(R,Z) = \Sigma_{\rm g}^-(R,Z) = \frac{1}{2}\,\Sigma_{\rm g}(R),
\ee
where
\be
\Sigma_{\rm g}(R) = 1700\,\left(\frac{R}{\rm AU}\right)^{-3/2}
~~~~~\rm{g\,cm^{-2}}
\label{eq-sigma}
\ee
is the total surface density.
Since the cosmic ray ionization rates of \Htwo\ and He are the same
to within about 20\%, we will take $\zeta_{\rm cr,0}$ to be the
rate for molecular hydrogen.
Values of the latter inferred from observations
of molecular clouds span about two orders of magnitude, from
$3 \times 10^{-18}$\,\sM\ to $4 \times 10^{-16}$\,\sM\ (Padovani \etal\ 2009 and
references therein).
We will adopt the value suggested by Dalgarno (2006),
\be
\zeta_{\rm cr,0} = 5 \times 10^{-17}\,\sM,
\ee
which is approximately the geometric mean of the extremes.

\begin{figure}
\epsscale{1.1}
\figurenum{2}
\plotone{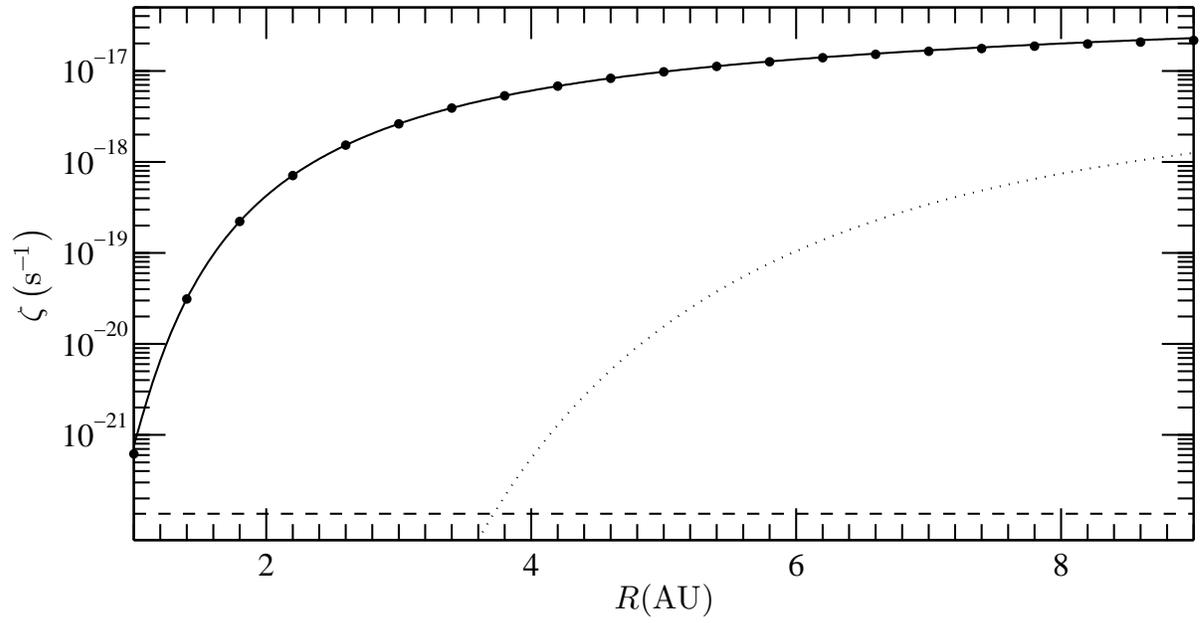}
\caption{
The midplane ionization rate per gas particle as a function 
of $R$ showing the contributions of cosmic rays (filled circles),
radioactivity (dashed), x rays (dotted) and the total rate (solid).
}
\label{fig-zeta}
\end{figure}

Our models also include ionization by stellar x rays and radionuclides.
For the former we adopt the fit of Turner \& Sano (2008) to calculations
of x-ray ionization by Igea \& Glassgold (1999):
\be
\begin{array}{rcl}
\zeta_{\rm xr}(R,Z) & = & \zeta_{{\rm xr},0}\,\left(\frac{R}{{\rm AU}}\right)^{-2}\,
\left(\frac{L_{\rm xr}}{2\times 10^{30}\,\ergsM}\right) \\
 & & \\
 & & \times \left\{
\exp\left[-\Sigma_{\rm g}^+(R,Z)\right]/\Sigma_{\rm xr} +
\exp\left[-\Sigma_{\rm g}^-(R,Z)\right]/\Sigma_{\rm xr}
\right\},\\
\end{array}
\label{eq-xray}
\ee
where $\Sigma_{\rm xr}=8.0$\,\gcmMM, $\zeta_{{\rm xr},0}=2.6\times 10^{-15}$\,\sM,
and the stellar x-ray luminosity, $L_{\rm xr}$, is a parameter (Table~\ref{table-ions}).
The rate of ionization by radioactivity depends strongly on the presence or
absence of short-lived radionuclides (SLRs) like $^{26}$Al.
It increases from $\zeta_{\rm ra} = 1.4 \times 10^{-22}$\,\sM\ 
if SLRs are absent to $\zeta_{\rm ra} = 7.6 \times 10^{-19}$\,\sM\ 
if SLRs are present with the abundances inferred for the early solar
nebula (Umebayashi \& Nakano 2009).
The fact that chondrules are a few Myr younger than the first solids
to condense from the solar nebula (the CAIs) suggests that the solar
nebula was a few Myr old when the asteroids formed, at least for the
parent bodies of chondritic meteorites.
Since the half life of $^{26}$Al is 0.7\,Myr,
we will neglect ionization by SLRs.
The midplane ionization rates are plotted vs.\ $R$ in Figure~\ref{fig-zeta}.

The ionization balance and magnetic diffusivities are sensitive to the presence of small 
(with radii $a$ less than a few \micron) dust grains.
Dust tends to sweep up electrons, the only particles that are well coupled
to \vecB\ at the densities of interest here (Section~\ref{sec-eta}).
This reduces the coupling between gas and magnetic field with a consequent increase
in the diffusivities.
For example Wardle (2007) showed that a standard interstellar population of
grains with $a=0.1$\,\micron\ would increase the diffusivities enormously, by $\sim 8$ orders
of magnitude in the midplane.
However grain growth can reduce the total dust surface
area\footnote{If the grains grow as spheres with fractal dimension $D_{\rm f}=3$. 
However the initial stages of grain growth generally produce fractal 
structures with
$D_{\rm f}<3$ (Blum \& Wurm 2008 and references therein). To be 
consistent with the grain model of Birnstiel et al.\ (2011) considered
below [see Equation~(\ref{eq-mdust})] we will set $D_{\rm f}=3$.}
to a point where the effect of grains on the diffusivities becomes negligible;
Wardle (2007) showed that (for nonfractal grains) this occurs when $a$ exceeds
a few \micron.
In this paper we are concerned with disk ages of a few Myr, which is longer than the time scale
for grain growth (Dullemond \& Dominik 2005).
The expectation that grain growth is significant at $t\sim$\,Myr is borne out by millimeter
observations of disks, which suggest that most of the dust is locked up in millimeter and larger-size particles
at about $t=2$\,Myr (Williams 2012 and references therein).
However Spitzer observations of the infrared features produced by amorphous (at $10$\,\micron) and 
crystalline ($23$\,\micron) silicates imply the presence of \micron-sized 
grains in numerous disks (e.g., Furlan \etal\ 2011).
Because these disks are optically thick the observed small grains lie
in the disk atmospheres, i.e., not in the midplane.
Nevertheless, it is important to know whether enough \micron-sized dust might reside
in the midplane to affect the diffusivities.

\begin{figure}
\figurenum{3}
\plotone{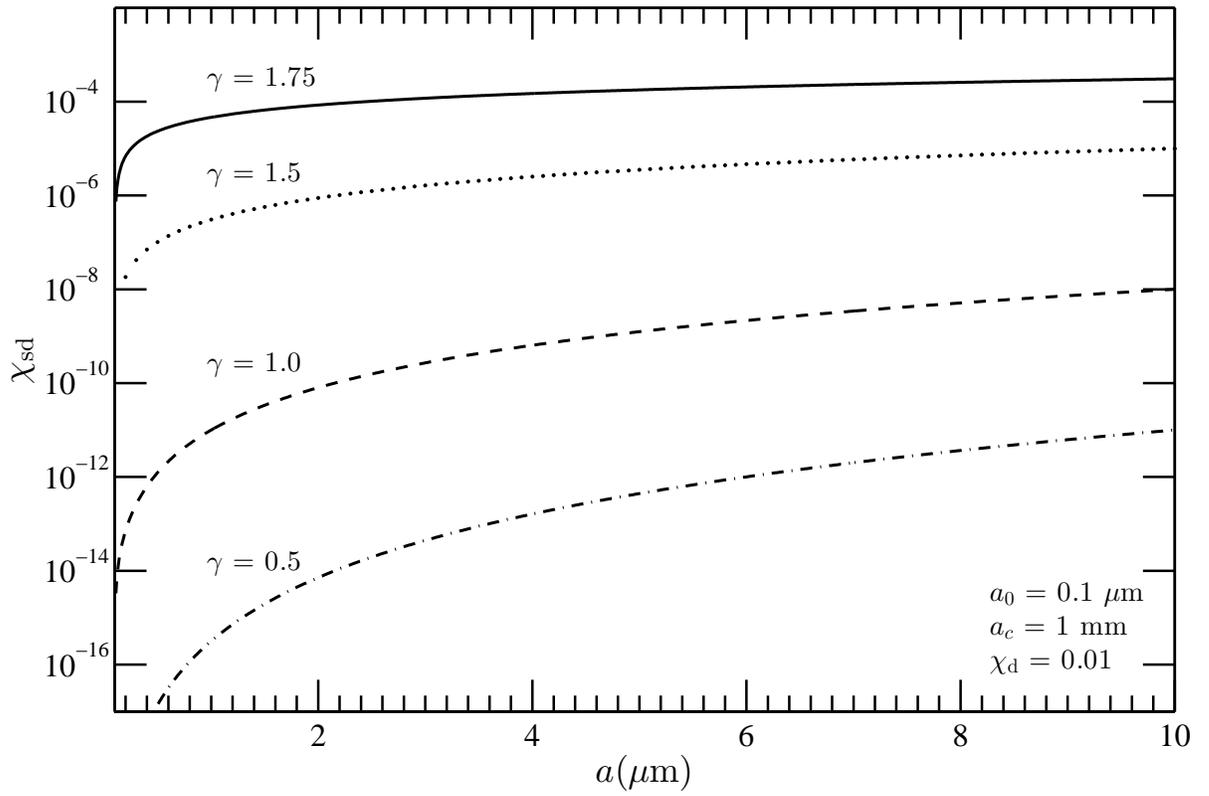}
\caption{
Plot of \chisd\ vs.\ grain radius, $a$ (see Equation~[\ref{eq-chisd}])
for a few values of the exponent in Equation~(\ref{eq-nmdust}).
}
\label{fig-chisd}
\end{figure}

To estimate the density of small dust at the midplane we exploited the calculations of Birnstiel et al.\ (2011),
who described the equilibrium size distribution established by coagulation and fragmentation.
Birnstiel et al.\ modeled a grain of radius $a$ as a constant-porosity sphere of mass
\be
m(a) = \frac{4}{3}\,\pi\,\rhos\,a^3,
\label{eq-mdust}
\ee
where \rhos\ is the bulk density of the porous solid.
At any given time the mass distribution extends from the monomer mass, $m_0$,
to some upper limit, $m_{\rm c}$, determined by the growth process.
For a range of models with different assumptions about the underlying physics they found
that the mass distribution can be approximated by a power law\footnote{Strictly
speaking, the distribution given in Equation~(\ref{eq-nmdust})
does not account for the effect of settling in disk.  However numerical
simulations performed by Birnstiel \etal\ (2011) including the effects of
settling and turbulent mixing find that the vertically integrated
dust distribution can be approximated by two power laws depending on whether
or not the radius of the grains are greater than a certain size (See their Section~3.2 and Figure~4).},
\be
n(m) = A\,m^{-\gamma}
~~~~~ m_0 < m < m_{\rm c},
\label{eq-nmdust}
\ee
where $n(m)\,dm$ is the number density of grains with masses in $(m,m+dm)$ and $\gamma$
lies in the range from about $0.5$ to $2$.
We set $m_{0}=m(a_{0})$ and $m_{\rm c}=m(a_{\rm{c}})$, where $a_{0}=0.1$\,\micron\ 
and $a_{\rm{c}}=1$\,mm, corresponding
to the coagulation of interstellar dust to sizes consistent with the millimeter observations
of disks.
The total abundance of dust with all sizes is characterized by \chid, where $\chid\rhog$ is
the total dust mass density and $\chid \sim 0.01$.
As a measure of how much {\it small}\/ dust is present we define
$\chisd(a)\rhog$ to be the mass density of dust with radii less than $a$.
Then it is easy to show that Equation~(\ref{eq-nmdust}) implies
\be
\chisd(a) = 
\left[ \frac{\left(a/a_0\right)^{3\left( 2 - \gamma \right)}-1}{\left(a_{\rm c}/a_0\right)^{3\left( 2 - \gamma \right)}-1} \right] \,\chid
~~~~~\mbox{if $\gamma \ne 2$}.
\label{eq-chisd}
\ee
Figure~\ref{fig-chisd} plots \chisd\ for a few plausible values of $\gamma$.
It is apparent that the theory allows a very broad range in the mass fraction of
micron-size and smaller grains, from $\sim 10^{-14}$ to $10^{-2}$.

\begin{figure}
\figurenum{4}
\plotone{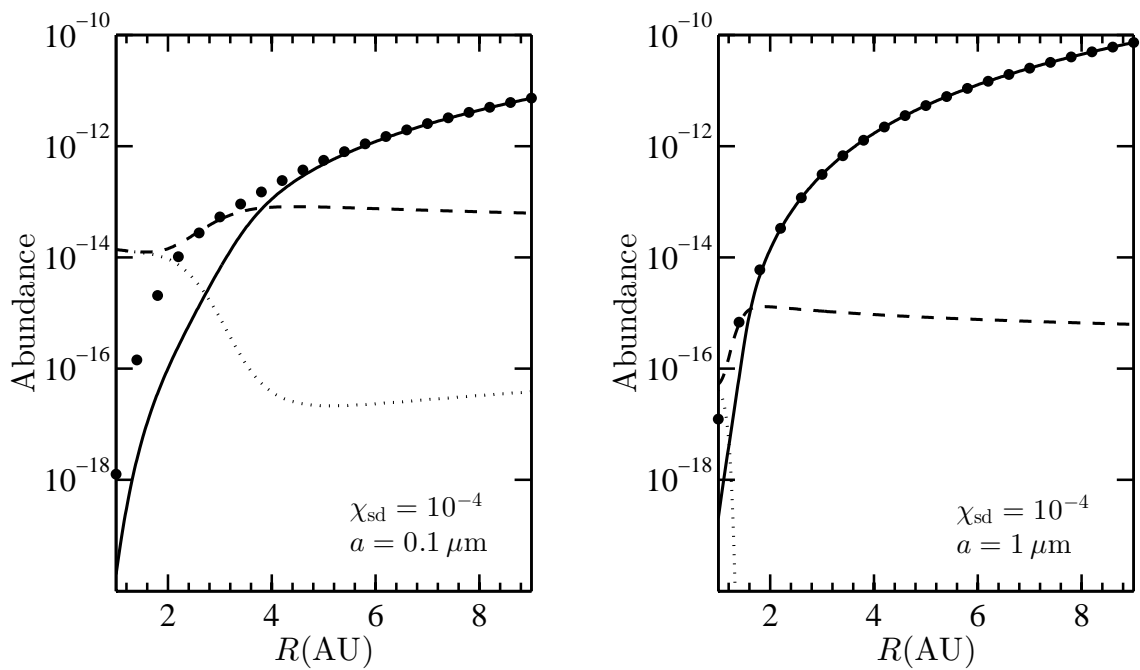}
\caption{
The abundances of ions (filled circles), electrons (solid curve), and charge contained
by single-sized 
dust grains (dotted curve = postively charged, dashed curve = negatively charged; See 
Equations~[\ref{eq-totdustposcharge}]--[\ref{eq-totdustneqcharge}]) 
relative to
$n_{\rm{H}}$ in the disk midplane are plotted versus distance from the central star.  The assumed
abundance of small dust and grain radii are indicated for each case.
}
\label{fig-dustabunds}
\end{figure}

Figure~\ref{fig-dustabunds} plots the abundances of ions, electrons, $\chi_{d+}$,
and $\chi_{d-}$ relative to $n_{\rm{H}}$ calculated using Okuzumi's 
model as a function of $R$ in the 
disk, assuming $\chisd=10^{-4}$. Here $\chi_{d+}$ and $\chi_{d-}$ describe the total
amount of positive and negative charge contained by dust grains, and are defined as
\be
\chi_{d+} = \sum_{Z>0} \frac{n_{\rm{d}}(Z)}{n_{\rm{H}}} \left| Z \right|
\label{eq-totdustposcharge}
\ee
and
\be
\chi_{d-} = \sum_{Z<0} \frac{n_{\rm{d}}(Z)}{n_{\rm{H}}} \left| Z \right|,
\label{eq-totdustneqcharge}
\ee
where $Z$ is
the charge of the grains in the units of elementary charge.  
We assume that all the dust grains are single-sized with radius $a$ 
and consider two cases: $a=0.1$\,\micron\ and $a=1$\,$\micron$.  When 
$a=0.1$\,\micron\, we find that positively and negative charged dust grains are the 
dominant charge carriers in the plasma for $R$~$\ltsim$~$2$\,AU.  For 
$2$~$\ltsim$~$R$~$\ltsim$~$4$\,AU, negatively charged dust grains still dominate 
over electrons while ions become the dominate carrier of positive charge.  When
$R$~$\gtsim$~$4$\,AU, ions and electrons become the dominant charged species.  If
the grain radius is increased to $a=1$\,\micron\,, we observe that the abundances
of ions and electrons exceed those of charged dust grains for all 
$R$~$\gtsim$~$1.5$\,AU. For 1~$\ltsim$~$R$~$\ltsim$~1.5\,AU, negatively charged dust
grains become more abundant than electrons in the plasma while ions continue to be
the dominate carrier of positive charge.

\begin{figure}
\figurenum{5}
\plotone{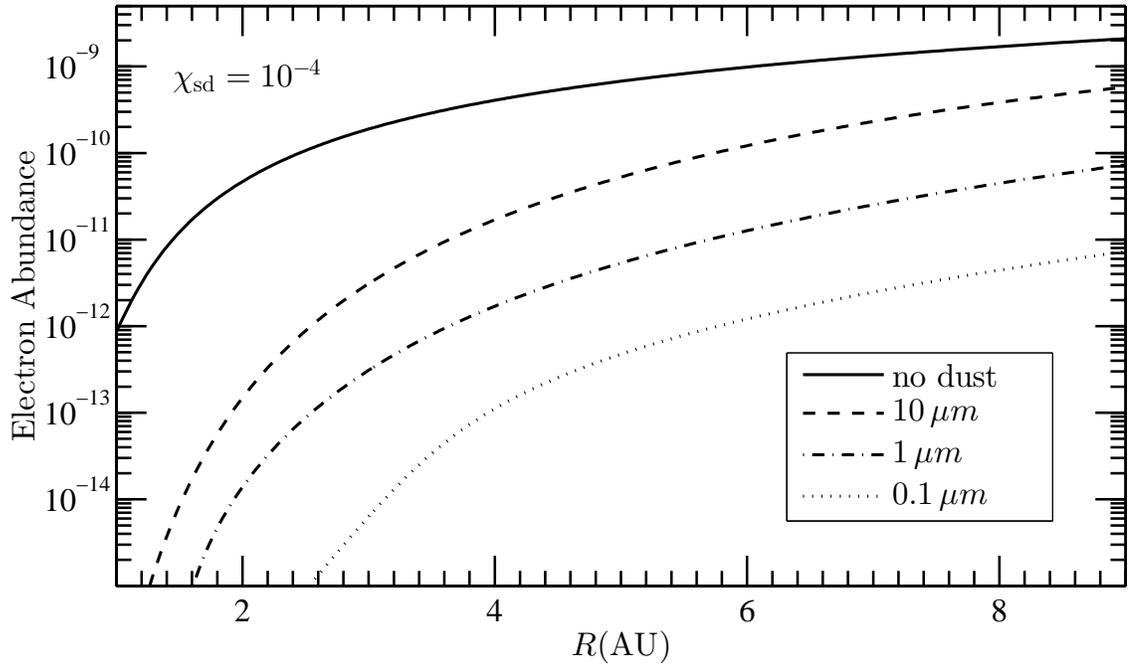}
\caption{
Variation of the electron abundance, \xe, with
$R$ including single-size dust grains with three
different radii indicated in the legend.
The abundance of small dust is
$\chi_{\rm sd}=10^{-4}$. 
}
\label{fig-xe-chisd1e-4}
\end{figure}

\begin{figure}
\figurenum{6}
\plotone{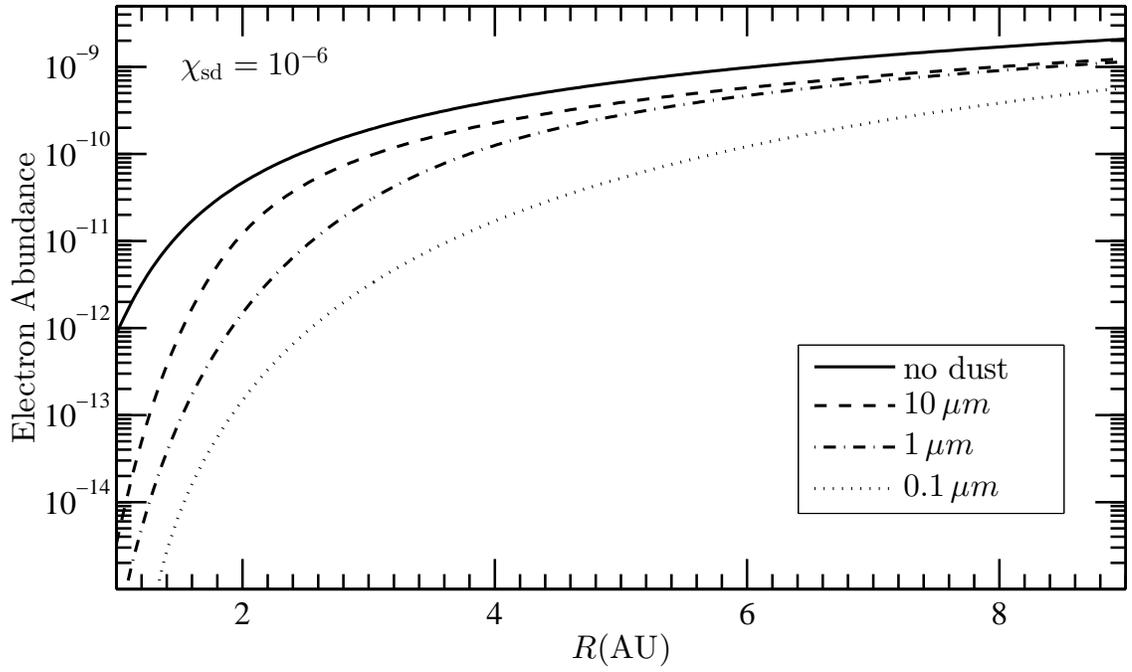}
\caption{
As in Figure~\ref{fig-xe-chisd1e-4}
but for $\chi_{\rm sd}=10^{-6}$. 
}
\label{fig-xe-chisd1e-6}
\end{figure}

Of particular interest to the flows considered in this paper is the electron
abundance,
\be
\xe \equiv n_{\rm e}/n_{\rm H},
\ee
since the electrons are the only particles that are strongly tied to the magnetic 
field (See Section~\ref{sec-eta}).  The effects of dust on the electron abundance
in the midplane of the disk are shown in Figures~\ref{fig-xe-chisd1e-4} and 
\ref{fig-xe-chisd1e-6}.  
The calculations were carried out for two values of $\chi_{\rm sd}$
toward the high end of the plausible range and
assumed single-size grains.
The electron abundance in a dust-free plasma is also shown for comparison.
The figures confirm that \xe\ is very sensitive to the abundance
of the smallest grains in the distribution if the latter are smaller
than a few \micron\ (Wardle 2007).
For example, the electron abundance is reduced by about three orders
of magnitude at $R=3$\,AU if $a=1$\,\micron\ and $\chisd=10^{-4}$
(Figure~\ref{fig-xe-chisd1e-4}).
However once the abundance of small dust falls to $\chisd=10^{-6}$
the effects of dust are small unless $a \le 0.1$\,\micron\ 
(Figure~\ref{fig-xe-chisd1e-6}).
We will calculate the diffusivities for two cases: a dust-free
plasma and a plasma with $\chisd=10^{-4}$, $a=1$\,\micron.
However it should be kept in mind that, if the power-law
grain size distribution is steeper than $\gamma=1.75$, our calculations
will underestimate the diffusivities by orders of magnitude.

\subsection{Viscosity and Magnetic Diffusivities}
\label{sec-eta}

The shear viscosity depends only on temperature.
We adopt the viscosity values calculated by Schaefer (2010),
which are well approximated by
\be
\alpha = 17 + 0.26\,\left(\frac{T}{\rm K}\right)
~~~~~\mu{\rm Poise}
\label{eq-alpha}
\ee
for both ortho- and para-\Htwo\ if $T<300$\,K.

The magnetic diffusivities are obtained from the charged particle abundances
as follows (Wardle 2007):
\be
\etao = \frac{c^2}{4\pi\sigma_{\rm o}},
\ee
\be
\etah = \frac{c^2}{4\pi\sigma_{\perp}}\,\frac{\sigma_{\rm h}}{\sigma_{\perp}},
\ee
and
\be
\etaa = \frac{c^2}{4\pi\sigma_{\perp}}\,\frac{\sigma_{\rm p}}{\sigma_{\perp}} ~-~\etao,
\ee
where 
\be
\sigma_{\rm o} = \frac{ec}{B}\,\sum_j\ n_j\,\left|Z_j\right|\,\beta_j
\label{eq-sigmao}
\ee
is the Ohmic conductivity,
\be
\sigma_{\rm h} = \frac{ec}{B}\,\sum_j\ \frac{n_j\,Z_j}{1+\beta_j^2}
\ee
is the Hall conductivity,
\be
\sigma_{\rm p} = \frac{ec}{B}\,\sum_j\ 
\frac{n_j\left|Z_j\right|\beta_j}{1+\beta_j^2}
\ee
is the Pedersen conductivity, and
$\sigma_{\perp}\equiv\sqrt{\sigma_{\rm h}^2+\sigma_{\rm p}^2}$.
The sums are over all charged species $j$ with mass $m_j$, number
density $n_j$, and charge $Z_je$.
We include electrons, ions, and single-size grains in various charge states.
In particular we include all grain charge states within four standard deviations of
the mean charge (see Okuzumi [2009]).

In addition to the abundance of each charged particle, the diffusivities depend
on its dimensionless Hall parameter, which 
depends in turn on its drag time, 
$\tau_{j{\rm n}}$. The latter is defined by
the drag force produced by friction with the neutral gas:
\be
\vecF_{jn} \equiv -m_j\,\left(\vecv_j-\vecv_{\rm n}\right)/\tau_{j{\rm n}}.
\ee
For the ion drag time we adopt Equation~\ref{eq-iondragt}
with $\sigin = 1.9 \times 10^{-9}$\,cm$^3$\,s$^{-1}$ and for the electrons we use
\be
\tauen = \frac{1}{n_{\rm g}\left<\sigma v\right>_{\rm en}}
\ee
with
\be
\left<\sigma v\right>_{\rm en} = \sigma_{\rm g}\,\left(\frac{128k_{\rm B}T}{9\pi m_{\rm e}}\right)^{1/2}
\ee
and $\sigma_{\rm g}=10^{-15}$\,cm$^2$ (Draine \etal\ 1983).
For the drag force on dust grains we use the expression given by Draine \& Salpeter (1979) in the limit of subsonic gas-grain drift.
It implies that
\be
\taudn = \frac{\sqrt{\pi}}{2}\,\left(\frac{\rho_{\rm s}}{\rho_{\rm g}}\right)\,\frac{a}{v_{\rm th}},
\label{eq-taudn}
\ee
where
\be
v_{\rm{th}} = \left( \frac{2 k_{\rm{B}} T}{\mu_{\rm{g}} m_{\rm{H}}} \right)^{1/2}.
\ee

\begin{figure}
\figurenum{7}
\plotone{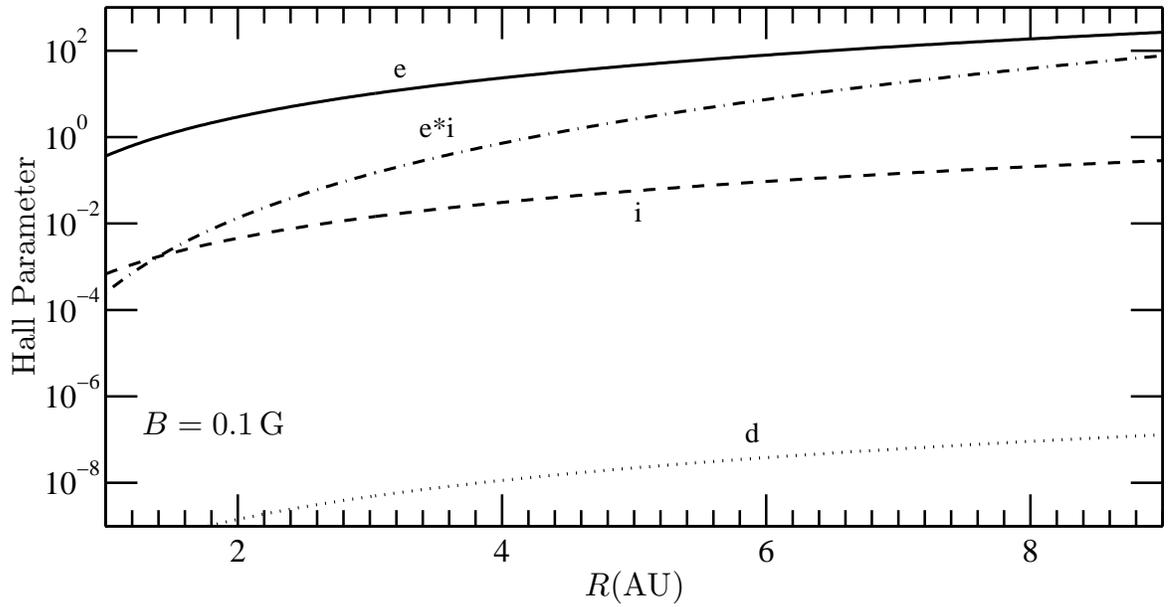}
\caption{
Hall parameters as a function of $R$ for the ions (i) and
electrons (e) and a dust grain with $a=1\,\micron$ and $Z=\pm 1$ (d).
The curve labeled i*e is the product of the ion and electron
Hall parameters.
Values are for the disk midplane.
}
\label{fig-Hall_Bsmall}
\end{figure}

\begin{figure}
\figurenum{8}
\plotone{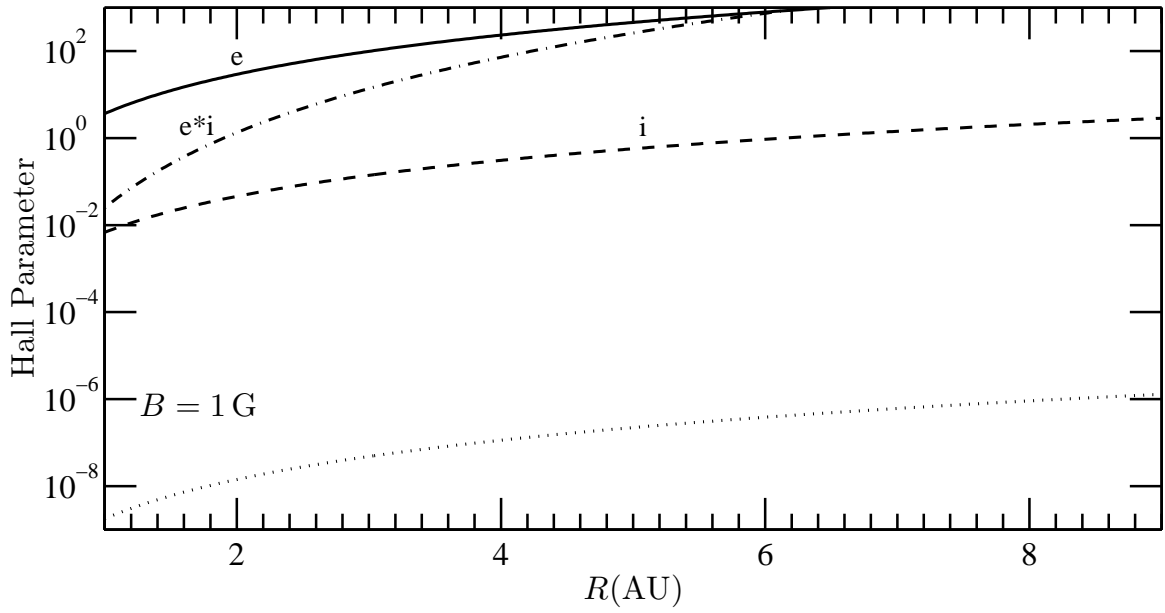}
\caption{
As in Figure~\ref{fig-Hall_Bsmall} but for $B=1$\,G.
}
\label{fig-Hall_Blarge}
\end{figure}

\begin{figure}
\figurenum{9}
\plotone{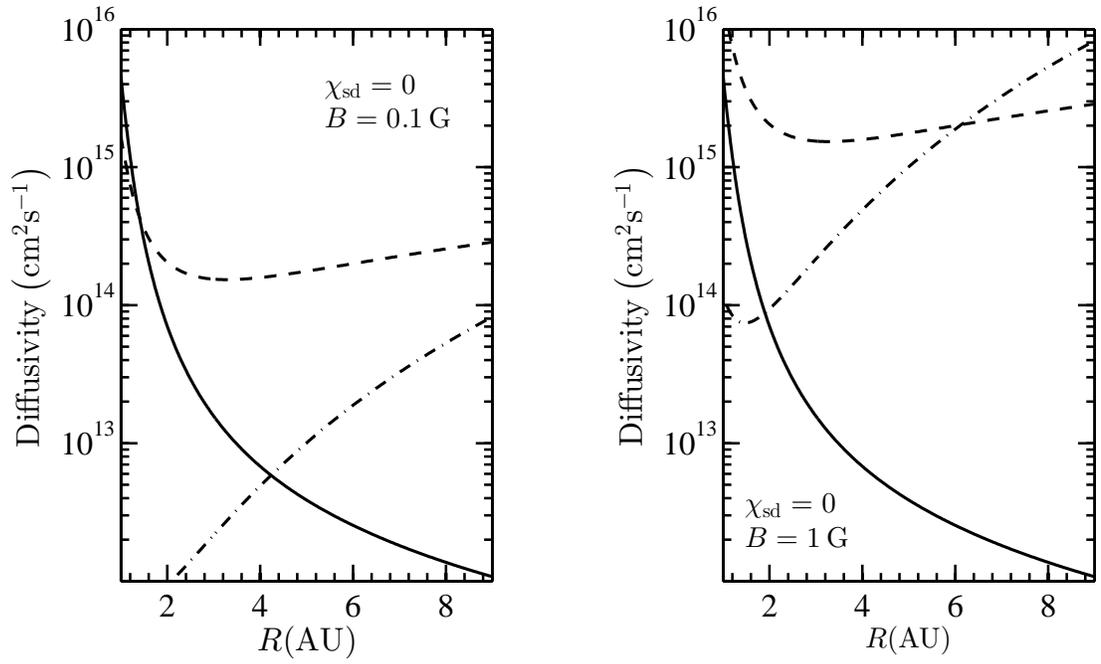}
\caption{
Variation of the Hall (dashed curve), Ohmic (solid),
and ambipolar (dash-dotted) diffusivities with $R$ for a dust-free plasma.
Results are plotted for two values of the magnetic field $B$.
Values are for the disk midplane.
}
\label{fig-eta_chisd0}
\end{figure}

\begin{figure}
\figurenum{10}
\plotone{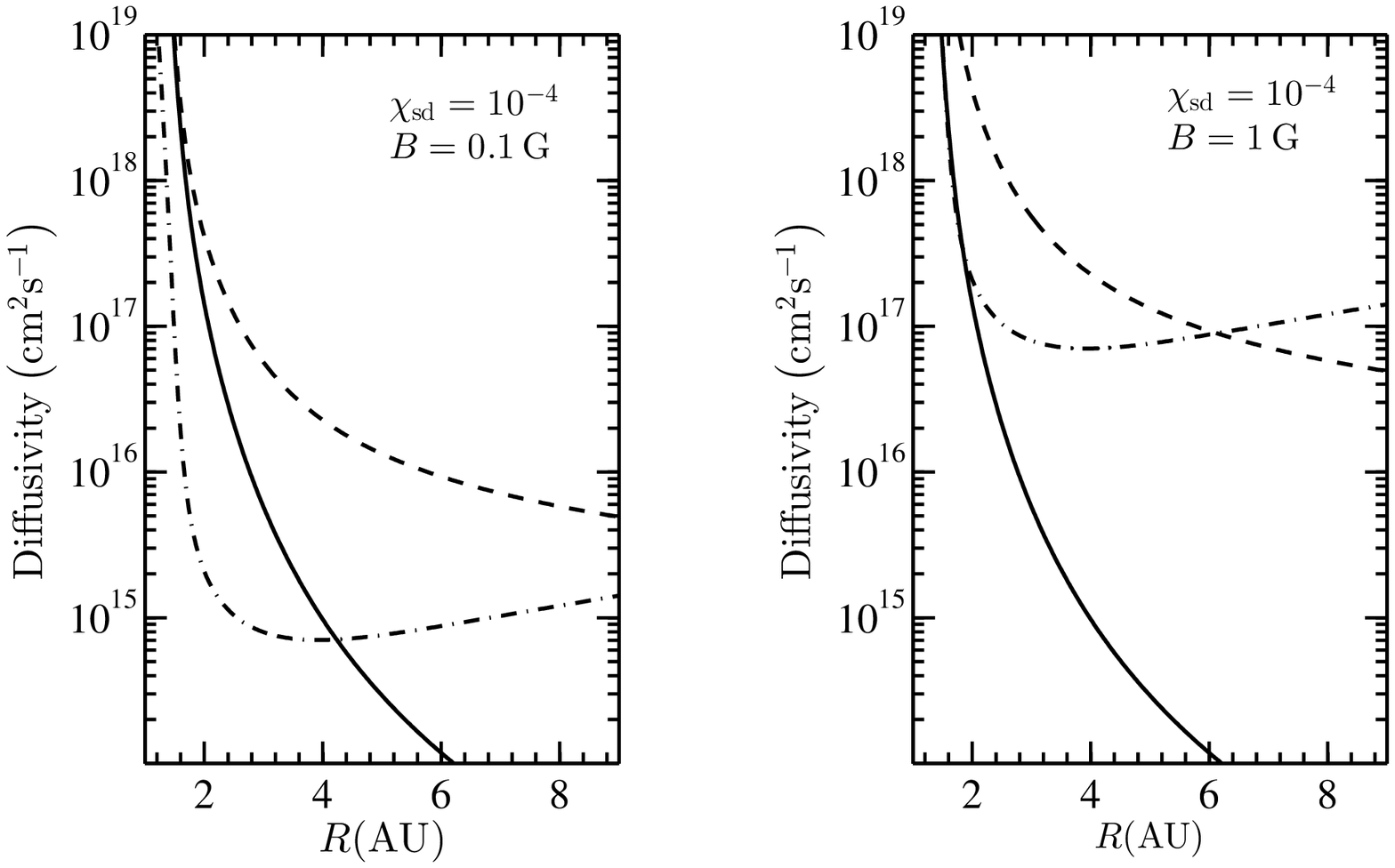}
\caption{
As in Figure~\ref{fig-eta_chisd0} but
for $\chisd=10^{-4}$.
The dust is a single-size distribution
with $a=1$\,\micron.
}
\label{fig-eta_chisd1e-4}
\end{figure}

The coupling between each charged particle and the magnetic field is 
characterized by $\beta_{j}$.
If $\beta_j \gg 1$ then species $j$ is well coupled to the magnetic field lines and moves with them.
If $\beta_j \gg 1$ for all charged species the magnetic field is frozen into the charged particles and we recover ideal
MHD.
The Hall parameters of the electrons, ions, and dust grains with $a=1$\,\micron\ are
plotted vs.\ $R$ in Figures~\ref{fig-Hall_Bsmall} and \ref{fig-Hall_Blarge} 
for two different values of $B$.
The model adopted here predicts $\betae \gg 1$ and $\betai \ll 1$ over the
region corresponding to the asteroid belt; this is the Hall regime where
the electrons move with \vecB\ but the ions do not.
Notice that the dust Hall parameter is extremely small so that the terms
corresponding to dust in the sums for the conductivities are negligible
if $a=1$\,\micron.
This will hold unless very small grains with $a$~$\ltsim$~$10$\,\AA\ are present
as free fliers in the plasma, a scenario we do not consider here.
Nevertheless, dust affects the diffusivities profoundly by reducing the
gas-phase abundances of the electrons and ions.
The diffusivities are plotted in Figures~\ref{fig-eta_chisd0} and \ref{fig-eta_chisd1e-4}
for $\chisd=0$ (no small dust), $\chisd=10^{-4}$, and two plausible values
of the magnetic field $B$.

\subsection{Another Ionization Model}
\label{sec-recomb}

To illustrate the dependence of the magnetic diffusivities on our 
specific disk model, we now compare our ionization and magnetic 
diffusivity calculations 
to another model described in the recent work of 
Dzyurkevich \etal\ (2013).  
Although the model used 
by Dzyurkevich \etal\ makes similar assumptions about the physical 
conditions (i.e., density, temperature, magnetic field) in the disk, they 
make significantly different assumptions regarding the ionization rates 
and properties of the dust grains.  
The most important difference between our models is the assumed 
fractal dimension $D_{\rm{f}}$ of the grains, which Dzyurkevich \etal\ set equal to 
2 corresponding to grains modeled as fluffy fractal aggregates.  In 
addition, Dzyurkevich \etal\ assume a smaller ionization rate (by about an
order of magnitude) and introduce the concept of a ``metal line" in 
the disk, beyond which metal ions freeze out and molecular ions become the 
dominant ionic species.  In order to make a quantitative comparison between models, 
we now calculate the abundances of charged particles and magnetic diffusivities as a 
function of $R$ for the model described by Dzyurkevich \etal\ (2013). The parameters
used in these calculations are listed in Table~\ref{table-Dz}, where we note that in
order to recreate the results described in their paper, we required a
total magnesium abundance relative to hydrogen nuclei in the disk of 
$2 \times 10^{-9}$, which is 100 times larger than the value quoted in 
their Appendix B.

\begin{figure}
\figurenum{11}
\plotone{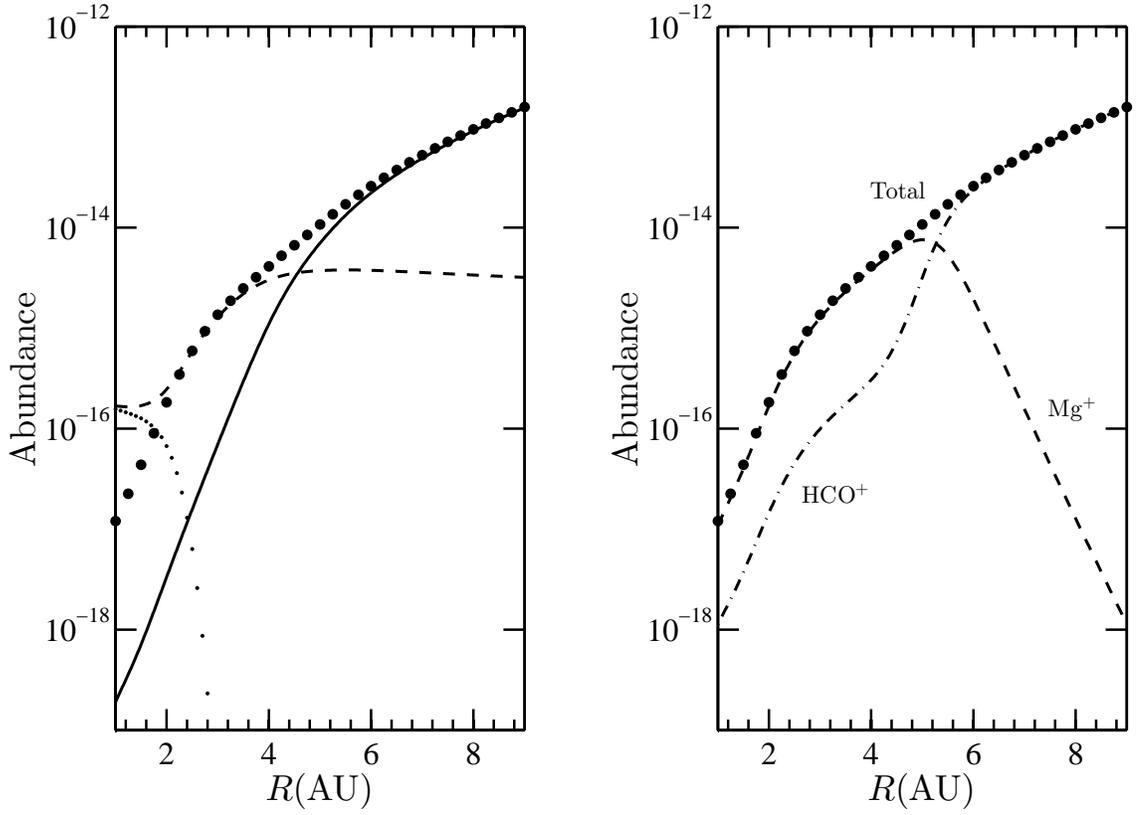}
\caption{Left panel: As in Figure~\ref{fig-dustabunds}, but for the
parameters described in Table~\ref{table-Dz}.  
Right panel: The number densities of
$\rm{Mg^{+}}$, $\rm{HCO^{+}}$, and their sum relative to the number
density of hydrogen nuclei in the disk.  Both panels are for
the model described in Dzyurkevich \etal\ (2013).
}
\label{fig-Dzabunds}
\end{figure}

\begin{figure}
\figurenum{12}
\plotone{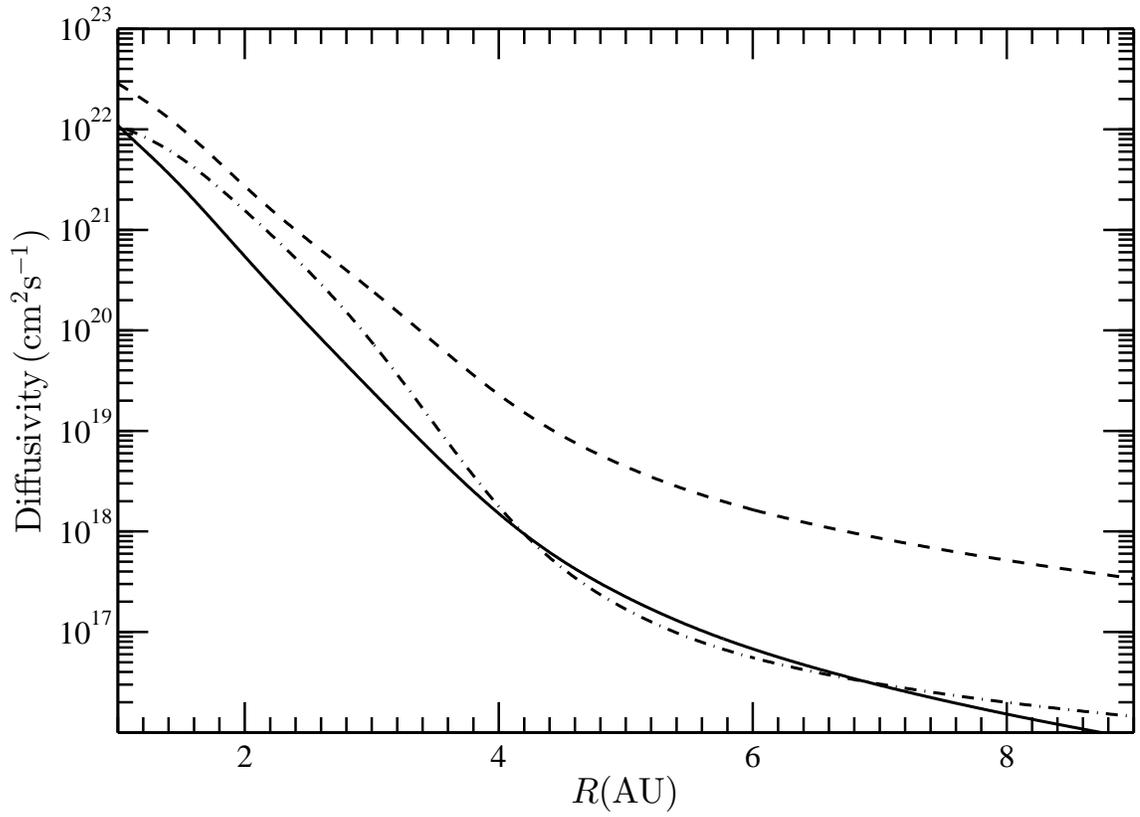}
\caption{As in Figure~\ref{fig-eta_chisd0}, but for
the model described in Dzyurkevich \etal\ (2013).
}
\label{fig-Dzdiffs}
\end{figure}

Figures~\ref{fig-Dzabunds}--\ref{fig-Dzdiffs} show the abundances of ions, electrons, 
and charge contained on dust grains, as well as the magnetic diffusivities in the 
disk midplane for the model described by Dzyurkevich \etal\ (2013).  In the left
panel of Figure~\ref{fig-Dzabunds}, we find that the dominant charge carriers are
positively- and negatively-charged dust grains for $R$~$\ltsim$~$2$~AU, ions and 
negatively-charged dust grains for $2$~$\ltsim$~$R$~$\ltsim$~$5$~AU, and ions and 
electrons for
$R$~$\gtsim$~$5$~AU.  For completeness we also plot the specific abundances 
of $\rm{Mg^{+}}$ and $\rm{HCO^{+}}$ in the right panel of Figure~\ref{fig-Dzabunds}, which
demonstrates the effect of the ``metal line" in this disk model. Comparing Figure~\ref
{fig-Dzabunds} with Figure~\ref{fig-dustabunds}, we find that although the ionization
is qualitatively similar, the total fractional ionization predicted by our model
exceeds that predicted by the model described by Dzyurkevich \etal\ by about 2
orders of magnitude.  As expected, this leads to a large difference in the predicted
values of the magnetic diffusivities.  Comparing Figure~\ref{fig-Dzdiffs} with
Figure~\ref{fig-eta_chisd1e-4} we find that although the Hall diffusivity is largest
for $R$~$\gtsim$~$2$~AU
in both models, in general the values of the magnetic diffusivities are larger
by $\sim 2$ orders of magnitude for the model described by Dzyurkevich \etal\ (2013).
The main effect of larger magnetic 
diffusivities in the disk is just to increase the thickness for the shear 
layer (See Section~\ref{sec-sheat}).

\section{Steady Flow Past an Infinite Slab}
\label{sec-slab}

\begin{figure}
\figurenum{13}
\plotone{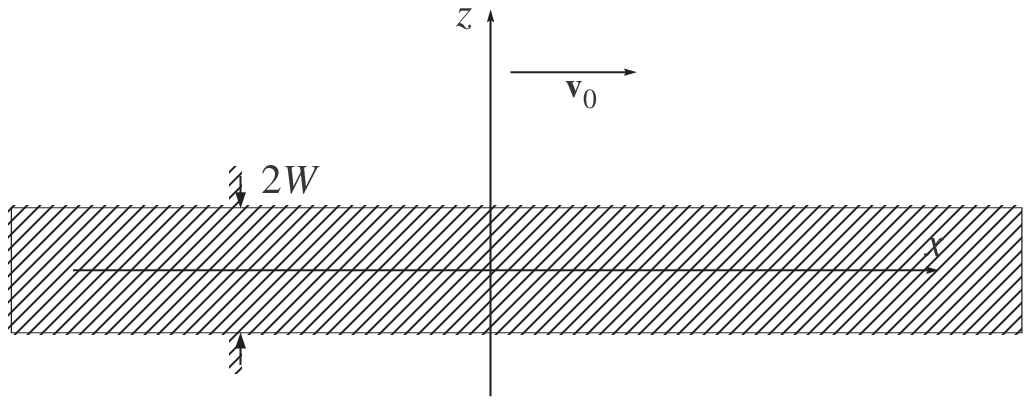}
\caption{
Flow past an infinite slab of thickness $2W$.
At the upper body surface the outward normal is $+\zhat$.
Far from the body the velocity is $\vecvzero = \vzero\,\xhat$.
We consider two cases, where the undisturbed magnetic field far from the body
is parallel ($\vecBzero = \Bzero\,\yhat$, ``parallel field geometry'')
and perpendicular ($\vecBzero = \Bzero\,\zhat$, ``perpendicular field geometry'')
to the body surfaces.
}
\label{fig-geometry}
\end{figure}

Given the complexity of the governing equations,
it seems inevitable that flows around bodies with realistic shapes
will have to be calculated numerically.
Here we consider a highly idealized problem--- 
steady flow past an infinite slab (Figure~\ref{fig-geometry})---
which can be studied analytically.
Although the geometry is unrealistic, the solutions illustrate how the
the electric field inside a body couples to the flow around it.
In this section we will assume that self heating of the shear flow is insignificant,
so that the temperature at each point in the flow equals the temperature
of the undisturbed plasma and the energy equation can be omitted;
the validity of this assumption is checked in Section~\ref{sec-sheat}.
Since mass is conserved identically in a steady shear flow,
it is only necessary to solve the momentum and induction equations
for \vecv, \vecBp, and \vecEp, plus
$\curl\vecEb=0$ and {\Ampere}'s Law for \vecEb\ and \vecBb.
 
\subsection{Parallel Field Geometry}
\label{sec-bpar}

We consider first the simplest possible case, where the ambient magnetic field
is parallel to the body surfaces: $\vecBzero=\Bzero\,\yhat$.
For the shear flow we assume that
\be
\vecv = [v_x(z),0,0],
\ee
\be
\vecBp = [0,B_0,0],
\ee
and that the pressure is constant.
Under these assumptions, the induction equation is satisfied identically and
the momentum equation reduces to a single ODE for $v_x$,
\be
\frac{d^{2} v_{x}}{d z^{2}} = 0,
\label{eq-parallelflowxmomentumeq}
\ee
with boundary conditions
\be
\lim_{z\to \infty} v_x = v_0
\label{eq-parallelflowvxinftybc}
\ee
and
\be
v_x(W) = 0.
\label{eq-parallelflowvxsurfbc}
\ee
Symmetry requires $v_{x}(z)$ to be an even function of $z$, 
so it is only necessary to calculate the solution explicitly for $z > 0$.

The electric field inside the body can be calculated exactly.
Outside the body the magnetic field is simply advected by the flow.
Since there is no compression or bending of the field lines, the magnetic field is uniform
with $\vecBp=\vecBzero$ everywhere.
It follows that the Ohmic, ambipolar, and Hall electric fields all vanish in the plasma and
hence that \vecEp\ equals the motional field.
Now the no-slip boundary condition implies that $\vecEp=0$ just outside each body 
surface.  We find the electric field just 
{\em inside}\/ each surface by applying the boundary conditions across the plasma sheath, 
whose thickness is neglected because it is of order meters
and thus much less than the dimensions of the shear layer.  Using 
Equations~(\ref{eq-etanbc}) and (\ref{eq-dnormbc}) with $\vecEp=0$, we find
\be
\vecE_{\rm{bs \pm}} = \mp\frac{4\pi\Sigma}{\epsb}\,\zhat
\label{eq-ebodypar}
\ee
where the upper and lower signs correspond respectively to the upper and lower slab surfaces.
The two electric fields in Equation~(\ref{eq-ebodypar}) are just
the electrostatic fields associated with electric charge in the plasma sheaths at
the upper and lower surfaces; their existence and potential consequences for heating
are not the subject of this investigation.
In any case, the sum of the ``sheath fields'' vanishes for the geometry considered here:
In order to satisfy $\curl\vecEb=0$ and $\grad\vdot\vecEb=0$ everywhere inside the body
each sheath field must be uniform (for an infinite slab), so that Equation~(\ref{eq-ebodypar})
is actually valid throughout the body interior.
But the sheath fields sum to zero so the total electric field inside the body is $\vecEb=0$.
We conclude that it is possible for induction heating to vanish when one accounts for
the interaction of a body with the flow around it.
Of course the fact that $\vecEb=0$ {\it exactly}\/ is a consequence of the extreme symmetry
of an infinite slab.
For realistic bodies of finite extent the electric field will not be zero, but instead
attain some value dependent on the nonzero magnetic field gradients associated with
departures from planar symmetry.

The electric field in the plasma must vary continuously with $z$, from \vecEzero\ at infinity
to zero at the slab surfaces.
One could calculate $\vecE_{\rm p}(z)$ in principle from the velocity solution by evaluating
\be
\vecE_{\rm p}(z) = -\frac{\vecv(z)}{c}\cross\vecBzero .
\ee 
However it is well known that the solution for \vecv\ does not exist
for an infinite plane: the solution to Equation~(\ref{eq-parallelflowxmomentumeq}) cannot
satisfy boundary conditions (\ref{eq-parallelflowvxinftybc})
and (\ref{eq-parallelflowvxsurfbc}) simultaneously because the shear layer
is infinitely thick.
Since the magnetic forces vanish for this geometry,
a rough picture of the variation can be obtained by considering
viscous flow past a semi-infinite flat plate, for which the shear layer
has finite thickness $\delta$.
Then the boundary condition at infinity can be replaced by
\be 
v_x(W+\delta) = v_{0}.
\label{eq-parallelflowvxsinftybcapprox}
\ee
From scaling arguments, Blasius (1908) inferred that
\be
\delta = \delta_{B} (x) \approx \left( \frac{\alpha x}{\rho v_{0}} \right)^{1/2}
\ee
for flow past a semi-infinite plate, where $x$ is the distance downstream from the leading
edge of the plate. 
The thickness of the shear layer varies with $x$ but the rate of change,
\be
\frac{d \delta_{B}}{dx} = \frac{1}{2} \left( \frac{\alpha}{\rho v_{0} x} \right)^{1/2},
\ee
goes to zero as $x\goesto\infty$.
Thus, $\delta$ becomes approximately independent of $x$ for large $x$
and the flow approximates flow past an infinite slab.
We will interpret $\delta$ in Equation~(\ref{eq-parallelflowvxsinftybcapprox})
to be $\delta_{B}(x_{0})$,  where $x_0$ is a large but unspecified value of $x$
for which the approximation holds to some desired precision. 
Then the solution to Equation~($\ref{eq-parallelflowxmomentumeq}$)
subject to the approximate boundary conditions ($\ref{eq-parallelflowvxsurfbc}$)
and ($\ref{eq-parallelflowvxsinftybcapprox}$) is
\be
v_{x} (z) \approx \left(\frac{z-W}{\delta - W}\right)\,\vzero
\ee
and the electric field in the plasma is
\be
\vecE_{\rm{p}}(z) \approx -\left(\frac{z-W}{\delta - W}\right) E_{0} \zhat.
\label{eq-eparapp}
\ee
That the electric field in Equation~(\ref{eq-eparapp}) has nonzero divergence
shows that the approximation for $\delta$ is unphysical.
However the nonzero macroscopic charge density implied by Gauss's Law is
\be
e (n_{\rm e}-n_{\rm i}) = \frac{v_0 B_0}{4 \pi c (\delta - W)} ,
\ee
which goes to zero as $\delta\to\infty$.

\subsection{Perpendicular Field Geometry}
\label{sec-bperp}

Now consider the case where the ambient magnetic field
is perpendicular to the slab faces, with $\vecBzero = \Bzero\,\zhat$.
Because the slab is infinite all variables depend only on $z$
and the electric field can be calculated trivially.
Far from the body, where magnetic field gradients vanish,
the electric field in the body frame is just the motional field
with components\footnote{Looking at the velocity and magnetic field solutions in
Equations~($\ref{eq-sol-vx}$)--($\ref{eq-sol-By}$), we note that the motional electric field in
the plasma also contains an unphysical non-zero $z$-component ($E_{\rm{p} \it{z}}$) 
in this case.
However $E_{\rm{p} \it{z}}$ is found to be of order $\sim \frac{v_{0} B_{1x,y}}{c}$,
which is $\ll E_{0}$ and is therefore neglected.}
\be
\vecEzero = \left[0,v_0B_0/c,0\right].
\label{eq-ezero-comp}
\ee
For steady flow the electric field at an arbitrary point in the
plasma\footnote{To avoid numerous disclaimers we imply henceforth that
``in the plasma'' means points in the plasma outside of plasma sheaths.}
must also satisfy $\curl\vecE_{\rm p}=0$ which, together with the fact that $\vecE_{\rm{p}}$ only depends on $z$, implies that $\vecE_{\rm p}=\vecEzero$ everywhere in the plasma.
Now the electric field inside the body is also uniform because
$\grad\vdot\vecE_{\rm b}=0$ and  $\curl\vecE_{\rm b}=0$.
Noting that $\vecE_{\rm p}$ is tangent to the slab surfaces, and that the
tangential component of \vecE\ does not change across the surfaces,
we see that $\vecE_{\rm b}=\vecEzero$ everywhere inside the body,
and hence that $\vecE=\vecEzero$ {\em everywhere}.
This simple example shows that it is possible to have electric
fields $\sim\!\vecEzero$ inside the body.
In the following subsections we show how this is physically possible 
by calculating the motional, Ohm, Hall, and ambipolar
contributions to \vecEp.

\subsubsection{The Body Interior}
\label{sec-perpi}

For infinite planar geometry the condition $\grad\vdot\vecB=0$ implies that \Bz\ is constant, so
\be
\Bz=\Bzero
~~~~ -\infty < z < +\infty.
\ee
We assume a solution of the form $\vecBb = \left[\Bonex(z),\Boney(z),B_0\right]$,
where the notation indicates that \Bonex\ and \Boney\ are regarded
as small perturbations: $\Bonex,\Boney \ll B_0$.
This is necessary because the $x$- and $y$-components of the magnetic field are small outside 
of the slab (Section~\ref{sec-perpe}) and it would not be possible
to satisfy the boundary conditions on \vecB\ at the slab
surfaces, in general, if the components of the magnetic field were small outside and large inside.
Because \vecEb\ is known, \Bonex\ and \Boney\ follow from {\Ampere}'s Law.
Writing out the $x$ and $y$ components of Equation~(\ref{eq-ampere}) gives
\be
\frac{d\Bonex}{dz} =  \frac{4\pi\mub\sigb E_y}{c}
\label{eq-amperebx}
\ee
and
\be
\frac{d\Boney}{dz} = -\frac{4\pi\mub\sigb E_x}{c},
\label{eq-ampereby}
\ee
where it has been assumed for convenience that \mub\ is constant
and \sigb\ is the electrical conductivity inside the body.
It is apparent from the symmetry of the problem that $\Bonex(z)$ and $\Boney(z)$ must
either be odd or even functions.
However even functions are excluded:
\vecBone\ is odd outside the body (Section~\ref{sec-perpe}) and
it would not be possible to satisfy the boundary conditions at the body surfaces in
general if \vecBone\ were even inside and odd outside.
The boundary conditions on Equations~(\ref{eq-amperebx})--(\ref{eq-ampereby}) are therefore
\be
\Bonex(0) = 0
\label{amperexbc}
\ee
and
\be
\Boney(0) = 0.
\label{eq-ampereybc}
\ee

The solution of Equations~(\ref{eq-amperebx})--(\ref{eq-ampereybc}) is
\be
\Bonex(z) = \frac{4\pi\mub\Ezero}{c}\,
\int_0^{z}\ \sigb\left(\zp\right)\,d\zp
\label{eq-bxbody}
\ee
and
\be
\Boney(z) = 0,
\label{eq-bybody}
\ee
where we have exploited the fact that $\vecE=\vecEzero$ everywhere.
In practice Equation~(\ref{eq-bxbody}) sets an upper limit on the conductivity, above which our
treatment of \Bonex\ as a perturbation would break down.
The conductivity appears inside the integral in Equation~(\ref{eq-bxbody}) to allow
for the sensitive dependence of $\sigma_{\rm{b}}$ on the temperature (Equation~[\ref{eq-sigarr}]), which varies
with $z$ if the body is heated significantly.


\subsubsection{The Shear Flow}
\label{sec-perpe}

Now consider the velocity and magnetic field in the plasma.
We seek solutions to Equations~(\ref{eq-momentum}) and (\ref{eq-induction}) of the form
\be
\vecv = \left[v_x(z), v_y(z), 0\right],
\ee
\be
\vecBp = \vecBzero + \left[\Bonex(z), \Boney(z), 0\right],
\ee
and $P = P(z)$, where we have anticipated that the magnetic field is deflected only slightly
and regard \Bonex\ and \Boney\ as infinitesimal perturbations.
Substituting the assumed solutions into Equations~(\ref{eq-momentum}) and (\ref{eq-induction}),
setting the time derivatives to zero, and retaining terms up to first order in \Bonex\
and \Boney\ yields six differential equations for the components of \vecv\ and \vecB.
However the $z$ component of the linearized induction equation is satisfied identically.
The $z$ component of the linearized momentum equation reduces to
an equation for the pressure gradient required to maintain hydrostatic
equilibrium in the $z$ direction; it decouples from the other equations
when the temperature dependence of the transport coefficients is neglected.
This leaves four ODEs for \vx, \vy, $B_{1x}$, and $B_{1y}$:
\be
\left(\etao+\etaa\right)\,\frac{d^2\Bonex}{dz^2} + \etah\,\frac{d^2\Boney}{dz^2} + \Bzero\,\frac{dv_x}{dz}= 0,
\label{eq-inductionx}
\ee
\be
-\etah\,\frac{d^2\Bonex}{dz^2} + \left(\etao+\etaa\right)\,\frac{d^2\Boney}{dz^2} + \Bzero\,\frac{dv_y}{dz} = 0,
\label{eq-inductiony}
\ee
\be
\frac{\Bzero}{4\pi}\,\frac{d\Bonex}{dz} + \alpha\,\frac{d^2v_x}{dz^2} = 0,
\label{eq-momentumx}
\ee
and
\be
\frac{\Bzero}{4\pi}\,\frac{d\Boney}{dz} + \alpha\,\frac{d^2v_y}{dz^2} = 0.
\label{eq-momentumy}
\ee
It follows from Equations~(\ref{eq-inductionx})--(\ref{eq-momentumy}) that
\vx\ and \vy\ are even functions of $z$ and that \Bonex\ and \Boney\ are odd.
Thus it is only necessary to calculate the flow explicitly for $z>0$.

Equations (\ref{eq-inductionx})--(\ref{eq-momentumy}) contain the
second derivatives of \vx, \vy, $B_{1x}$, and $B_{1y}$ so that
eight boundary conditions are required.
At the body surface the velocity satisfies the no-slip condition,
\be
v_x(W) = 0,
\ee
\be
v_y(W) = 0,
\ee
and the tangential components of the magnetic field satisfy
Equation~(\ref{eq-htanbc}):
\be
\frac{\Bonex\left(W^-\right)}{\mub} = \frac{\Bonex\left(W^+\right)}{\mup},
\label{eq-bxbc}
\ee
and  
\be
\frac{\Boney\left(W^-\right)}{\mub} = \frac{\Boney\left(W^+\right)}{\mup}.
\label{eq-bybc}
\ee
Far from the body the velocity approaches the free-stream velocity,
\be
\lim_{z\rightarrow +\infty} v_x = \vzero,
\ee
\be
\lim_{z\rightarrow +\infty} v_y = 0,
\ee
and the magnetic field perturbations approach constant (infinitesimal) values:
\be
\lim_{z\rightarrow +\infty} \Bonex = {\rm const.},
\ee
and
\be
\lim_{z\rightarrow +\infty} \Boney = {\rm const}.
\ee
For a body of finite size \Bonex\ and \Boney\ would {\it vanish}\/ at infinity.
However our infinite planar slab acts as an infinite current sheet;
thus it produces a magnetic field whose magnitude is constant everywhere outside the body.

\begin{figure}
\figurenum{14}
\plotone{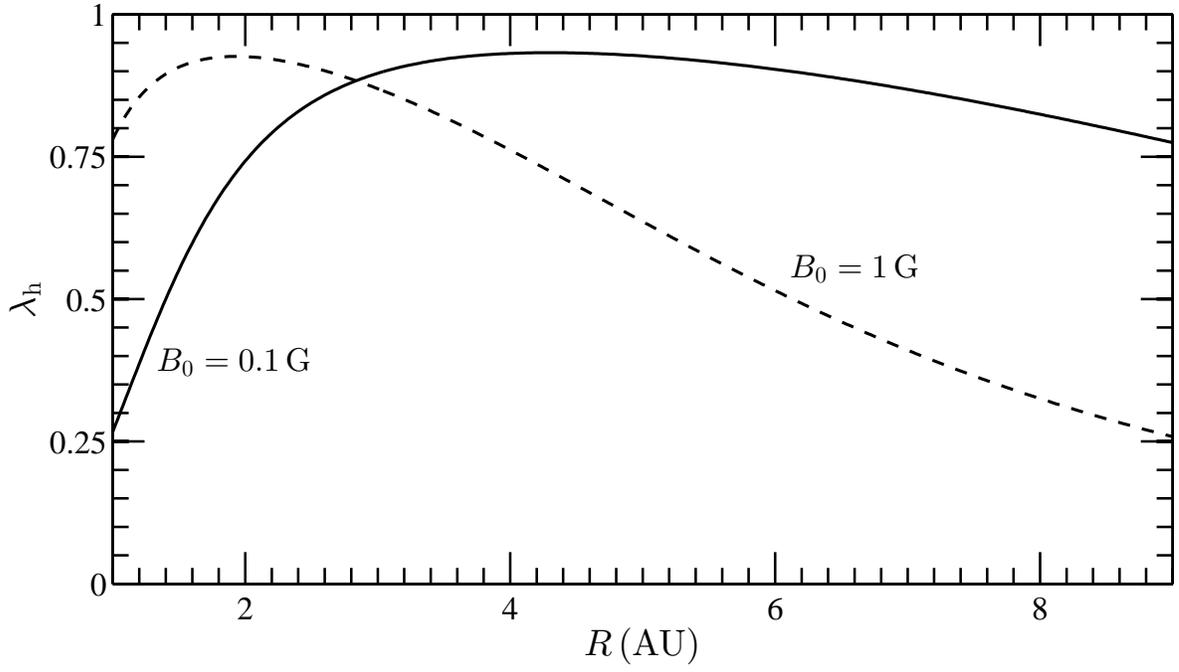}
\caption{
The parameter \lamh\ defined in Equation~(\ref{eq-def-lambda}) for a dust-free 
plasma is plotted vs distance from the central star.
The two curves correspond to different $B_0$ values as indicated.
}
\label{fig-lambdah}
\end{figure}

\begin{figure}
\figurenum{15}
\plotone{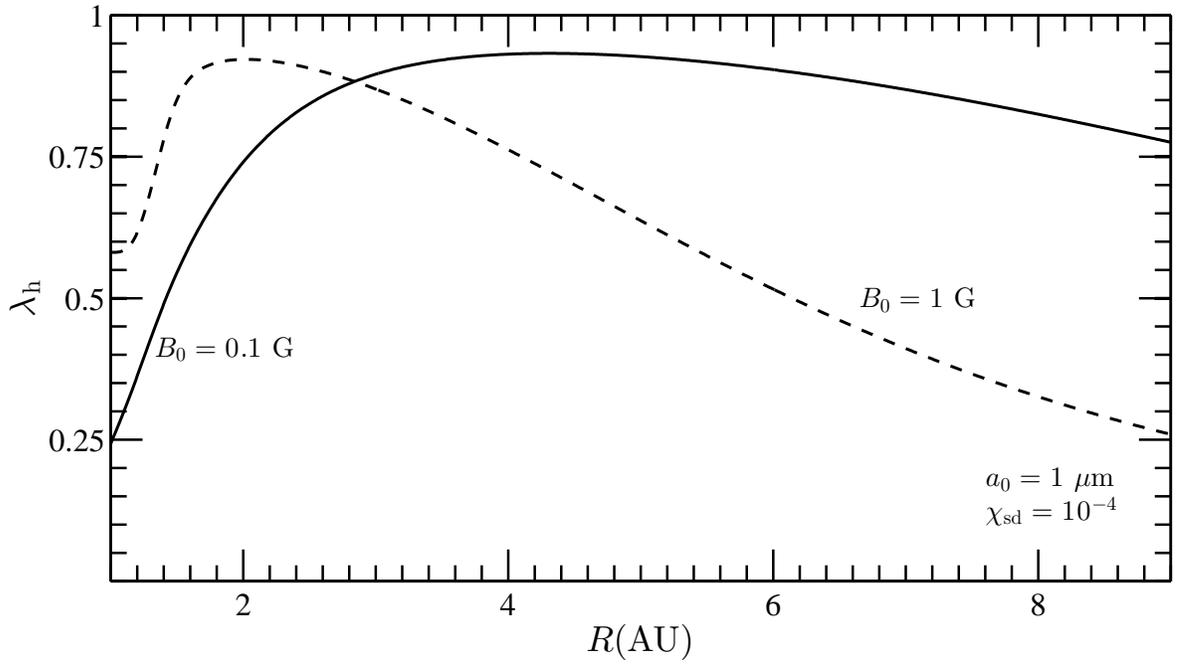}
\caption{
As in Figure~\ref{fig-lambdah}, but with $\chi_{sd} = 10^{-4}$ and
$a = 1$~\micron.
}
\label{fig-lambdahdust}
\end{figure}

We introduce the dimensionless position $\zbar\equiv z/\Lsf$, 
velocity, $\vecu=\vecv/\vzero$, and magnetic induction,
$\vecb\equiv\vecBone/\Bstar$, where
\be
\Bstar \equiv v_0\,\left(\alpha/\eta\right)^{1/2}
\label{eq-def-Bstar}
\ee
is the natural unit of magnetic induction.
In dimensionless units Equations~(\ref{eq-inductionx})--(\ref{eq-momentumy}) become
\be
\left(1-\lamh\right)\bxpp + \lamh\bypp + \uxp = 0,
\label{eq-d-inductionx}
\ee
\be
-\lamh\bxpp + (1-\lamh)\bypp + \uyp = 0,
\label{eq-d-inductiony}
\ee
\be
\bxp + 4\pi\uxpp = 0,
\label{eq-d-momentumx}
\ee
and
\be
\byp + 4\pi\uypp = 0,
\label{eq-d-momentumy}
\ee
where primes indicate derivatives with respect to \zbar.
Apart from the boundary conditions the exterior
solution depends on just one dimensionless parameter,
\be
\lamh \equiv \frac{\etah}{\etah+\etao+\etaa},
\label{eq-def-lambda}
\ee
which varies from zero (no Hall effect) to unity
(Hall effect only, no Ohmic dissipation or ambipolar diffusion).
Figures~\ref{fig-lambdah}--\ref{fig-lambdahdust} show the value of
\lamh\ as a function of $R$ for two cases: the dust-free case, and the
case where $\chi_{sd} = 10^{-4}$ and the dust is assumed to be
single-sized with $a = 1$~$\mu$m.  We point out that
in both cases the shear flows are ``Hall dominated'' in the sense that 
$\lamh > \frac{1}{2}$ for most radii of interest.\footnote{Although
the Hall effect dominates the dynamics of the flows studied here, 
this may not be the case for other flows, particularly in
the context of the magnetorotational instability (MRI; Balbus \& Hawley 
1991).  For example,
3D numerical MHD simulations done by Sano \& Stone (2002) show that the Hall
effect may be neglected when determining the condition for MRI turbulence,
even though the value of the Hall diffusivity may exceed that of the Ohmic 
and ambipolar diffusivities.} 

\begin{figure}
\figurenum{16}
\plotone{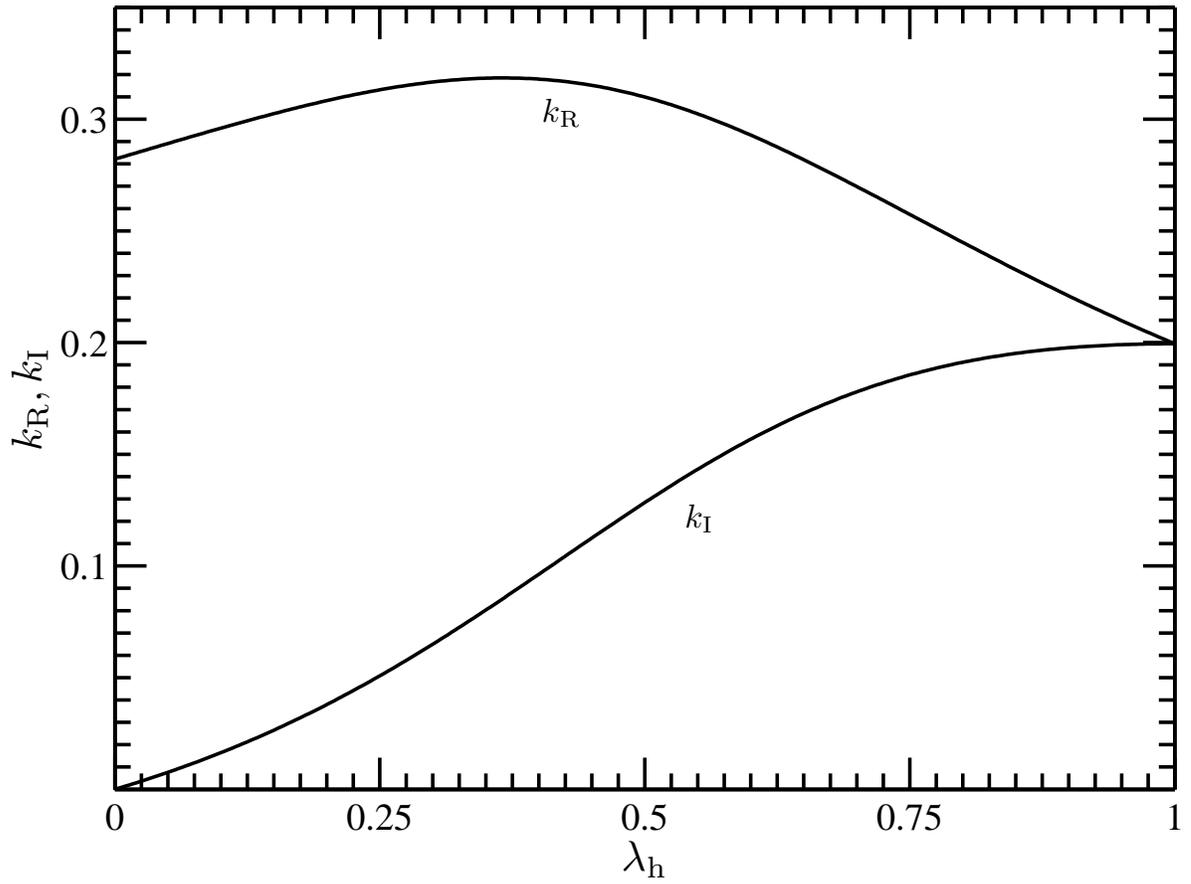}
\caption{
Dimensionless quantities \kR\ and \kI\ defined in eqs.~(\ref{eq-krdef})
and (\ref{eq-kidef}) plotted vs.\ the dimensionless parameter \lamh. 
}
\label{fig-kplot}
\end{figure}

Equations (\ref{eq-d-inductionx})--(\ref{eq-d-momentumy}) are solved
in Appendix~\ref{app-soln}, where we show that
\be
v_x = \left[1-\cos\kI\xi\ e^{-\kR\xi}\right]~\vzero,
\label{eq-sol-vx}
\ee
\be
v_y = \sin\kI\xi\ e^{-\kR\xi}~\vzero,
\label{eq-sol-vy}
\ee
\be
\Bonex = 
\Bonex\left(W^+\right) +
\sqrt{\frac{4\pi}{D}}\,\left[\,
\cos\theta/2-\cos\left(\kI\xi-\theta/2\right)\,e^{-\kR\xi}
\,\right]\,\Bstar ,
\label{eq-sol-Bx}
\ee
and
\be
\Boney = \sqrt{\frac{4\pi}{D}}\,\left[\,
\sin\theta/2 + \sin\left(\kI\xi-\theta/2\right)\,e^{-\kR\xi}
\,\right]\,\Bstar ,
\label{eq-sol-By}
\ee
where
\be
\xi \equiv \frac{z-W}{\Lsf}
\label{eq-xidef}
\ee
is dimensionless height above the body surface and
\be
\Bonex\left(W^+\right) = \frac{4\pi\mup\Ezero}{c}\,
\int_0^{W}\ \sigb\left(\zp\right)\,d\zp.
\ee
The solution depends on \lamh\ via the dimensionless quantitities
\be
\kR\left(\lamh\right) \equiv \frac{\cos\theta/2}{\sqrt{4\pi D}}
\label{eq-krdef}
\ee
and
\be
\kI\left(\lamh\right) \equiv \frac{\sin\theta/2}{\sqrt{4\pi D}}
\label{eq-kidef}
\ee
(Figure~\ref{fig-kplot}), where
\be
\theta \equiv \arctan\left(\frac{\lamh}{1-\lamh}\right)
\label{eq-thetadef}
\ee
and
\be
D \equiv \sqrt{ \lamh^2 + \left(1-\lamh\right)^2}.
\label{eq-Ddef}
\ee
Notice that $0\le\theta\le\pi/2$ and $\kR,\kI \ge 0$.  Now that the flow velocity 
and magnetic field solutions are known, the motional, Ohm,
Hall, and ambipolar components of the electric field can be found by evaluating
Equations~(\ref{eq-eparts})--(\ref{eq-eamb}).  These fields will be discussed in the
next section.

\section{Discussion}
\label{sec-disc}

\subsection{Electric Fields in Multifluid Shear Flows}
\label{sec-efanddust}

As demonstrated in the previous section, the solutions for the velocity, magnetic, 
and electric fields depend only on the transport coefficients $\alpha$, $\etah$, 
$\etao$, and $\etaa$.  Although the shear viscosity $\alpha$ 
does not depend on the amount of small dust in the disk, dust may
profoundly affect the 
diffusivities (see Section~\ref{sec-eta}, 
Figures~\ref{fig-eta_chisd0}--\ref{fig-eta_chisd1e-4}).  In order to 
illustrate the effects of dust on the components of the electric field we will 
consider two cases: (1) the limit where all the dust is locked into millimeter or
larger grains and their is no small dust present in the disk ($\chisd=0$) 
and (2) a case where the abundance of small grains is taken to be 
$\chisd=10^{-4}$ and all the small dust is the same size with a radius 
of $a=1$~$\mu$m.

\begin{figure}
\figurenum{17}
\plotone{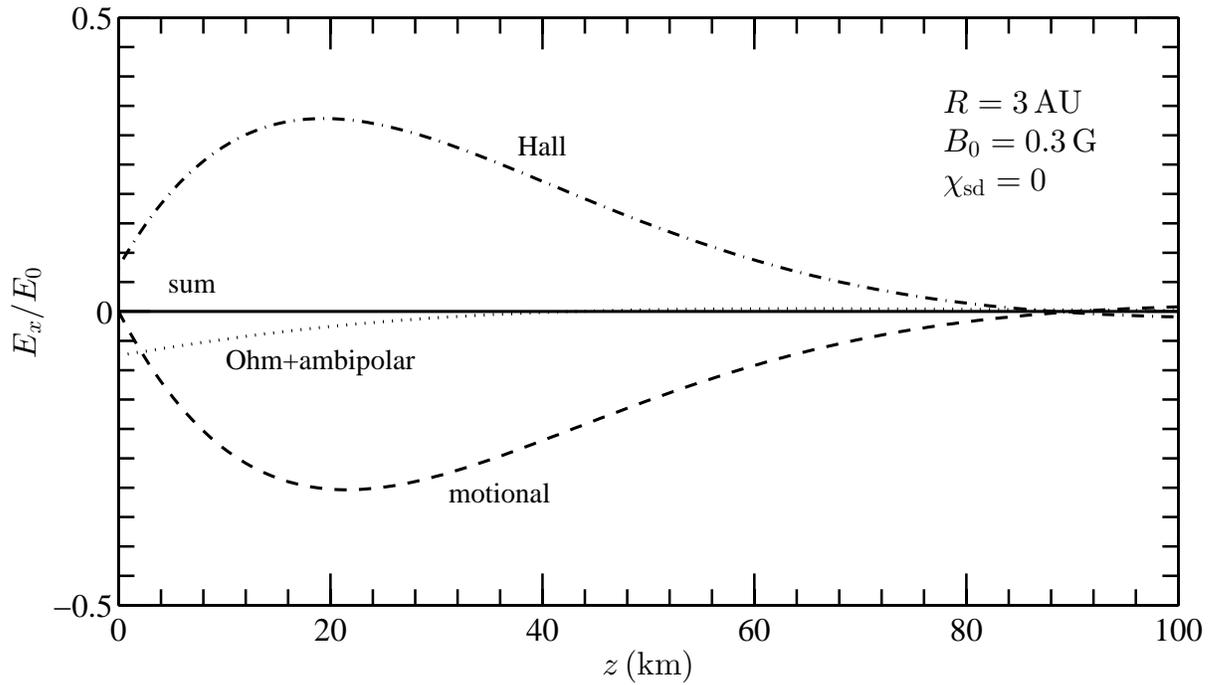}
\epsscale{1.1}
\caption{
The x-component of the electric field, \vecEp, in units of \Ezero\ as a function of
height $z$ above the body surface.
The values of $R$ and $B_0$ are indicated.
The motional, Ohm+ambipolar, and Hall fields are 
indicated.
The curve labeled ``sum'' is the total electric field.
}
\label{fig-efieldxnd}
\end{figure}

\begin{figure}
\figurenum{18}
\plotone{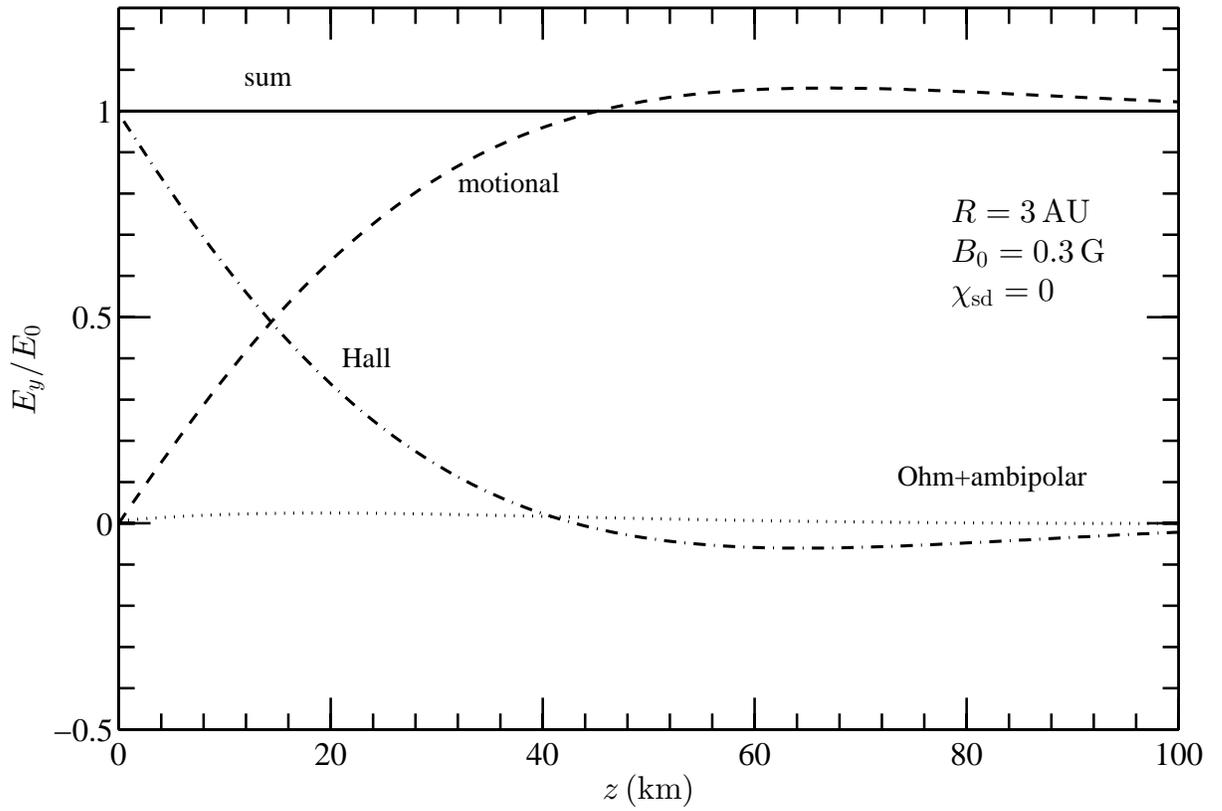}
\epsscale{1.1}
\caption{
As in Figure~\ref{fig-efieldxnd} but for the $y$ component
of the electric field.
}
\label{fig-efieldynd}
\end{figure}

Figures~\ref{fig-efieldxnd}--\ref{fig-efieldynd} 
describe the electric field in a perpendicular shear flow when no small dust 
($\chi_{\rm{sd}}$=0) is present in the disk.  These electric fields were computed 
using our disk model (Section~\ref{sec-tcoeff}) where the 
diffusivities were calculated at $R=3$\,AU assuming $\vecB_{0}=0.3$\,G.  The 
plots are independent of \vzero\ because \vecE\ is plotted in units of \Ezero.  
The motional, Ohm+ambipolar\footnote{The sum of the Ohm and ambipolar E fields 
is plotted because, for perpendicular geometry, they are proportional to one 
another.} and Hall electric fields are indicated and the total field \vecEp\ is 
also plotted.  The figures show the inevitable result that $\vecEp=\vecEzero$.
What is interesting about the figure is how this result comes about.
The Ohm and ambipolar fields are relatively small.
The motional field dominates $\left(\vecEm \approx \vecEzero\right)$
far from the body and the Hall field dominates $\left(\vecEh \approx 
\vecEzero\right)$ close to the body. For a body of realistic shape the four 
contributions to the electric field would obviously not ``conspire'' to sum to 
\vecEzero\ as in Figures~\ref{fig-efieldxnd}--\ref{fig-efieldynd}.
However since the magnitudes of the various magnetic field gradients should
not be very sensitive to body shape (for large bodies), it seems clear that real 
shear flows will have electric fields $\sim\Ezero$.

\begin{figure}
\figurenum{19}
\plotone{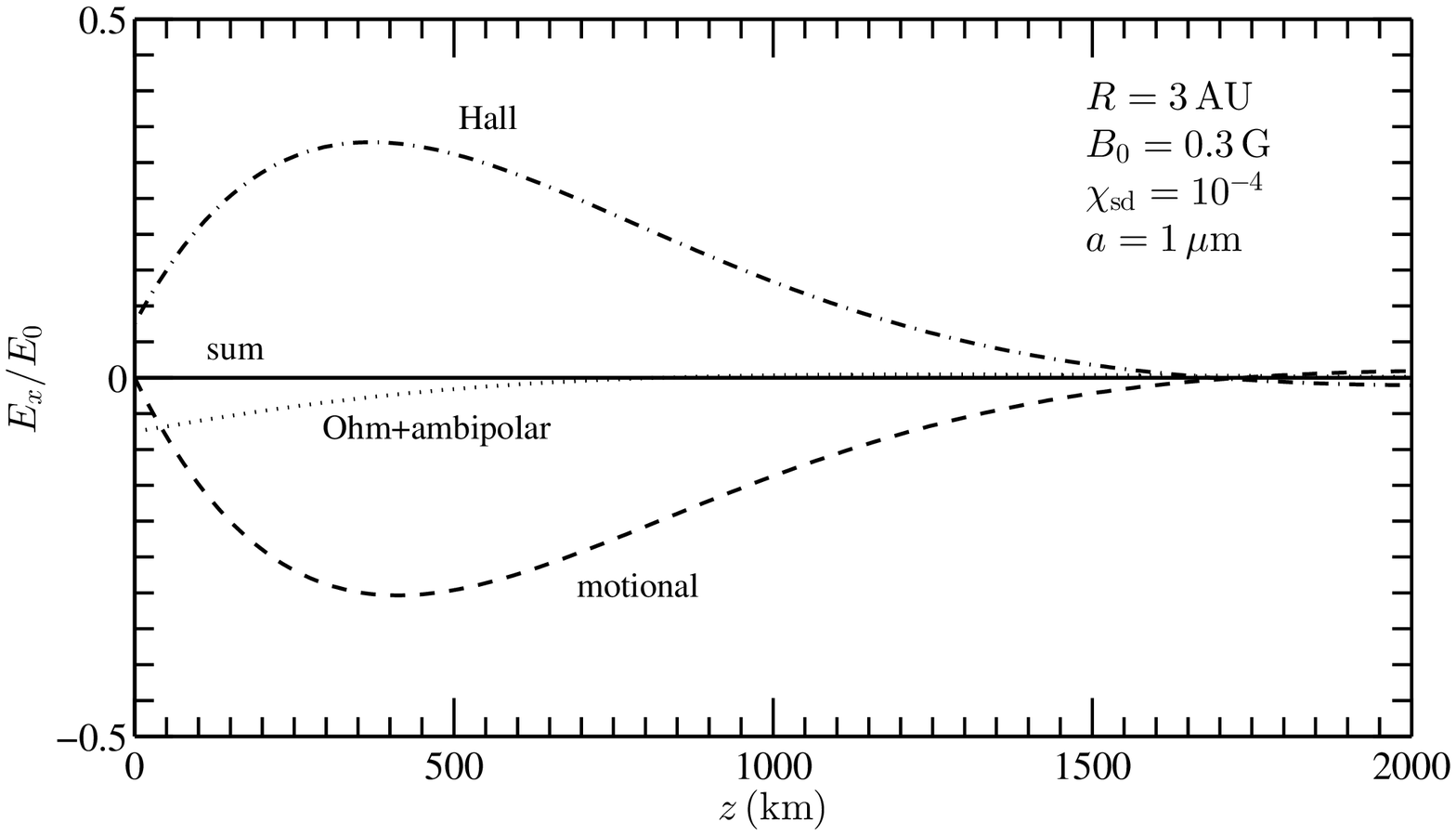}
\epsscale{1.1}
\caption{
As in Figure~\ref{fig-efieldxnd} but
for $\chisd=10^{-4}$.
The dust is a single-size distribution
with $a=1$\,\micron.
}
\label{fig-efieldxyd}
\end{figure}

\begin{figure}
\figurenum{20}
\plotone{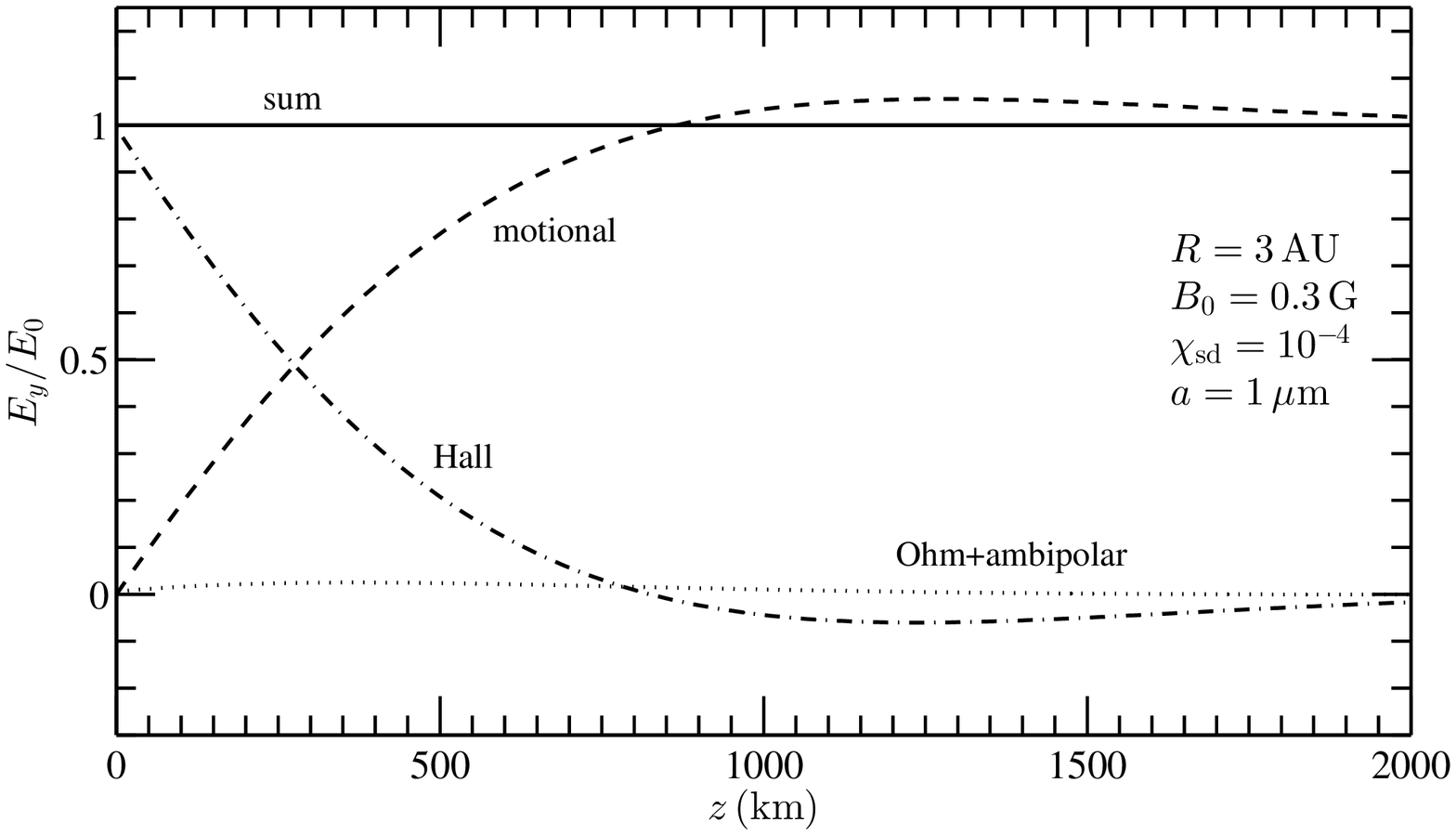}
\epsscale{1.1}
\caption{
As in Figure~\ref{fig-efieldynd} but
for $\chisd=10^{-4}$.
The dust is a single-size distribution
with $a=1$\,\micron.
}
\label{fig-efieldyyd}
\end{figure}

If small dust is present in the disk, the diffusivities and thus the electric 
field will be affected.  Figures~\ref{fig-efieldxyd}--\ref{fig-efieldyyd} describe
the electric field for the case analogous to 
Figures~\ref{fig-efieldxnd}--\ref{fig-efieldynd} except with $\chisd=10^{-4}$ and $a=1$\,$\mu$m. Comparing 
Figures~\ref{fig-efieldxyd}--\ref{fig-efieldyyd} with 
Figures~\ref{fig-efieldxnd}--\ref{fig-efieldynd}, it is
evident that even with small dust with abundances as high as $\chisd=10^{-4}$ 
present in the disk, the flow remains Hall-dominated at $R=3$\,AU with a total 
electric field $\vecEp=\vecEzero$.  However dust does produce a dramatic effect on 
the characteristic length scale of the shear layer $L_{\rm{sf}}$, which grows from 
$\sim 10$\,km in the absence of small dust to $\sim 100$\,km when small dust is 
present.  This effect is important to note because our approximation of the asteroid 
as an infinite plane is only valid when $L_{\rm{body}} \gg L_{\rm{sf}}$.  Taking $L_{\rm{body}} \sim 
100$\,km for a typical large asteroid, we find that our solutions should provide a 
decent approximation in a dust-free plasma, however their validity should be 
questioned when \micron-sized\ dust is present in the abundances discussed 
above.  In the extreme case that $\sim~0.1$ \micron\ dust 
grains are present
in the midplane, the magnetic diffusivities will be even greater 
corresponding to
an even larger characteristic length scale of shear layer and rendering our 
approximation of an asteroid as an infinite plane completely unrealistic.
In this case numerical calculations of the flow over a finite body would
be required.

\subsection{The Importance of Self Heating}
\label{sec-sheat}

In Section~\ref{sec-bperp} we solved the equations of motion for a multifluid 
shear flow on the assumption that heating of the flow by viscous dissipation and 
ion-neutral scattering is negligible.  This will be a good assumption if the 
heating rates are $\ll\Lambda$, where $\Lambda$ is the radiative cooling rate of 
the plasma.  

\subsubsection{Radiative Cooling}
\label{subsec-radiativecooling}

\begin{figure}
\figurenum{21}
\plotone{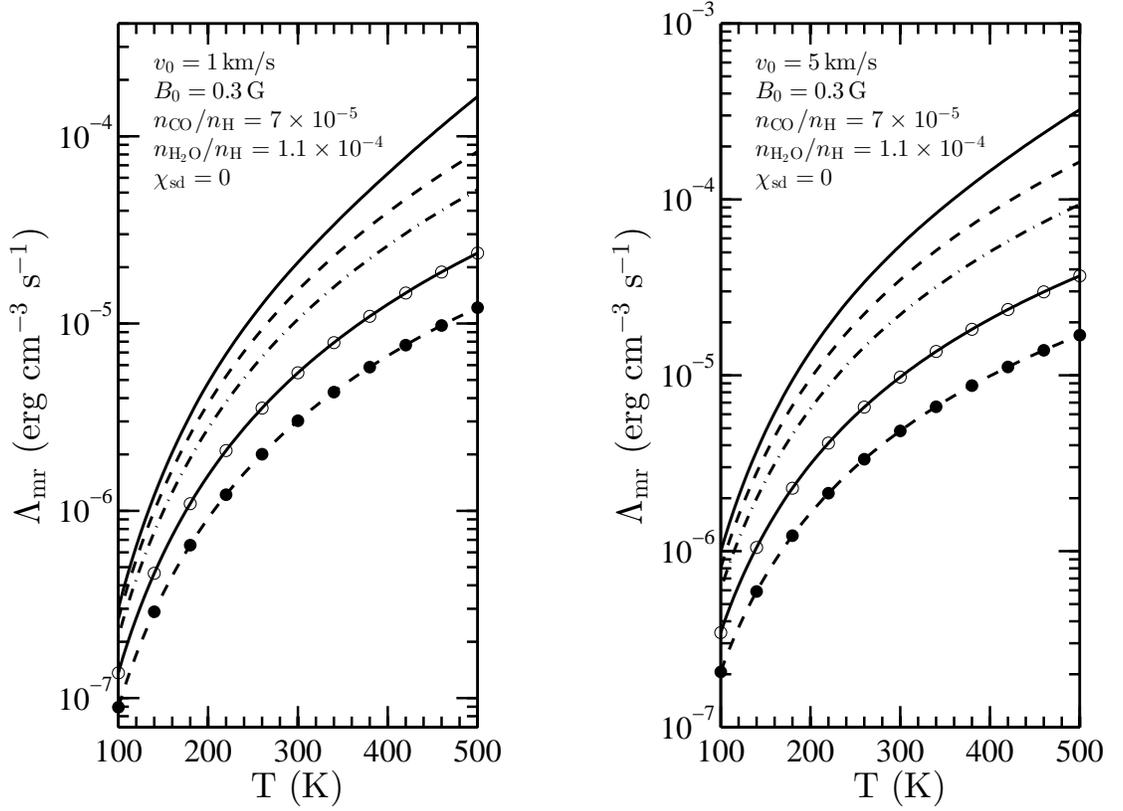}
\caption{
 The rate of radiative cooling by H$_2$, CO, and H$_2$O 
molecules plotted vs. $\!$the gas temperature, computed using the cooling rate 
function of Neufeld \& Kaufman (1993).  Each curve represents the cooling rate
a specific distance from the central star as follows: $R=2$\,AU (solid),
$R=3$\,AU (dashed), $R=4$\,AU (dashed-dotted), $R=6$\,AU (solid with open circles),
and $R=8$\,AU (dashed with closed circles).  The free-stream flow velocity, ambient
magnetic field, molecular abundances, and small grain abundance are indicated.
}
\label{fig-rcoolnd}
\end{figure}

\begin{figure}
\figurenum{22}
\plotone{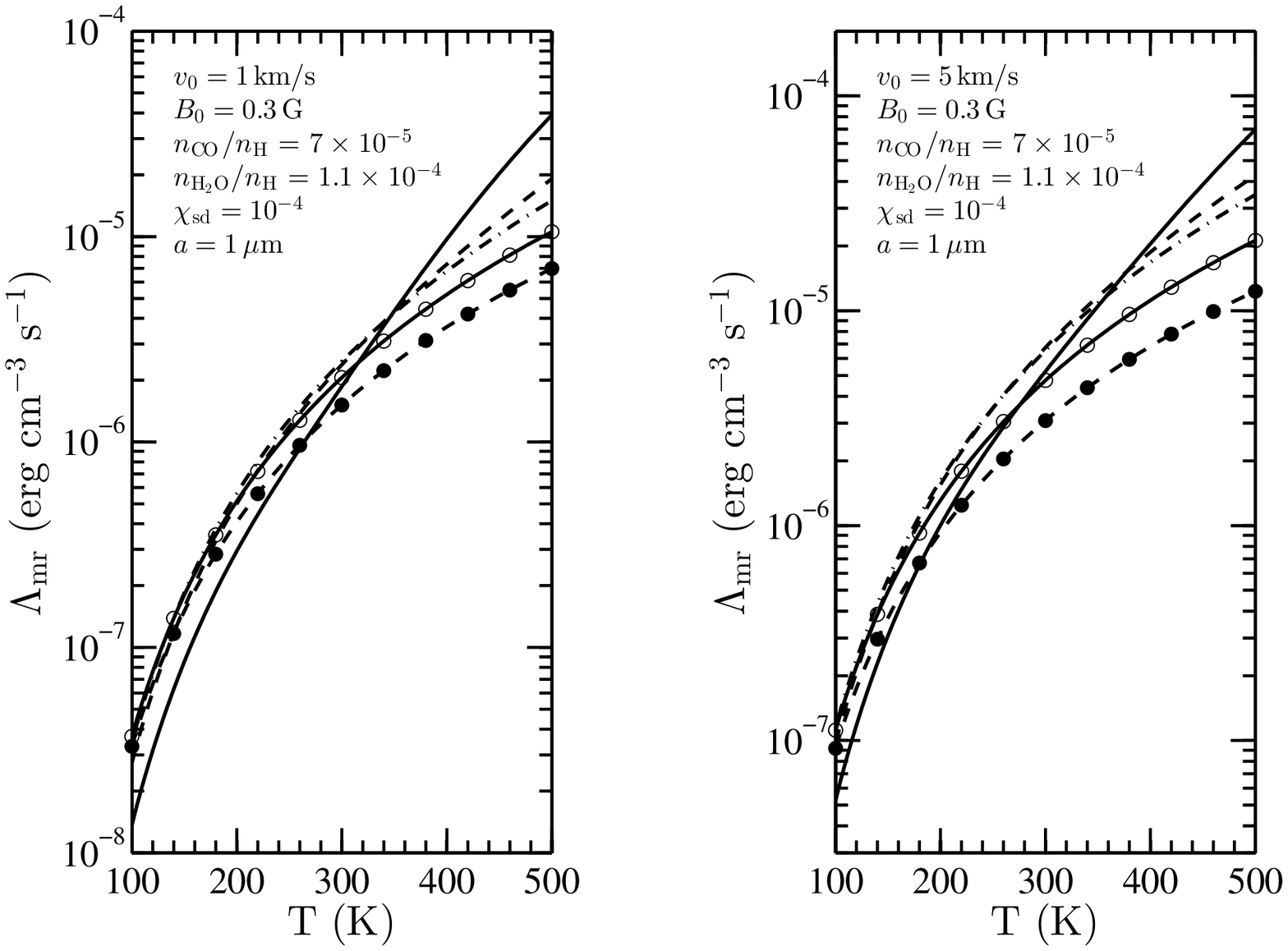}
\caption{
As in Figure~\ref{fig-rcoolnd} but
for $\chisd=10^{-4}$.
The dust is a single-size distribution
with $a=1$\,\micron.
}
\label{fig-rcoolyd}
\end{figure}

\begin{figure}
\figurenum{23}
\plotone{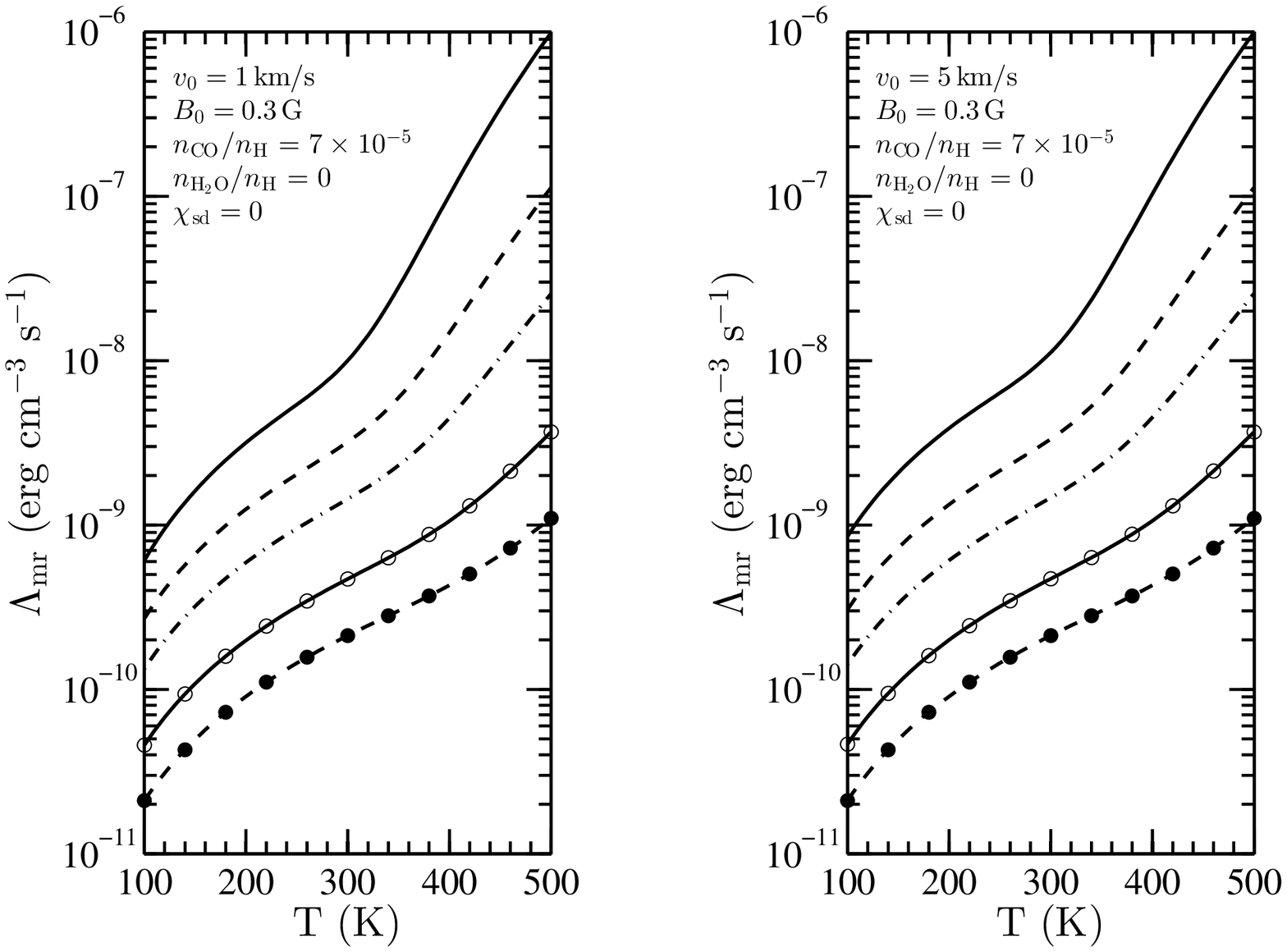}
\caption{
As in Figure~{\ref{fig-rcoolnd}}, but with $n_{\rm{H_{2}O}}/n_{\rm{H}}=0$.
}
\label{fig-rcoolnwnd}
\end{figure}

\begin{figure}
\figurenum{24}
\plotone{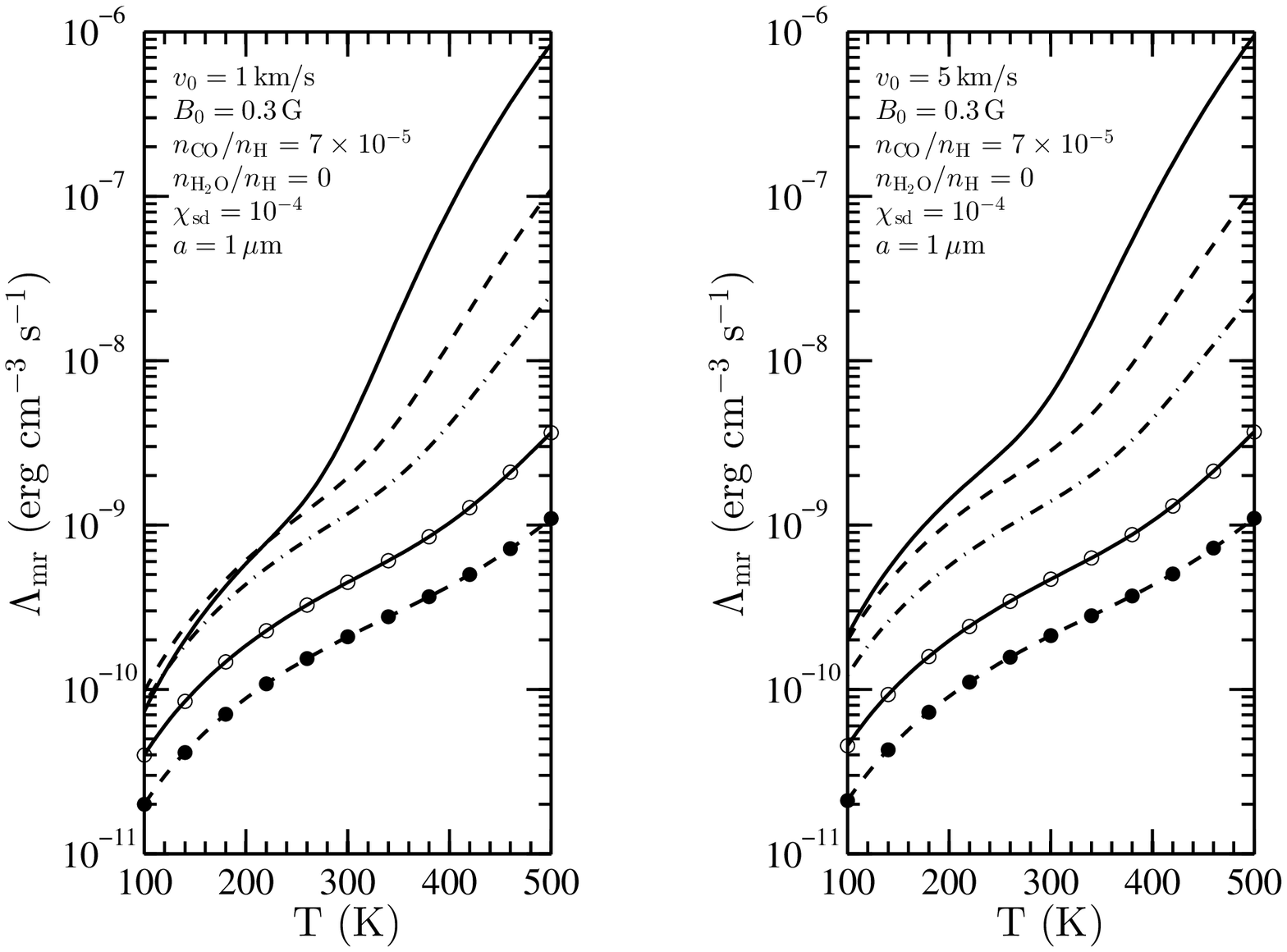}
\caption{
As in Figure~{\ref{fig-rcoolyd}}, but with $n_{\rm{H_{2}O}}/n_{\rm{H}}=0$.
}
\label{fig-rcoolnwyd}
\end{figure}

In the absence of dust, radiative cooling is mainly due
to rotational and vibrational transitions of 
H$_{\rm{2}}$, CO, and H$_{\rm{2}}$O molecules.  We calculate the
cooling rate due to molecular line emission by interpolating the 
results given by Neufeld $\&$ Kaufman~(1993) at several values of $R$.  These cooling
rates depend on the velocity gradients in the plasma, which determine the probability
that an emitted photon has of escaping (See Neufeld $\&$ Kaufman 1993).
Since the viscous dissipation
and ion-neutral scattering heating rates are greatest at the asteroid surface (See
Equations~[\ref{eq-gvis}] and [\ref{eq-gin}] below), 
we use the
velocity gradients in the shear flow 
\be
\frac{d\vecv}{dz} = 
\sqrt{\left(\frac{dv_{x}}{dz}\right)^{2} + \left( \frac{dv_{y}}{dz} \right)^{2}}
\ee
at the body surface ($\xi=0$).  The radiative cooling rates also depend strongly on the
abundance of water vapor in the gas, which may freeze out.  Since the location of the 
snow line in 
protoplanetary disks is still a topic of active research (e.g., Martin $\&$ Livio 2012
and references therein), 
we assume that the abundance of water $n_{\rm{H_{2}O}}/n_{\rm{H}}$ is constant for
all $R$ and calculate the cooling rate for two cases:  
$n_{\rm{H_{2}O}}/n_{\rm{H}}=1.1\times10^{-4}$ (Figures~\ref{fig-rcoolnd}--\ref{fig-rcoolyd})
corresponding to a standard abundance 
water in the gas phase, and $n_{\rm{H_{2}O}}/n_{\rm{H}}=0$ (Figures~\ref{fig-rcoolnwnd}--\ref{fig-rcoolnwyd}) assuming that all the water 
has frozen out.

\begin{figure}
\figurenum{25}
\plotone{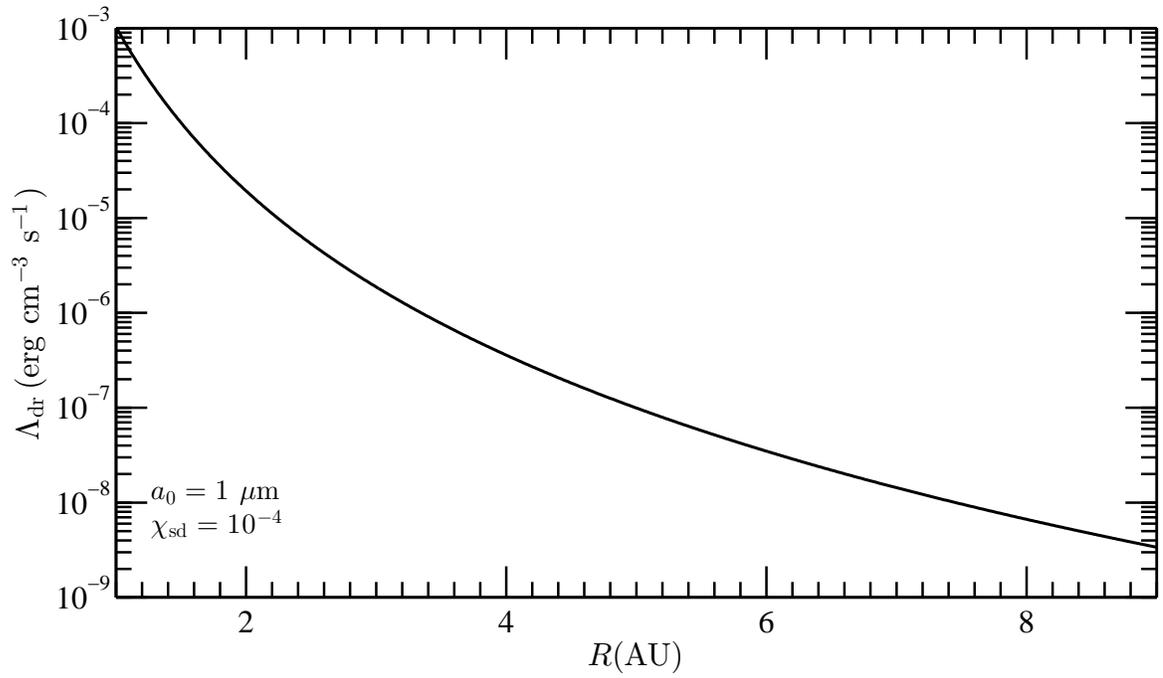}
\caption{
The cooling rate due to thermal emission from dust grains is plotted vs. distance $R$ from the central star.  The dust is a single-size distribution with $a = 1$~$\micron$ and $\chi_{\rm{sd}} = 10^{-4}$.
}
\label{fig-cooldte}
\end{figure}

If dust is present, its thermal emission will also contribute to the 
cooling of the plasma.  It is easy to show that the shear layer is optically thin to the dust 
thermal emission, and thus the dust radiative cooling rate is
\be
\Lambda_{\rm{dr}} = 4 \pi n_{\rm{d}} a^{2} \langle Q_{\rm{abs}} \rangle
\sigma_{\rm{sb}} T^{4}_{\rm{d}},
\ee
(Draine 2011) where $\sigma_{\rm{sb}}$ is the Stefan-Boltzmann constant,
$T_{\rm{d}}$ is the dust temperature, and $\langle Q_{\rm{abs}} \rangle$ is the dust 
grains' Planck-averaged emission efficiency with
\be
\langle Q_{\rm{abs}} \rangle \approx  
0.13 \left( \frac{a}{1\,\micron} \right)
\left( \frac{T}{100\,\rm{K}} \right)^{2} ~~~~~ (\rm{silicates}).
\ee
The cooling rate due to dust thermal emission is plotted as a 
function of $R$ in the disk in Figure~\ref{fig-cooldte}, assuming that the
dust grains are a single-size distribution with $a=1$~\micron\ and 
$\chi_{\rm{sd}}=10^{-4}$.

\subsubsection{Self Heating Rates}
\label{subsec-selfheatingrates}

The heating rate due to viscous dissipation 
can be calculated by evaluating Equation~\ref{eq-gvisc} using the velocity 
solution given in Equations~\ref{eq-sol-vx}--\ref{eq-sol-vy}. For the 
shear flow considered 
in Section~\ref{sec-bperp}, the components
of the viscous stress tensor become
\be
\tau_{xy} = \tau_{yx} = 0,
\ee
\be
\tau_{yz} = \tau_{zy} = \alpha \frac{d v_{y}}{dz},
\ee
\be
\tau_{zx} = \tau_{xz} = \alpha \frac{d v_{x}}{dz},
\ee
and
\be
\tau_{xx} = \tau_{yy} = \tau_{zz} = 0. 
\ee
Substituting these stresses into Equation~($\ref{eq-gvisc}$) gives
\be
\Gamma_{\rm{visc}} = \alpha \left[ \left( \frac{d v_{x}}{dz} \right)^{2} + \left( \frac{d v_{y}}{dz} \right)^2 \right].
\ee
Using the solutions found in Section~\ref{sec-bperp} to calculate the derivatives of $v_{x}$ and $v_{y}$ and
substituting them into the equation above gives
\be
\Gamma_{\rm{visc}} = \frac{ v_{0}^{2} B_{0}^{2}}{4 \pi D \eta} e^{-2 k_{\rm{R}} \xi} .
\label{eq-gvis}
\ee

In addition to the viscous dissipation heating rate, the heating rate in
the shear flow due to friction between the ion and neutral fluids can also be calculated.
Substituting the assumed forms of the plasma velocity, magnetic, and electric fields
described in Section~\ref{sec-bperp} into Equation~($\ref{eq-Wardleiondriftv}$), the ion-neutral velocity
differences are found to be
\be
v_{\rm{i} \it{x}} - v_{x} = \frac{\beta_{\rm{i}}}{1 + \beta_{\rm{i}}^{2}} 
\left( \beta_{\rm{i}} v_{0} - \beta_{\rm{i}} v_{x} + v_{y} \right) ,
\ee
\be
v_{\rm{i} \it{y}} - v_{y} = \frac{\beta_{\rm{i}}}{1 + \beta_{\rm{i}}^{2}} 
\left( v_{0} - v_{x} - \beta_{\rm{i}}v_{y} \right) ,
\ee
and
\be
v_{\rm{i} \it{z}} - v_{z} \approx 0.
\ee
The ion-neutral frictional heating rate is
\be
\Gamma_{\rm{in}} \approx \frac{n_{\rm{i}} n_{\rm{n}} 
\langle \sigma v \rangle_{\rm{in}}  \mu_{\rm{in}} v_{0}^{2}
\beta_{\rm{i}}^{2}}{1 + \beta_{\rm{i}}^{2}} e^{-2 k_{\rm{R}} \xi}.
\label{eq-gin}
\ee

Upon inspection of Equations~(\ref{eq-gvis}) and (\ref{eq-gin}), it is clear
that the heating rates due to viscous dissipation and ion-neutral scattering
depend strongly on the free-stream velocity of the plasma $v_{0}$ and are
largest at the asteroid surface ($\xi=0$).  In order to assess the importance 
of self heating, we consider flows with $v_{0}=1$\,km/s and $v_{0}=5$\,km/s, 
which could correspond to flows driven by the relative motions between the
asteroid and the gas on an eccentric orbit (Morris et al. 2012) or a passing
shock wave in the disk.  As done in the previous section, we present two cases: 
the dust-free case, and the case where $\chisd=10^{-4}$ with $a=1$\,$\mu$m.

\begin{figure}
\figurenum{26}
\plotone{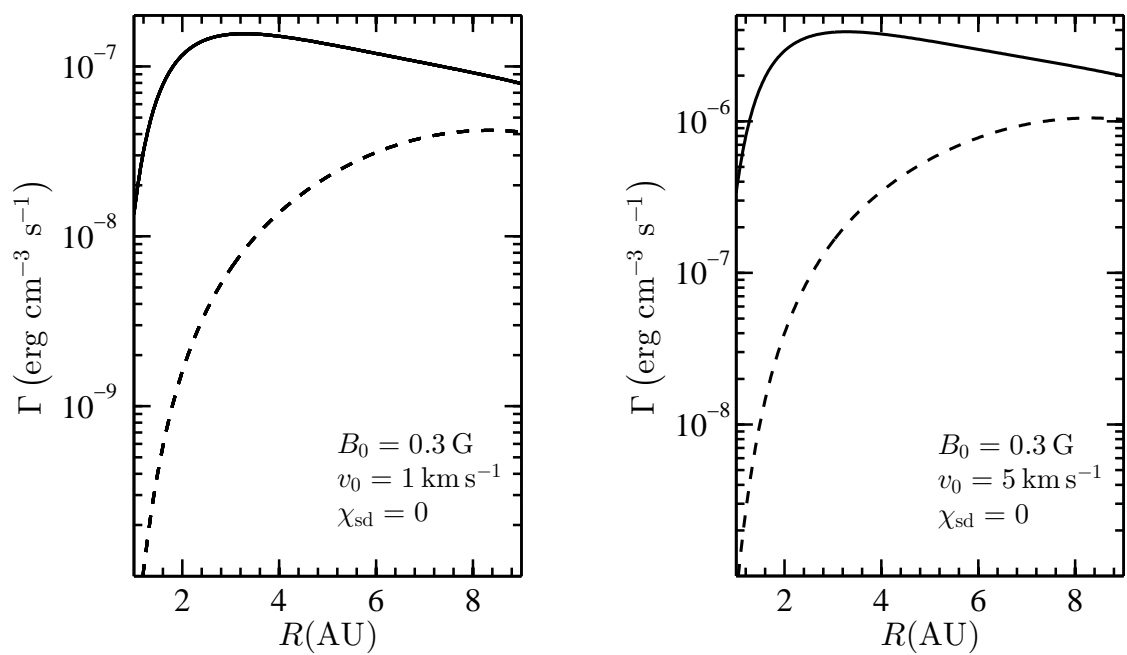}
\caption{
The volumetric heating rate $\Gamma$ due to viscous dissipation (solid curve) and
ion-neutral scattering (dashed curve) in the plasma at the body surface ($\xi=0$)
are plotted versus distance from the central star.  The values of the free-stream
velocity, ambient magnetic field, and small dust abundance are indicated.
}
\label{fig-gvdinnd}
\end{figure}

Figure~\ref{fig-gvdinnd} describes the heating rates due to viscous dissipation
and ion-neutral scattering at the body surface $\xi=0$ as a function of $R$ in
a dust-free plasma.  The left and right panels describe flows with $v_{0}=1$\,km/s
and $v_{0}=5$\,km/s respectively, and assume $B_{0}=0.3$\,G.  For both flows we
note that the heating rate due to viscous dissipation dominates that of ion-neutral
scattering at all distances $R$ of interest.  Comparing the heating rates (Figure~\ref{fig-gvdinnd})
to the radiative cooling rates when gas-phase water is present in the abundance
$n_{\rm{H_{2}O}}/n_{\rm{H}} = 1.1\times10^{-4}$
(Figure~\ref{fig-rcoolnd}), we observe that for flows with $v_{0}=1$\,km/s the 
cooling rate exceeds the heating rate for $R$~$\ltsim$~4\,AU.  For 
6~$\ltsim$~$R$~$\ltsim$~8\,AU, the heating and cooling rates become comparable.  
If the free-stream flow velocity is increased to $v_{0}=5$\,km/s, 
the cooling 
rate still dominates the heating rate for $R$~$\ltsim$~$2$\,AU.  Once $R \approx 
3$\,AU, the heating and cooling rates become comparable.  For 
$R$~$\gtsim$~$4$\,AU, we find that the heating rates start to exceed the cooling
rate.  Thus we conclude that self heating of the plasma is insignificant for flows
with $v_{0}=1$\,km/s at $R$~$\ltsim$~6\,AU.  However for faster flows with free-stream
velocities $\gtsim$~$5$\,km/s, significant self heating of the plasma to temperatures 
above ambient (Equation~\ref{eq-tdisk}) may become important, particularly at 
$R$~$\gtsim$~3\,AU.  If gas-phase water is absent in the disk, the heating rate is found
to exceed the cooling rate (Figure~\ref{fig-rcoolnwnd}) at all $R$ for both $v_{0}=1$ 
and $5$\,km/s, implying that significant self heating of the plasma would occur.

\begin{figure}
\figurenum{27}
\plotone{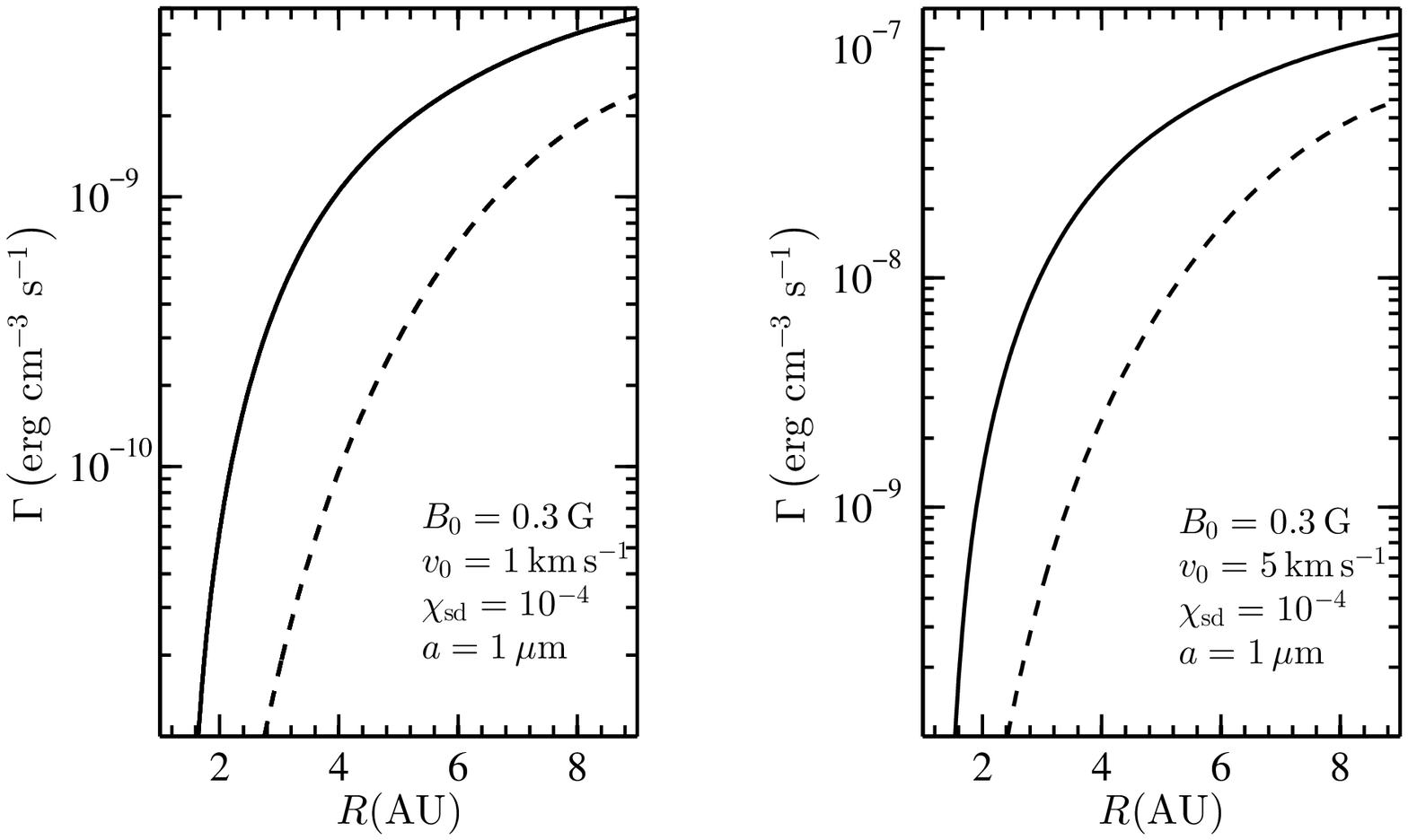}
\caption{
As in Figure~\ref{fig-gvdinnd} but
for $\chisd=10^{-4}$.
The dust is a single-size distribution
with $a=1$\,\micron.
}
\label{fig-gvdinyd}
\end{figure}

Figure~\ref{fig-gvdinyd} shows the analogous heating rates in the case were single-sized
small dust grains are present in the disk with an abundance $\chisd=10^{-4}$ and radius 
$a=1$\,$\mu$m.  Once again we note that viscous dissipation dominates the heating
at all considered values of $R$. Comparing the heating rates 
(Figure~\ref{fig-gvdinyd}) to the cooling rates by dust thermal emission
(Figure~\ref{fig-cooldte}) and molecular line emission (Figures~\ref{fig-rcoolyd}) when a standard abundance of gas-phase water is present in the disk, we find that the cooling rate exceeds the heating rate at all $R$ for $v_{0} = 1$~km/s.  Under these conditions, dust emission dominates the total radiative cooling rate of the plasma for $R$~$\ltsim$~$4$~AU, but becomes insignificant for $R$~$\gtsim$~$8$~AU.  For faster flows with $v_{0}=5$~km/s, the cooling rate is again found to exceed the heating rate for $R$~$\ltsim$~$8$~AU.  In this case dust thermal emission again dominates the cooling for $R$~$\ltsim$~$3$~AU, but becomes smaller than the rate of cooling by molecular line emission for $R$~$\gtsim$~$4$~AU.  If gas-phase water is absent, the cooling rate becomes dominated by thermal dust emission at all $R$ of interest for both $v_{0} = 1$ and $5$~km/s.  Comparing the heating rates to the cooling rate due to dust emission, we find that the cooling rate exceeds the heating rate everywhere for $v_{0} = 1$~km/s.  For faster flows with $v_{0} = 5$~km/s, the cooling rate still exceeds the heating rate for $R$~$\ltsim$~$4$~AU.  However for $R$~$\gtsim$~$6$~AU, the heating rate is found to exceed the cooling rate, implying that significant heating of the plasma would occur.


\subsection{Chiral Asymmetry and the Hall Effect}
\label{sec-chiral}

\begin{figure}
\figurenum{28}
\plotone{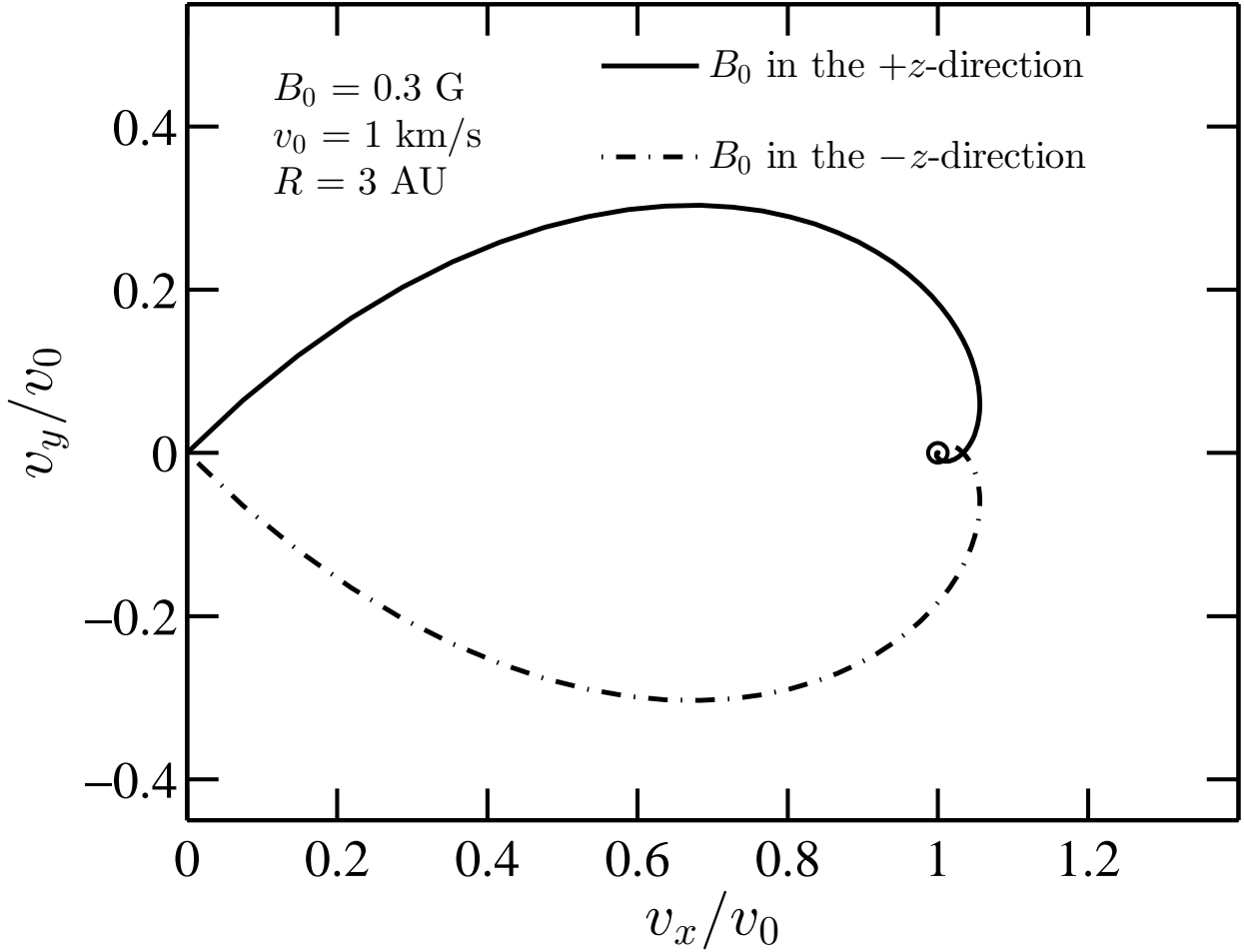}
\caption{
The neutral bulk plasma velocity profile for the flow calculated in 
Section~\ref{sec-bperp}
is plotted in the $x$-$y$ plane
for two different orientations of the magnetic field with $\vecv_{0}\,=\,v_{0}\,\hat{\vecx}$ and the
indicated parameters.  The solid and dot dashed curves correspond to
$\vecB_{0} = B_{0} \hat{\vecz}$ and $\vecB_{0} = -B_{0} \hat{\vecz}$ respectively.
Each point on the curves represents the velocity at a certain height above the body surface,
with the black circle at the point (1,0) indicating the edge of the shear
layer ($\xi = \pm \infty$, $\vecv = \vecv_{0}$), and the intercept at (0,0) indicating
the body surface ($\xi = 0$, $\vecv = 0$).
}
\label{fig-chiralvfield}
\end{figure}

As noted by Wardle\ (2007), the Hall effect is capable of imparting a handedness to 
multifluid MHD.  Using our velocity solutions calculated in Section~\ref{sec-bperp}, we 
demonstrate this feature and the resulting chiral asymmetry in the plasma flow.
Outside the shear layer the neutral velocity
is just the free-stream velocity $\vecv_{0}$.
Inside the shear layer, however, the velocity vector rotates.
As shown in Figure~\ref{fig-chiralvfield}, if the unperturbed magnetic field $\vecB_{0}$
points in the $+z$-direction the velocity rotates counterclockwise in the $x$-$y$ plane and acquires a component $v_{y}$, which is positive
near the body surface.  However, if $\vecB_{0}$ points in the $-z$-direction, the velocity
rotates clockwise through the same angle in the $x$-$y$ plane and acquires a component $v_{y}$,
which is negative close to the body.  
As is clearly seen in Figure~$\ref{fig-chiralvfield}$, these rotations define a chiral asymmetry 
in the neutral velocity profile which depends on the relative orientation of $\vecv_{0}$ and $\vecB_{0}$.  

The observed chiral behavior of the flow can be explained by considering the motions of the neutral and charged particles.
Since electromagnetic forces do not act directly on the neutral (bulk) plasma and the
electron-neutral collisional force is neglected, the only force that can be responsible for producing
the neutral velocity component $v_{y}$ is drag between ions and neutral particles.
In the absence of the Hall effect, the ions and electrons flow together in the $x$-direction.  
However if the Hall effect is present, the Lorentz force causes the ions and electrons to move relative to
one another in the $x$-direction.  Assuming macroscopic charge neutrality ($n_{\rm{i}} = n_{\rm{e}}$) in the flow, 
this relative motion between the ions and electrons defines a component of the current density in the $x$-direction given by
\be
J_{x} \equiv e n_{i} \left( v_{\rm{i} \it{x}} - v_{\rm{e} \it{x}} \right).
\ee  
This component of the current density is associated with a gradient in the $y$-component of the magnetic field through \Ampere 's Law
\be
J_{x}  = - \frac{c}{4 \pi } \frac{d B_{y}}{dz},
\label{eq-xcurrentchiraldiscussion} 
\ee
which produces a $y$-component of the magnetic tension force
\be
\left[\frac{1}{4 \pi} \left( \vecB \cdot \nabla \right) \vecB \right]_{y} 
\approx \frac{B_{0}}{4 \pi} \frac{d B_{y}}{dz}.
\ee
For weakly ionized plasmas the ion-neutral drag force can generally be written as
\be
\rho \rho_{\rm{i}} \gamma_{\rm{i}} \left( \vecv_{\rm{i}} - \vecv \right) \approx \frac{\vecJ}{c} \times \vecB,
\label{eq-balbus}
\ee 
(e.g., Balbus 2011), where $\gamma_{\rm i} \equiv 1/\left(\rhoi\,\tauin\right)$.
Substituting Equation~($\ref{eq-xcurrentchiraldiscussion}$) into Equation~($\ref{eq-balbus}$) and
looking at the $y$-component of the equation gives
\be
\frac{B_{0}}{4 \pi} \frac{dB_{y}}{dz} - \rho \rho_{\rm{i}} \gamma_{\rm{i}} \left( v_{\rm{i} \it{y}} - v_{y} \right) \approx 0 ,
\ee
which shows that as the magnetic tension force pushes the ions in the $y$-direction, they drag the
neutral particles with them.  Thus, when the free-stream velocity of the neutral particles
($\vecv_{0}$) points in the $+x$-direction and the undisturbed magnetic field ($\vecB_{0}$) points in
the $+z$-direction, magnetic tension due to the Hall effect forces the ions to move and drag the
neutral particles with them in the $+y$-direction, causing the flow to rotate in the counterclockwise
direction; while if the direction of $\vecB_{0}$ is reversed and now points in the $-z$-direction, magnetic tension due again to the Hall effect forces the ions and neutral particles to move in the $-y$-direction,
causing the flow to rotate in the clockwise direction.  As shown by the solutions in Section~\ref{sec-bperp}, this component
of the neutral velocity outside of the plane defined by the directions of $\vecv_{0}$ and $\vecB_{0}$ can be
large and thus cannot necessarily be treated as a perturbation in MHD planar shear flow analysis. 
 
\subsection{Electrodynamic Heating}
\label{sec-edheat}

Although the motional electric field decreases to zero at the surface of a large body, Section~\ref{sec-bperp}
shows that it is nevertheless possible to have interior electric fields $\sim\vecEzero$, i.e., 
electric fields comparable to those invoked in classical induction heating.
It is therefore interesting to inquire whether significant heating may ensue.
To emphasize the fact that the interior electric field is due ultimately to the
dynamical interaction of the body with surrounding plasma, we will refer to
the heating process as ``electrodynamic heating.''

The benchmark for electrodynamic heating is the rate of heating by short-lived radionuclides (SLRs).
We will use the rate of heating by $^{26}$Al (Urey 1955), which  is known to be the 
dominant species (e.g., Ghosh et al.\ 2006). Recent work on $^{60}$Fe suggests it may 
also
contribute to within an order of magnitude of $^{26}$Al (Castillo-Rogez et al.\ 
2007); however we ignore its contribution here because we are only interested in
making order of magnitude estimates.
The volumetric heating rate is generally given by
\be
\Gamma_{26}(t) = \Gamma_{26,0}\,\exp\left(-0.966t/{\rm Myr}\right),
\label{eq-gam26}
\ee
where $t$ is the time elapsed since CAI formation and
\be
\Gamma_{26,0} = 5.33\times 10^{-4}\,
\left(\frac{\rho_{\rm{s}}}{3000\,{\rm kg}\,{\rm m}^{-3}}\right)\,
\left(\frac{x_{\rm Al}}{1\,{\rm wt}\%}\right)\,
\left(\frac{f_{26,0}}{5\times 10^{-5}}\right)  
~~~\hbox{W\,m$^{-3}$},
\label{eq-alscale}
\ee
where
$\rho_{\rm{s}}$ is the density of the solid material,
$x_{\rm Al}$ is the total (including all isotopes) aluminum abundance,
and $f_{26,0}$ is fraction of all aluminum in $^{26}$Al at the time of CAI formation
(Lee et al.\ 1976; MacPherson et al.\ 1995, 2010).
This expression assumes that the half-life of $^{26}$Al is $7.17\times 10^5$\,yr
and that each decay delivers 3.12\,MeV of thermal energy to the solid (Castillo-Rogez et al.\ 2009).  

Also of interest is the total thermal energy deposited in the body per unit volume, $\Delta E_{26}$,
obtained by integrating expression (\ref{eq-gam26}) over time.
If a body is assumed to accrete instantaneously at time $t=t_0$, then
\be
\Delta E_{26} = 2 \times 10^{10}\,
\left(\frac{\rho_{\rm{s}}}{3000\,{\rm kg}\,{\rm m}^{-3}}\right)\,
\left(\frac{x_{\rm Al}}{1\,{\rm wt}\%}\right)\,
\left(\frac{f_{26,0}}{5\times 10^{-5}}\right)\,
\exp\left(-0.966t_0/{\rm Myr}\right)
~~~\hbox{J\,m$^{-3}$}.
\ee
In comparing different heating mechanisms, $\Delta E_{26}$ is probably the more
useful benchmark, because the time scale for $^{26}$Al to completely decay is
much shorter than the time scale for thermal energy to diffuse through the body.
For a body of size $L_{\rm{body}}$ the diffusion time is
\be
\tau_{\rm dif} \equiv \frac{L_{\rm{body}}^2}{\kappa} > 300\,
\left(\frac{L_{\rm{body}}}{100\,{\rm km}}\right)\,{\rm Myr},
\ee
where $\kappa < 10^{-6}$\,m$^2$\,s$^{-1}$ is the thermal diffusivity
and the numerical value describes a variety of chondritic materials (Yomogida \& Matsui 1983).
Of course,
thermal diffusion is always important in a thin layer just below the body surface,
however the thickness of this layer is $\ll L_{\rm{body}}$ for times $\ll\tau_{\rm dif}$ 
(e.g., see Figure~4 of Ghosh \& McSween 1998).

For comparison, the rate of electrodynamic heating is
\be
\Gamma_{\rm ed} < \sigma_{\rm b}\,E_0^2,
\label{eq-edheatupperlimit}
\ee
where the inequality means that this is an upper limit for reasons discussed below.
If we assume that the temperature dependence of the conductivity obeys the Arrhenius law given in Equation~(\ref{eq-sigarr}),
then
\be
\Gamma_{\rm ed} < 1.11\times 10^{-2}\,
\left(\frac{\sigma_0}{10^{12}\,{\rm s}^{-1}}\right)\,
\left(\frac{v_0}{{\rm km\,s}^{-1}}\right)^2\,
\left(\frac{B_0}{0.1\,{\rm G}}\right)^2\,
\exp\left(-1161\,{\rm K}/T\right)
~~~\hbox{W\,m$^{-3}$}.
\label{eq-gamed}
\ee
Equation~(\ref{eq-gamed}) assumes an activation energy, $E_{\rm a}=0.1$\,eV, and
$\sigma_0$ consistent with measurements for a few chondritic meteorites (Schwerer et al.\ 1971; Brecher 1973);
these are plausible values but highly uncertain.
The total energy deposited in the body by electrodynamic heating in a time $\Delta t$ is just
\be
\Delta E_{\rm ed} < 4 \times 10^{11}\,
\left(\frac{\sigma_0}{10^{12}\,{\rm s}^{-1}}\right)\,
\left(\frac{v_0}{{\rm km\,s}^{-1}}\right)^2\,
\left(\frac{B_0}{0.1\,{\rm G}}\right)^2\,
\left(\frac{\Delta t}{{\rm Myr}}\right)\,
\exp\left(-1161\,{\rm K}/T\right)
~~~\hbox{J\,m$^{-3}$}.
\label{eq-deltaeed}
\ee

\begin{figure}
\figurenum{29}
\plotone{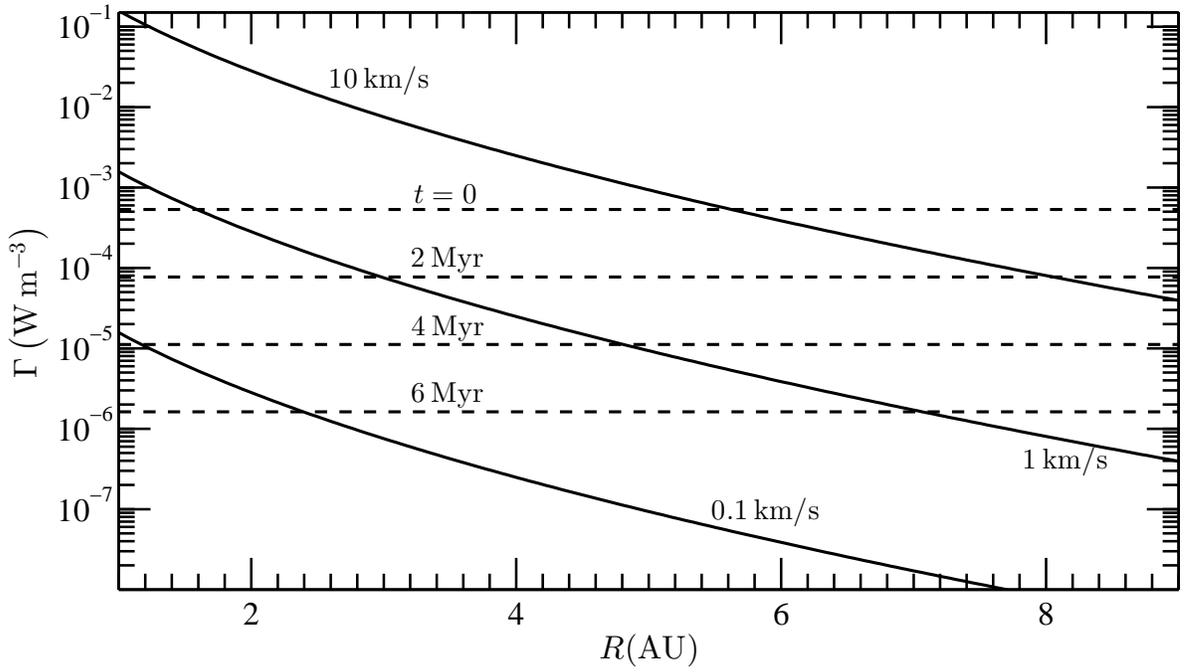}
\caption{
Heating rates vs.$\!$ heliocentric distance.  Dashed curves: 
$^{26}$Al heating, where the labels indicate the time since CAI 
formation.  Solid curves:  electrodynamic heating, where the labels 
indicate $v_0$.  The electrodynamic heating rate depends on heliocentric 
distance via the temperature dependence of the conductivity (see text).
}
\label{fig-astgamma}
\end{figure}

\begin{figure}
\figurenum{30}
\plotone{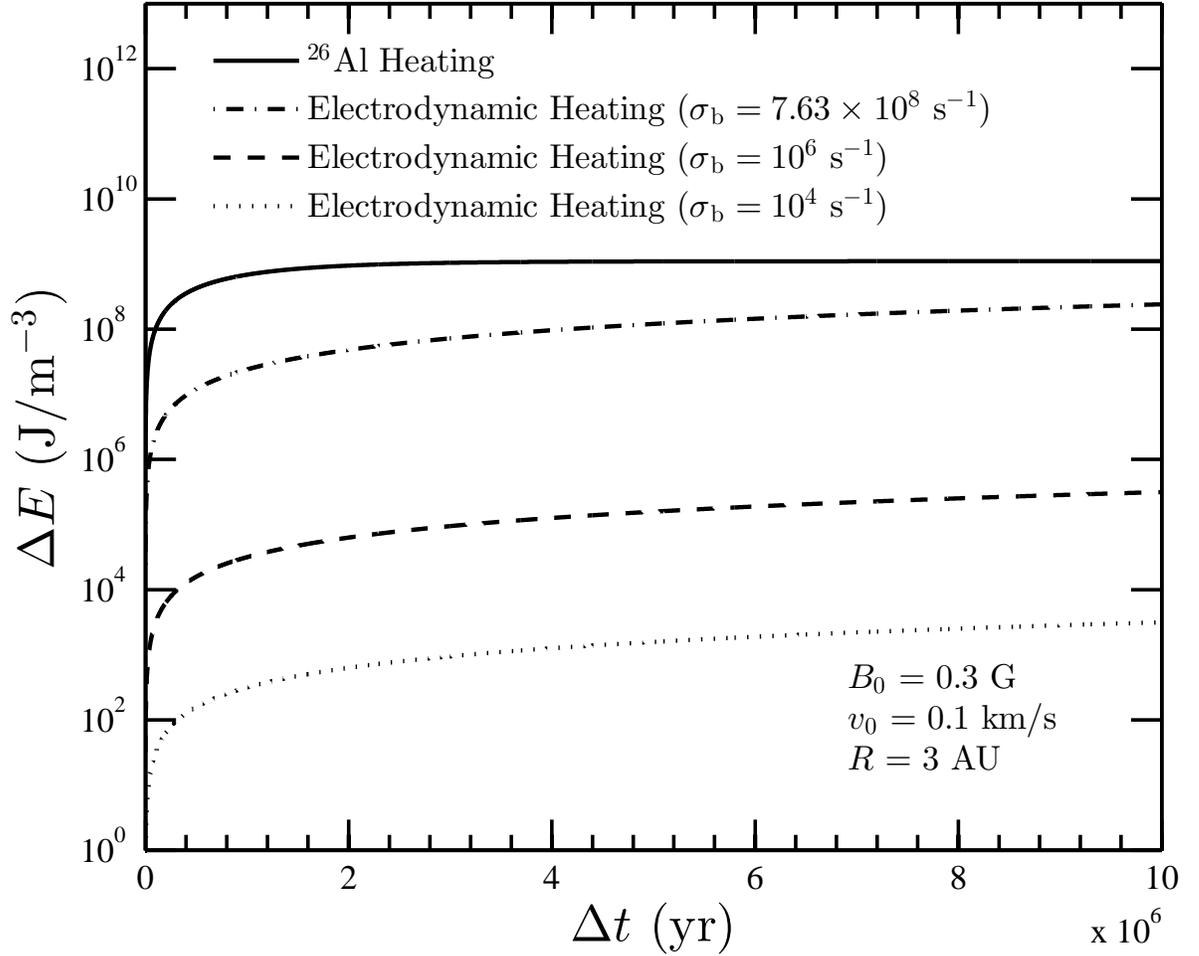}
\caption{ 
The total amount of energy deposited in our model asteroid body per volume by both 
$^{26}$Al and electrodynamic heating driven by the orbital motions 
($v_{0} = 0.1$\, km/s, $B_{0} = 0.3$\,G) of the asteroid is plotted versus time 
relative to the time of accretion.  Curves are shown for several different plausible 
values of the asteroid's electrical conductivity.
}
\label{fig-energydep}
\end{figure}

\begin{figure}
\figurenum{31}
\plotone{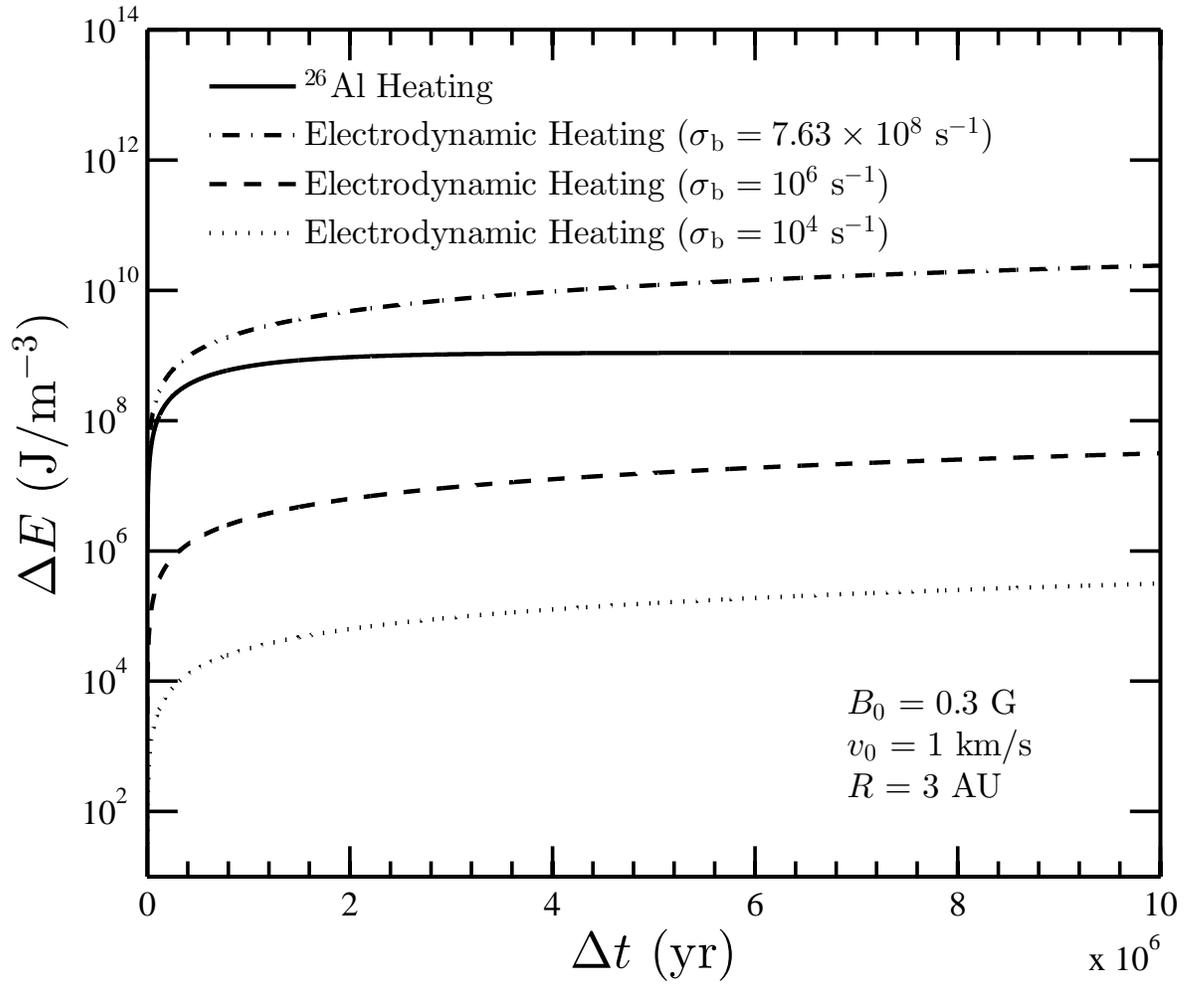}
\caption{ 
As in Figure~\ref{fig-energydep} but with $v_{0}=1$\,km/s.
}
\label{fig-energydepv1}
\end{figure}

The heating rates due to both the decay of $^{26}$Al and electrodynamic heating are plotted in
Figure~\ref{fig-astgamma}, with the unperturbed magnetic field strength once again taken to be 
$B_{0} = 0.3$\,G.  Several electrodynamic heating curves are shown corresponding to plasma flows 
driven by the body's orbital motions ($v_{0} \sim 0.1$ km/s), passing shock waves ($v_{0} \sim 10$ km/s), 
and an intermediate value ($v_{0} \sim 1$ km/s).  
Although larger electrodynamic heating rates are observed for faster flows such as the ones driven 
by shock waves, it is unlikely that these flows could produce heating inside the body comparable 
to $^{26}$Al because they would need to operate for time periods of order $\Delta t \ \gtsim \ 10^{3}$ yr, 

A more likely scenario is that electrodynamic heating comparable to $^{26}$Al is driven by 
the orbital motions of the asteroids relative to the gas because this mechanism should 
operate for a much longer time period of order $\Delta t \sim 10$ Myr.   
This scenario can be tested quantitatively by comparing the total energy deposited by orbital 
motion driven electrodynamic heating to $^{26}$Al heating in a simple model asteroid.  
Assuming that the model asteroid body accreted instantaneously at a time $t_{0}$ = 2.85 Myr 
after the the formation of the CAIs (Ghosh \& McSween 1998) and has the density, $x_{\rm{Al}}$, 
and $f_{26,0}$ values given in Equation~(\ref{eq-alscale}), the total energy deposited by each mechanism 
is plotted as a function of time relative to the $t_{0}$ in Figures~\ref{fig-energydep}--\ref{fig-energydepv1}.  We consider two values of the free-stream
velocity: $v_{0}=0.1$\,km corresponding to the relative velocity calculated by
Weidenschilling (1997a), and $v_{0}=1$\,km/s corresponding to relative motion between
the gas and a body on an eccentric orbit (Morris et al.\ 2012).
For simplicity we neglect the temperature dependence of the body's electrical conductivity and instead 
use a range of constant conductivity values, the largest of which is determined by substituting the 
ambient plasma temperature at $R=3$ AU (Equation~[\ref{eq-tdisk}]) into Equation~(\ref{eq-sigarr}).  
As the plot shows, electrodynamic heating is only competitive with $^{26}$Al when the body's electrical 
conductivity is of order $\sim 10^{9}$\,s$^{-1}$ for $v_{0}=0.1$\,km/s, which is a high but plausible value for some asteroids.  For $v_{0}=1$\,km/s, comparable 
electrodynamic heating may be possible if the electrical conductivity of the body is
of order $\sim10^{7}$\,s$^{-1}$, which is probably more plausible, but still on the high end. 
For bodies with smaller electrical conductivities, electrodynamic heating will not be significant.

It is very important to stress that the above analysis only describes possible \emph{upper limits} 
on electrodynamic heating.  Since the current which drives electrodynamic heating must leave the body, 
pass through a magnetized plasma sheath at each body surface, and close through the  bulk plasma, the 
electrodynamic heating rate will not generally be given by Equation~(\ref{eq-edheatupperlimit}), but instead 
depend on the entire current circuit.  In order to determine the limiting effects of the plasma on the 
current, the dynamics of the ions and
electrons in the magnetic sheaths and the resistive properties of the specific path taken by the current through the bulk
plasma must be taken into account.  The dynamics of charged particles in magnetic sheaths have been investigated
numerically using two-fluid (ions and electrons) models for simple geometries (Tskhakaya et al.\ 2005; 
Pandey et al.\ 2008 and references therein), and similar calculations for either side of the body are required
in order to establish the effects these sheaths would have on the total current density flowing through the body.
Limitations on the current due to both sheath and bulk plasma effects in a solar wind have been calculated by
Srnka (1975) for the induction heating mechanism described in Sonett et al.\ (1970).
However we leave as future work the analogous calculation for weakly ionized protoplanetary disks.
Once the entire current circuit can be completely described, more accurate heating rates and temperature profiles produced
inside the body by MHD shear flows can be calculated.

\section{Summary}
\label{sec-summ}

In this paper we reexamined the ``unipolar induction'' heating mechanism developed by Sonett et al. for primitive bodies in weakly ionized protoplanetary disks.
We were motivated by the fact that some asteroids once experienced
thermal conditions that were conducive to a rich prebiotic chemistry,
including the production of amino acids with chiral asymmetries of
the type observed in terrestrial proteins.
Understanding whether and how primitive bodies are/were heated is
therefore an important issue for astrobiology.

According to current wisdom, asteroids in the solar nebula
were heated by the radioactive decay of short-lived radionuclides (SLRs),
principally $^{26}$Al.
However the SLR model requires a finely tuned dependence of asteroid accretion times on heliocentric distance, which have not yet been shown to produce both the correct mass distribution and heating gradient across the asteroid belt. 
Furthermore, the existence of other heating mechanisms may be crucial
for the viability of prebiotic chemistry in other disks, where the
probability of the SLR scenario is only $\sim  10^{-3}$--$10^{-2}$ (Ouellette et al. 2010).  For these reasons it seems appropriate to explore other heating mechanisms. 

Our principal results are:

\begin{enumerate}

\item
We pointed out a subtle conceptual error in the induction heating
mechanism as originally conceived, in consequence of which
the electric field inside the body was calculated incorrectly.

\item
We described the steps required to calculate the electric field
correctly for bodies of arbitrary shape moving through weakly ionized plasmas
of the type expected in protoplanetary disks including dust.

\item
We presented a highly idealized example which demonstrated
that it is possible for the electric field inside the body to vanish.
Under these circumstances there is no heating.

\item
We presented another highly idealized example which demonstrated that
large electric fields are possible, in the sense that \vecE\ is
comparable to the field predicted by classical induction.

\item
We demonstrated that the main effect of micron-sized dust grains on the flow and electric fields is to increase the thickness of the shear layer
from $\sim$~10 to $\sim$~100 km.

\item
We assessed the possible importance of heating by viscous dissipation and ion-neutral 
scattering in the shear flow.  If there is no small dust in the disk, we find that 
heating
may be significant for flows with $v_{0}$~$\gtsim$~5\,km/s at 
$R$~$\gtsim$~3\,AU when gas-phase water is present in the disk, and
$v_{0}$~$\gtsim$~1\,km/s at all $R$ values considered when water is absent. 
When micron-sized dust is present in the disk with an abundance of $\chisd=10^{-4}$,
we find that heating is insignificant for $v_{0}$~$\ltsim$~$5$~km/s
at $R$~$\ltsim$~$8$~AU when gas-phase water is present and at
$R$~$\ltsim$~$6$~AU when gas-phase water is absent.

\item
We demonstrated an interesting property of shear flows around
primitive bodies in protoplanetary disks, namely, that the velocity field is
chirally asymmetric.  As pointed out by Wardle\ (2007), the existence of the 
asymmetry is due to the Hall effect
and its sense depends on the orientation of the ambient magnetic field relative
to the primitive body's velocity.  The {\it significance}\/
of the asymmetry, if any exists, remains to be determined.

\item
We discovered a new ``electrodynamic heating'' mechanism and quantitatively compared its heating rate with the
rate of heating produced by the decay of $^{26}$Al.  We found that electrodynamic heating can only be
competitive with $^{26}$Al heating in asteroids with relatively high but plausible electrical
conductivities of order $\sim 10^{7}$-$10^{9}$~$\rm{s^{-1}}$ depending on the value of $v_{0}$; however we stress that this is an upper bound on the heating and listed a series
of problems which must be solved in order to assert this unambiguously.

\end{enumerate}

\acknowledgments

The authors are grateful to the anonymous referee for a thorough,
constructive report which greatly improved the paper.
We also acknowledge helpful comments by Mike Gaffey,
Satoshi Okuzumi, Heather Watson, and Stuart Weidenschilling.
This work was supported by the New York Center for Astrobiology,
a member of the NASA Astrobiology Institute, under grant
{\#}NNA09DA80A.

\appendix
\section{Solution for Perpendicular Field Geometry}
\label{app-soln}

\subsection{Formal Solution}

Equations~(\ref{eq-d-inductionx})--(\ref{eq-d-momentumy}) are a set of coupled, linear ODEs
for the velocity and magnetic field derivatives.
In matrix form they become
\be
\vecy^{\prime} = \matrA\vdot\vecy,
\label{eq-odey}
\ee
where
\be
\vecy \equiv \left[\bxp,\byp,\uxp,\uyp\right]^{t},
\ee
is the vector of unknowns,
\be
\matrA \equiv \left(
\begin{array}{cccc}
  0    &    0   & h_{11} & h_{12}  \\
  0    &    0   & h_{21} & h_{22}  \\
-\eps  &    0   &    0   &    0    \\
  0    & -\eps  &    0   &    0    \\
\end{array}
\right),
\ee
is the coupling matrix,
$\eps\equiv 1/4\pi$,
\be
\matrh \equiv \frac{1}{D^{2}} \left(
\begin{array}{cccc}
-\left(1-\lamh\right) &         \lamh         \\
       -\lamh         & -\left(1-\lamh\right) \\
\end{array}
\right),
\ee
and
\be
D^2 = \left(1-\lamh\right)^2 + \lamh^2 .
\ee

The solution of Equation~(\ref{eq-odey}) is
\be
\vecy = \sum_{m}\ C_m\,\vecphi_{m}\,e^{k_m \zbar},
\label{eq-formal}
\ee
where $\left\{k_m\right\}$ are the four eigenvalues of \matrA,
$\left\{\vecphi_{m}\right\}$ are the corresponding eigenvectors,
and the expansion coefficients $\left\{C_m\right\}$ are determined by the boundary
conditions.
Two of the eigenvalues have positive real parts and are
excluded by the boundary conditions at $z=+\infty$.
The two physically admissible eigenvalues are
\be
\km = -\kR - i\kI
\ee
and
\be
\kp = -\kR + i\kI
\ee
where \kR\ and \kI\ are functions of \lamh\ defined in eqs.~(\ref{eq-krdef})--(\ref{eq-kidef}).
The corresponding eigenvectors are
\be
\vecphi_{-} = \left[ -4\pi\km,  4\pi i\km,  1,  -i  \right]^{t}
\ee
and
\be
\vecphi_{+} = \left[ -4\pi\kp, -4\pi i\kp,  1,  i  \right]^{t}.
\ee
The formal solution therefore reduces to
\be
\vecy = \Cm\,\vecphi_{-}\,e^{\km\zbar} + \Cp\,\vecphi_{+}\,e^{\kp\zbar}.
\label{eq-twotermsoln}
\ee

\subsection{Boundary Conditions}

To apply the boundary conditions it is useful to rewrite
Equation~(\ref{eq-twotermsoln}) in the equivalent form
\be
\vecy = \Am\ \vecphi_{-}\ \exp\left(\km\xi\right)
      + \Ap\ \vecphi_{+}\ \exp\left(\kp\xi\right)
\ee
in terms of the redefined expansion coefficients \Am\ and \Ap\
and the dimensionless distance $\xi$ from the body surface
[see Equation~(\ref{eq-xidef})].
Writing out the third and fourth components gives
\be
\uxp = \Am\ \exp\left(\km\xi\right) 
     + \Ap\ \exp\left(\kp\xi\right)
\ee
and
\be
\uyp = -i\Am\ \exp\left(\km\xi\right)
       +i\Ap\ \exp\left(\kp\xi\right).
\ee
Integrating the last two equations and applying the boundary conditions at
$\xi=\infty$ yields the velocity solution,
\be
\ux = 1 + \frac{A_-}{k_-}\ \exp\left(\km\xi\right)
        + \frac{A_+}{k_+}\ \exp\left(\kp\xi\right)
\ee
and
\be
\uy = -\frac{iA_-}{k_-}\ \exp\left(\km\xi\right)
    +  \frac{iA_+}{k_+}\ \exp\left(\kp\xi\right).
\ee
The constants \Am\ and \Ap\ are determined by the boundary conditions
at $\xi=0$ with the result
\be
\ux = 1 - \frac{1}{2}\exp\left(\km\xi\right)
        - \frac{1}{2}\exp\left(\kp\xi\right)
\ee
\be
\uy = \frac{i}{2}\ \exp\left(\km\xi\right)
    - \frac{i}{2}\ \exp\left(\kp\xi\right).
\ee
Writing out $k_{\pm}$ in terms of \kR\ and \kI\ and taking the real parts
gives eqs.~(\ref{eq-sol-vx})--(\ref{eq-sol-vy}).

With \ux\ and \uy\ known, the magnetic field solution can be found by
integrating
\be
\bxp = 4\pi\uxpp
\ee
and
\be
\byp = 4\pi\uypp
\ee
and applying the boundary conditions, (\ref{eq-bxbc})
and (\ref{eq-bybc}).
Expressing the result in dimensional units
gives Equations~(\ref{eq-sol-Bx}) and (\ref{eq-sol-By}).



\clearpage
\begin{deluxetable}{clc}
\tablecaption{Parameters in Transport Coefficient Calculation\label{table-ions}}
\tablewidth{0pt}
\tablehead{
\colhead{Symbol}  & \colhead{Definition} & \colhead{Value(s)}
}
\startdata
$u_{\rm i}$      & Average\udg of mean ion thermal speed      & Equation~\ref{eq-uion}         \\
$\beta_{\rm{r}}$          & Average\udg recombination rate coefficient & Equation~\ref{eq-mgrr}         \\
$\zeta$          & Average\udg\udg ionization rate            & See text.                 \\
$L_{\rm xr}$     & Stellar x-ray luminosity                   & $2\times 10^{30}\,\ergsM$ \\
$\zeta_{\rm ra}$ & Rate of ionization by radionuclides        & $1.4\times 10^{-22}\,\sM$ \\
\enH             & Number density of H nuclei                 & From disk model           \\
$T$              & Gas temperature                            & From disk model           \\
\chisd           & Mass fraction, small grains                & $\le 10^{-4}$             \\
$D_{\rm f}$      & Dust fractal dimension                     & $3$                       \\
$\mu_{\rm i}$    & Ion mass/H atom mass                       & $24$                      \\
$S_{\rm i}$      & Sticking efficiency, ion-grain collision   & $1$                       \\
$S_{\rm e}$      & Sticking efficiency, e$^-$-grain collision & $0.3$                     \\
\rhos            & Bulk density of dust material              & $1.4$\,\gcmMMM              \\
\enddata
\tablenotetext{\dagger}{Over all ionic species. See text.}
\tablenotetext{\dagger\dagger}{Over all gas-phase species. See text.}
\end{deluxetable}

\clearpage
\begin{deluxetable}{clc}
\tablecaption{Parameters in Dzyurkevich \etal\ (2013) Model \label{table-Dz}}
\tablewidth{0pt}
\tablehead{
\colhead{Symbol}  & \colhead{Definition} & \colhead{Value(s)}
}
\startdata
$\chi_{\rm{sd}}$  & Abundance of small dust					  & $10^{-4}$ \\	   
$u_{\rm i}$      & Average\udg of mean ion thermal speed      & Equation~\ref{eq-uion}         \\
$\beta_{\rm{rMg^{+}}}$  & $\rm{Mg^{+}}$ recombination rate coefficient & $3 \times 10^{-11} / \sqrt{T}$~$\rm{cm^{3} s^{-1}}$         \\
$\beta_{\rm{rHCO^{+}}}$  & $\rm{HCO^{+}}$ recombination rate coefficient & $3 \times 10^{-6} / \sqrt{T}$~$\rm{cm^{3} s^{-1}}$         \\
$L_{\rm xr}$     & Stellar x-ray luminosity                   &  $10^{29}\,\ergsM$ \\
$\zeta_{\rm cr}$ & Rate of ionization by cosmic rays        & See 
Dzyurkevich \etal\ (2013) \\
$\zeta_{\rm xr}$ & Rate of ionization by x rays        & See 
Dzyurkevich \etal\ (2013) \\
$\zeta_{\rm ra}$ & Rate of ionization by radionuclides        & $7\times 10^{-19} \left(\chi_{\rm{sd}}/10^{-2}\right)\,\sM$ \\
\enH             & Number density of H nuclei                 & From disk model           \\
$T$              & Gas temperature                            & From disk model           \\
$a_{0}$          & Monomer radius                             & 0.1 \micron \\

$N$                & Number of monomers/aggregate               & $400$              \\
$D_{\rm f}$      & Dust fractal dimension                     & $2$                       \\
$\mu_{\rm mg}$   & Mg mass/H atom mass                       & $24$                      \\
$\mu_{\rm HCO}$  & HCO mass/H atom mass                      & $29$                      \\
$S_{\rm i}$      & Sticking efficiency, ion-grain collision   & $1$                       \\
$S_{\rm e}$      & Sticking efficiency, e$^-$-grain collision & $0.3$                     \\
\rhos            & Bulk density of dust material              & $1.4$\,\gcmMMM              \\
$n_{\rm{Mg}}$    & Total number density of Mg nuclei/$n_{\rm{H}}$& $2\times 10^{-9}$\udg\udg      \\
$B$          & Magnetic field strength                    & See 
Dzyurkevich \etal\ (2013) \\
\enddata
\tablenotetext{\dagger}{Over $\rm{Mg^{+}}$ and $\rm{HCO^{+}}$. See text.}
\tablenotetext{\dagger\dagger}{100 times greater than the
abundance quoted in Appendix B of Dzyurkevich \etal\ (2013).}
\end{deluxetable}


\clearpage

\clearpage

\clearpage

\clearpage

\clearpage

\clearpage

\clearpage

\clearpage

\clearpage

\clearpage

\clearpage

\clearpage

\clearpage

\clearpage

\clearpage

\clearpage

\clearpage

\clearpage

\clearpage

\clearpage

\clearpage

\clearpage

\clearpage

\clearpage

\clearpage

\clearpage

\clearpage

\clearpage

\clearpage

\clearpage

\clearpage

\end{document}